\documentclass[onecolumn,10pt]{article}
\usepackage[utf8]{inputenc}
\usepackage[T1]{fontenc}
\usepackage[margin=2cm]{geometry}
\usepackage{amsmath, amssymb, booktabs, algorithmicx, algpseudocode, tabularx, enumitem, longtable}

\usepackage[numbers,sort&compress]{natbib}

\usepackage{graphicx}%
\graphicspath{{./figures/}}
\usepackage{amsmath,amssymb,amsfonts,amsthm}%
\usepackage{xurl}
\usepackage{hyperref}
\usepackage{xcolor}%

\usepackage{tikz,lmodern}
\usetikzlibrary{positioning, shapes, arrows.meta, calc, decorations.pathreplacing}
\usetikzlibrary{matrix,arrows.meta, positioning, fit, backgrounds}

\usepackage{enumerate}
\usepackage{caption}
\usepackage{subcaption}

\usepackage{authblk}
\usepackage{placeins}

\newcommand{\ket}[1]{\ensuremath{\left|#1\right\rangle}} 
\newcommand{\bra}[1]{\ensuremath{\left\langle#1\right|}} 
\newcommand{\braket}[2]{\ensuremath{\left\langle#1|#2\right\rangle}}

\let\vec\mathbf
\newtheorem{proposition}{Proposition}[section]
\newtheorem{lemma}[proposition]{Lemma}
\theoremstyle{remark}
\newtheorem{remark}[proposition]{Remark}
\theoremstyle{plain}
\usepackage{orcidlink}
\providecommand{\orcidlink}[1]{}
\begin{document}
\title{From Theory to Practice: Analyzing Variational Quantum Power Method for Quantum Optimization of QUBO Problems}
\author{Ammar Daskin\orcidlink{0000-0002-1497-5031}\thanks{adaskin25@gmail.com}}
\affil{Department of Computer Engineering, Istanbul Medeniyet University, Uskudar, Istanbul, Turkiye}

\date{}

\maketitle
\begin{abstract}
The variational quantum power method (VQPM) replaces the coherent power iteration $(I+\mathcal{U})^k$ for a given unitary matrix $\mathcal{U}$ by one variational layer plus a controlled application of $\mathcal{U}$ per iteration, followed by measurements that set the angle values of the gates for the next iteration. The method also includes qubit locking, which fixes a qubit once its two outcome probabilities differ by more than a threshold to speed up convergence. In this paper, we give a comprehensive theoretical and numerical analysis of VQPM for quadratic unconstrained binary optimization (QUBO) problems, the problems also targeted by the Quantum Approximate Optimization Algorithm (QAOA), by studying its convergence, robustness, and qubit locking mechanisms.
In numerical experiments, we observe that VQPM without locking finds the optimum as often as QAOA with a good classical optimizer (e.g. L-BFGS-B or Adam). However, VQPM requires many more iterations (growing like $n^2$ or faster with the number of parameters $n$), which can be reduced by using the qubit locking mechanism. The qubit locking mechanism, in turn, lowers the accuracy considerably, reducing the success rate at $n=16$ from 0.79 to 0.58.

In addition, we note that when the first layer of VQPM is defined through single qubit gates (no entangling gates), the algorithm can be simulated classically: every measurement shot, and hence every qubit-locking decision, can be reproduced by classical sampling at $O(n^2)$ cost per shot. This classical equivalent of VQPM allows us to run it up to $n=100$ qubits and see the algorithmic performance for large-scale problems. In these large-scale experiments, the algorithmic success rate falls to 0.24 without locking and to at most 0.05 with locking. Since the locking depends on probability differences, on noisy devices, it requires careful tuning and many shots. We also note that simulated annealing gives better results than VQPM and QAOA, achieving a success rate near 0.98 on the instances whose optimum is known.  
\end{abstract}

\section{Introduction}


\begin{figure*}[ht]
    \centering
\begin{tikzpicture}[
    node distance=5mm,
    title/.style={font=\large\bfseries},
    circuit/.style={draw,thick,minimum width=3cm,minimum height=1.5cm},
    qgate/.style={draw,fill=blue!20,minimum size=6mm},
      Ugate/.style={draw,fill=blue!20,minimum size=24mm},
    measure/.style={fill=gray!30,shape=rectangle,minimum size=6mm},
    arrow/.style={-Stealth,thick},
    highlight/.style={fill=yellow!20,rounded corners}
]
\begin{scope}[local bounding box=left]
    \foreach \y in {0,1,2,3} {
        \draw[thick] (0,\y) -- (6,\y);
        \node[left] at (0,\y) {q\tiny{\y}};
    } 
    \node[qgate] (h0) at (1,3) {H};
    \node[qgate] (h1) at (5,3) {H};
    \node[measure] (m0) at (6,3) {M};
    \node[Ugate] (crz) at (4,1) {$\mathcal{U}$};
    \draw[thick]  (4,3) -- (crz);
    \foreach \y in {0,1,2} {
        \node[qgate,fill=blue!30] (ry\y) at (2,\y) {$R_y(\theta_{\y})$};
        \node[measure] (m\y) at (6,\y) {M};
    }
    \foreach \y in {0,1,2} {
        \node[draw,fill=red!20,scale=0.75] at ($(m\y)+(0.75,0)$) {lock?};
    }   
    \node[title,below=0.5cm of left] {VQPM Circuit};
\end{scope}

\begin{scope}[local bounding box=center,xshift=10cm, yshift=3cm]
    \node[title] (workflow) {VQPM Workflow with Locking};
    
    \node[draw,fill=blue!20,minimum width=2cm] (init) at (0,-1) {Initial State};
    \node[draw,fill=blue!20,minimum width=2cm] (apply) at (0,-2) {Apply $V = I+\mathcal{U}$};
    \node[draw,fill=blue!20,minimum width=2cm] (measure) at (0,-3) {Measure};
    \node[draw,fill=blue!20,minimum width=2cm] (lock) at (0,-5) {Lock Qubits};
    
    \draw[arrow] (init) -- (apply);
    \draw[arrow] (apply) -- (measure);
    \draw[arrow] (measure) -- (lock);
    \draw[arrow] (lock) -- ++(3,0) |- node[pos=0.25,right] {Repeat} (init);
    
    \node[highlight, text width=5cm] at (0,-4) {Stop when a string has probability $\ge 1/2$};
\end{scope}
\end{tikzpicture}
    \caption{Left: the VQPM circuit for three register qubits and the control qubit. Right: the VQPM loop. Each iteration prepares the product state, applies the controlled $\mathcal{U}$ between two Hadamard gates on the control qubit, measures all qubits, and locks the qubits whose two outcome probabilities differ by more than the threshold $p_{diff}$; the loop repeats until a string has probability at least $1/2$ or all qubits are locked.}
    \label{fig:Circuit Summary of VQPM}
 \end{figure*}
For a given matrix of coefficients $Q$, the problem of finding a binary vector $\vec{x}$ that minimizes $\mathbf{x}^T Q \mathbf{x}$ is called quadratic binary optimization \cite{brandao2022faster}. The unconstrained formalization of this framework, the so-called quadratic unconstrained binary optimization (QUBO) \cite{glover2022quantum}, is used to formulate many complex optimization problems including NP-hard problems such as the maximum graph cut (max-cut) problem \cite{dunning2018works} and other non-linear binary optimization problems \cite{anthony2017quadratic}. It is known that these problems can be formulated to represent energy functions. For instance, a max-cut problem with edge weights $c_{ij}$ can be formulated as finding the lowest energy-eigenvalue of the following Hamiltonian \cite{poljak1995solving,kolmogorov2004energy}:
\begin{equation}
 \mathcal{H} = -\sum_{(i,j)} c_{ij} (x_i - x_j)^2.
\end{equation}
Therefore, the classical algorithms for a max-cut problem can be used to solve the systems that are represented by these Hamiltonians. 
Similarly, quantum eigenvalue-related algorithms can be used to solve these problems by slight bit value mappings and modification of the Hamiltonian to include different interaction terms (equivalent to correlating different binary values). In its most generic form, a QUBO problem for quantum algorithms can be written as: 
\begin{equation}
    \mathcal{H}= \sum q_{ii}z_i + \sum_{(i,j)\ i< j} q_{ij} z_i z_j.
\end{equation}
Here, $z_i\in\{0,1\}$ represents a parameter of the optimization and $q_{ij}$ are the matrix elements of the matrix $Q$; only the upper triangle of $Q$, diagonal included, enters. We write $E(x)=\sum_iq_{ii}x_i+\sum_{i<j}q_{ij}x_ix_j$ for the energy of a bit string $x$, which is the eigenvalue of $\mathcal{H}$ for the basis state $\ket{x}$. This Hamiltonian can be easily mapped into quantum circuits by using the quantum phase gates:
 \begin{equation}
  R_{z_{ij}} =  \begin{bmatrix}
        1&0\\
        0&e^{iq_{ij}}
    \end{bmatrix}.
\end{equation}
This gate is controlled by the qubit $i$ and acts on the qubit $j$. For $i=j$, this acts as a single phase gate. Therefore, a full quantum circuit representing the Hamiltonian requires $O(n^2)$ phase gates for a given coefficient matrix $Q$ of dimension $n^2$ representing the correlations of $n$ parameters.

When we have this quantum circuit, equivalently, we have a \emph{diagonal} unitary matrix $\mathcal{U} = e^{i\mathcal{H}}$ whose eigenstates encode the different combinations of the parameters $z_i$ and whose eigenvalue phases encode the energy value of the associated combination (the fitness value for an instance in the solution space).  
In other words, this circuit, the eigenspace of $\mathcal{U}$, encodes the whole solution space for the considered QUBO problem.  
Here, note that while $\mathcal{U}$ is diagonal, its dimension is exponential ($2^n$) in the number of parameters $n$, so writing it down as a matrix takes exponential space. Applying it to a basis state, however, only needs the energy of that state, which takes $O(n^2)$ operations; together with the product-state input of VQPM, this makes VQPM for QUBO problems classically simulable (Remark~\ref{rem:classical}).
Since the circuit includes only a polynomial number of gates, any quantum eigenvalue algorithm can be considered to obtain the minimum eigenphase and its associated eigenstate.

While classical methods are successful in optimizing many QUBO problems with thousands of parameters related to practical applications today \cite{anand2017comparative}, QUBO is NP-hard in general, so no classical algorithm is known that solves every instance efficiently, and quantum approaches are studied as possible alternatives. In addition, although there are other early quantum optimization algorithms \cite{hogg2000quantum}, the adiabatic quantum computation (AQC) \cite{farhi2000quantum,van2001powerful,aharonov2008adiabatic} along with the quantum annealing \cite{finnila1994quantum}, quantum phase estimation \cite{kitaev1995quantum} and quantum approximate optimization algorithm (QAOA) \cite{farhi2014quantum} have been successfully applied to real-world problems, which show the practicalities of these algorithms. For some problems QAOA comes with a performance guarantee: for MaxCut on 3-regular graphs, depth $p=1$ already gives an approximation ratio of at least 0.6924~\cite{farhi2014quantum}. QUBO formulations of practical problems, such as network routing~\cite{silva2022qubo}, are a common target of these algorithms.  
While these algorithms provide a workable and practically usable framework for solving complex optimization problems, having versatile algorithms helps overcome hardware limitations and further refine existing algorithms and inspire new applications that can unlock quantum advantage in the real-world optimization landscape.

Variational quantum power method (VQPM) \cite{daskin2021combinatorial} was proposed to solve combinatorial optimization problems by using the quantum version \cite{daskin2018quantum} of the classical power iteration \cite{golub13}. 
For a given Hermitian matrix $\mathcal{H}$ with a dominant eigenvalue $|\lambda_{d}|$, which is greater than any other eigenvalue, and an initial vector $\ket{\psi_0}$ whose overlap with the eigenvector $\ket{d}$ associated with this eigenvalue is non-zero ($\braket{d}{\psi_0} \neq 0$), it is known that if we apply $\mathcal{H}$ to the initial vector $k$ times recursively, the final vector $\ket{\psi_k}$ converges to the ``dominant" eigenvector $\ket{d}$.

It is also known that the unitary matrix $\mathcal{U}=e^{i\mathcal{H}}$ has the same eigenbasis, with eigenvalues $e^{i\lambda_j}$.  
That means if the eigenbasis of $\mathcal{H}$ of dimension $2^n$ is defined by the vectors $\{\ket{j}\}$ with $0 \leq j < 2^n$, we get:
\begin{equation}
    \mathcal{U}\ket{j} = e^{i\lambda_j}\ket{j}.
\end{equation}
There are also different implementations of this algorithm on quantum computers.  
Below, we will review the quantum version of the algorithm by following Ref.\cite{daskin2018quantum}.  Then, we will analyze its variational version \cite{daskin2021combinatorial} for the QUBO problems and show numerical results with different settings. The circuit summary of the overall approach is given in Fig. \ref{fig:Circuit Summary of VQPM}. 

\paragraph{Contributions.} This paper makes four contributions.
\begin{itemize}
\item[(i)] \emph{Classical VQPM.} For QUBO problems, VQPM has an exact classical equivalent: every measurement shot, and therefore every locking decision, can be sampled classically at $O(n^2)$ cost per shot (Remark~\ref{rem:classical}). To first order, the VQPM update is a gradient method on the mean energy of the product distribution of the qubits, in the logits of the qubit probabilities (Sec.~\ref{sec:update}). The gradient method itself is known (mirror descent, Hopfield--Tank networks); what we add is the equivalence, which we use for the analysis. This classical form explains how the algorithm behaves, and it lets us run VQPM up to $n=100$, far beyond the reach of a quantum-circuit simulation.
\item[(ii)] \emph{The update and locking.} The VQPM update never lowers the control probability and settles at strings where every single bit flip raises the energy (Sec.~\ref{sec:VQPM}). Its signs follow first the first-iteration probability differences and then the lowest eigenvectors of the coupling matrix, and locking reads these signs at a time set by the threshold $p_{diff}$. Locking therefore speeds the update up, but it does not steer it toward the optimum (Sec.~\ref{sec:locksigns}).
\item[(iii)] \emph{Comparison.} We compare VQPM with QAOA under four classical optimizers and with simulated annealing on 50 to 100 instances per $n$ up to $n=16$, and follow VQPM to $n=100$ with its classical equivalent (Sec.~\ref{sec:comparison}). Without locking and with exact marginals, VQPM finds the optimum as often as QAOA with a good optimizer up to $n=16$, but with about 100 times more iterations than with locking; with locking it falls behind; and beyond $n=16$ it declines in both forms, while simulated annealing keeps reaching the best strings we know.
\item[(iv)] \emph{Device requirements and cost.} We measure the shots per iteration, the circuit size and the gate error rates that VQPM needs, and compare the cost per instance of all methods (Sec.~\ref{sec:noise}).
\end{itemize}

Sec.~\ref{sec:qpi} reviews quantum power iteration. Sec.~\ref{sec:VQPM} describes VQPM, its update, qubit locking and dynamic locking thresholds. Sec.~\ref{sec:update} gives the classical equivalent of VQPM and shows which signs locking reads. Sec.~\ref{sec:comparison} compares VQPM with QAOA and simulated annealing, Sec.~\ref{sec:noise} studies finite shots, circuit size, noise and cost, and Sec.~\ref{sec:conclusion} concludes with the limitations. Appendix~\ref{alg:1} gives the pseudocode and the parameters, and Appendices~\ref{app:theory} and~\ref{app:sampling} the proofs and the classical simulation of VQPM up to $n=100$. The code and the data are available online\footnote{\url{https://github.com/adaskin/vqpm}}; the README of the repository lists the script and the data behind each figure and table.

\subsection{Quantum power iteration (QPI)}
\label{sec:qpi}
Quantum power iteration uses a control register in the Hadamard basis to apply the operator $\mathcal{U}$ when it is in the \ket{1} state. This can be simply depicted as:
\begin{align}
\frac{\left(\ket{0}\ket{j}+\ket{1}\ket{j}\right)}{\sqrt{2}} \ \\ \ \xrightarrow[\text{on 2nd register}]{\text{controlled-U}} \ \ \  \frac{\left(\ket{0}\ket{j}+\ket{1}\mathcal{U} \ket{j}\right)}{\sqrt{2}}.
\end{align}
After applying the second Hadamard gate to the first qubit and measuring 0 on it, the following normalized, collapsed quantum state is produced:
\begin{align}
  \frac{\ket{0}\left(\ket{j}+\mathcal{U} \ket{j}\right) + \ket{1}\left(\ket{j}-\mathcal{U} \ket{j}\right)}{2} \ \\ \ \xrightarrow[\text{on 1st}]{\text{measure \ket{0}}} \ \ \ \frac{(I + \mathcal{U}) \ket{j}}{\|(I + \mathcal{U})\ket{j}\|}.
\end{align}
Note that measuring \ket{1} is analogous, but it generates $(I-\mathcal{U})$.  
Similarly to classical power iteration, if we start with an initial vector \ket{\psi_0} and repeat the above procedure $k$ times, we get the following unnormalized state:
\begin{equation}
\label{eq:qpi}
    \ket{\psi_k} \propto (I + \mathcal{U})^k \ket{\psi_0}.
\end{equation}

\begin{figure*}[h]
    \centering
    \begin{subfigure}[t]{0.5\linewidth}
    \includegraphics[width=1\linewidth]{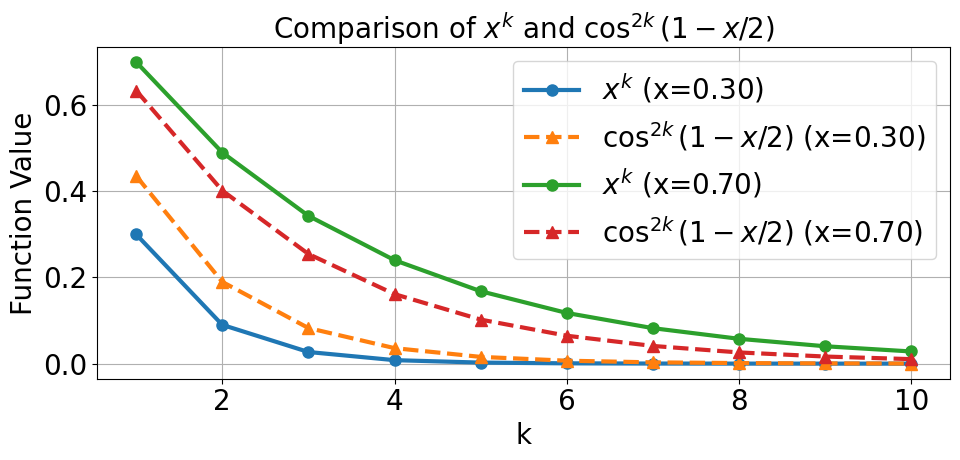}
    \caption{$\cos^{2k}(x/2)$ vs $x^k$.}  
    \end{subfigure}\hfill
    \begin{subfigure}[t]{0.5\linewidth}
    \includegraphics[width=1\linewidth]{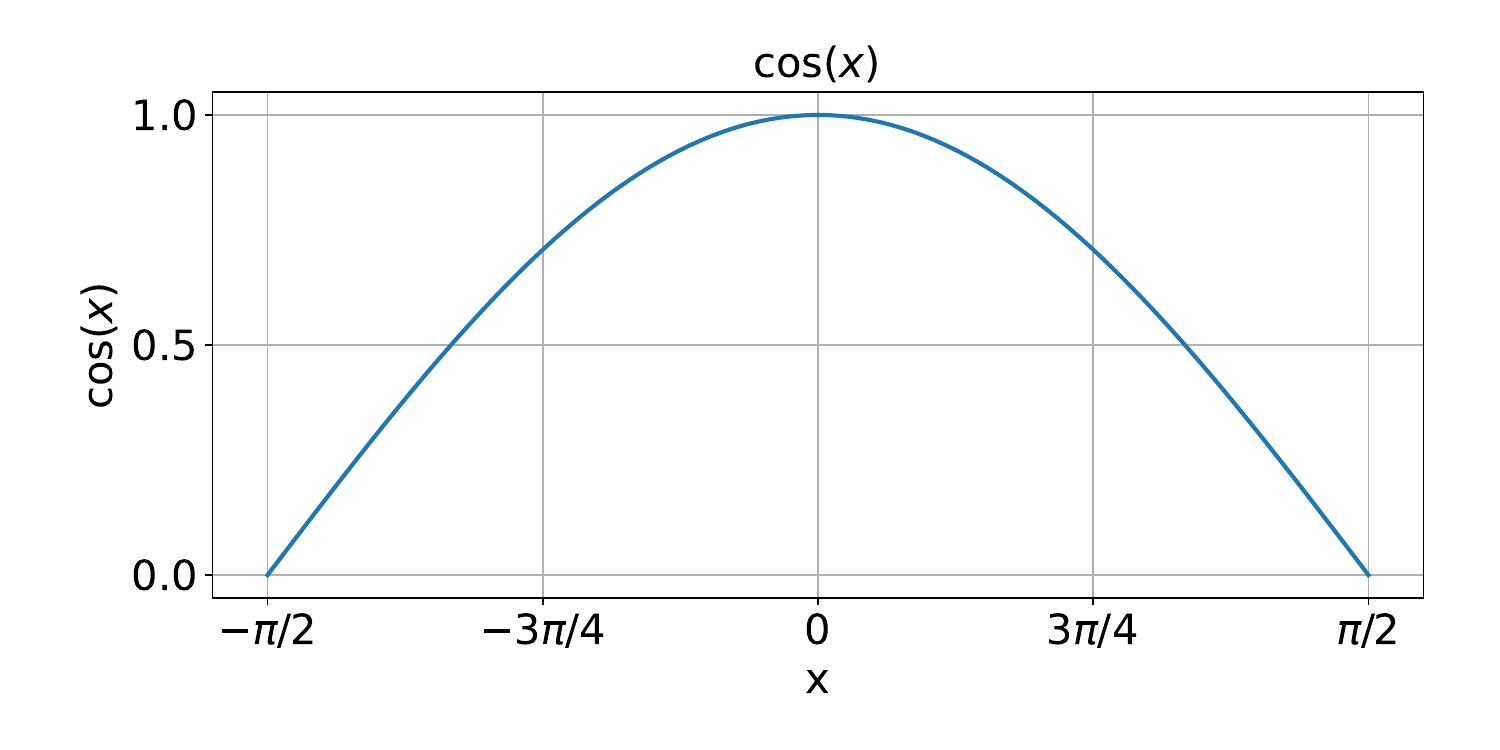}
    \caption{$\cos(x)$ in $[-\pi/2, \pi/2]$.}  
    \end{subfigure}
    \caption{Cosine function and its power. For maximization and minimization problems, the phases should be scaled based on $\cos(x)$ output. In this paper, we have used $[0, \pi/2]$ to find the minimum positive eigenvalue of $\mathcal{H}$.}
    \label{fig:cosx}
\end{figure*}

The normalization of this quantum state occurs because of the probabilistic nature of the quantum system when we measure the first qubit. This recursion and normalization process involves the eigenvalues of the shifted operator $(I+\mathcal{U})$. Since the identity matrix is just a diagonal shift for $\mathcal{U}$ with eigenvalues $e^{i\lambda_j}$, the eigenvalues of $I + \mathcal{U}$ are:
\begin{equation}
1 + e^{i\lambda_j} = 2\cos\left(\frac{\lambda_j}{2}\right)e^{i\lambda_j/2}.
\end{equation}
The magnitude of these eigenvalues is $2|\cos(\lambda_j/2)|$. If all eigenvalues of $\mathcal{H}$ lie in the range $[0, \pi/2]$, this magnitude decreases as $\lambda_j$ increases (this is depicted in Fig.\ref{fig:cosx}). Therefore, it gives larger values for the smallest $\lambda_j$, which we will denote by $\lambda_d$, and $\ket{d}$ will represent its associated eigenvector.

For QUBO problems this window is guaranteed by the phase map we use. With $S=\sum_{i\le j}|q_{ij}|$, every energy satisfies $|E(x)|\le S$, so
\begin{equation}
\label{eq:phasemap}
\lambda(x)=\frac{\pi}{4}+\frac{\pi}{4}\,\frac{E(x)}{S}
\end{equation}
places all $2^n$ eigenphases in $[0,\pi/2]$, keeps the order of the energies and costs $O(n^2)$ operations, with no spectral information (Lemma~\ref{lem:window}). Any window inside $[0,\pi)$ would keep the minimum-energy state dominant, because $\cos^2(\lambda/2)$ decreases on $[0,\pi)$. The upper end $\pi/2$ keeps the probability of measuring the control qubit in $\ket{0}$ at least $1/2$ for every input; a wider or shifted window raises the contrast between nearby phases but lowers this probability (Remark~\ref{rem:window}). On our instances the energies fill only about a fifth of the window (Table~\ref{tab:app-phase}). We also compared Eq.~\eqref{eq:phasemap} with three other maps in VQPM runs on 100 instances per $n$ for $n=8$ to 16, with three decimals and $p_{diff}=0.01$: a map that uses the tighter bounds $L\le E(x)\le U$, with $L$ and $U$ the sums of the negative and of the positive coefficients, $\lambda=\frac{\pi}{2}(E-L)/(U-L)$, which has twice the slope of Eq.~\eqref{eq:phasemap} (Remark~\ref{rem:window}); the same map stretched to $[0,0.9\pi]$; and Eq.~\eqref{eq:phasemap} shifted by $\pi/2$. Eq.~\eqref{eq:phasemap} is best or tied up to $n=12$. The map with the tighter bounds is better at $n=16$ (found-optimum rate 0.47 against 0.38) and needs fewer iterations. The stretched and shifted maps are worse at every $n$ (0.21 to 0.43), because almost all qubits lock within 2 to 5 iterations. With three decimals, part of the advantage of the tighter bounds may come from their larger first-iteration differences, which survive the rounding (Appendix~\ref{app:large}).

\textbf{Convergence: } After each iteration of the algorithm, the natural normalization process scales the matrix eigenvalues. This can be seen more clearly if we write the matrix in its eigenbasis:
\begin{equation}
\begin{split}
    \frac{\left(I + \mathcal{U}\right)\ket{\psi_0}}{{\|(I + \mathcal{U})\ket{\psi_0}\|}} = & \sum_j \frac{1 + e^{i\lambda_j}}{\mathcal{N}}\ket{j}\bra{j} \ket{\psi_0}
    \\ = & \sum_j \alpha_j \frac{1 + e^{i\lambda_j}}{\mathcal{N}} \ket{j}.
\end{split}
\end{equation}
Here, $\alpha_j$ represents the overlap $\braket{j}{\psi_0}$ and $\mathcal{N}$ is a positive value representing the norm in the denominator. Since $\mathcal{N}$ is at most $|1+e^{i\lambda_d}|$, the main contribution to the error comes from the eigenvector \(\ket{s}\) of the second-smallest eigenvalue $\lambda_s$ of the Hamiltonian $\mathcal{H}$. Therefore, the convergence rate can be estimated by taking the magnitude of its coefficient, which simply gives the following ratio:
\begin{equation}
\label{eq:ratio}
R = \frac{\cos(\lambda_s/2)}{\cos(\lambda_d/2)}.
\end{equation}
After $k$ iterations, the overlap of the collapsed state with the target eigenstate \(\ket{d}\) becomes proportional to the $k$th power of this ratio, \(R^k\). Note that the power of the cosine decreases slower than the power of a similar value.  
However, the success probability scales with \(R^{2k}\). This is shown in Fig.\ref{fig:cosx}, where we show how both the classical and the quantum versions require similar asymptotic numbers of iterations. This demonstrates how probabilities associated with smaller eigenvalues vanish rapidly as \(k\) increases.

Furthermore, on quantum computers the weight of the other eigenvectors does not have to vanish. It is enough that the dominant eigenvector \ket{d} carries a probability that is not too small, for example of order $1/\mathrm{poly}(n)$, since it is then found by repeating the measurement. This also covers degenerate minimum energies. Let $A$ be the initial probability on the minimum-energy eigenvectors (for the uniform start, their number divided by $2^n$), and let $R^2$ be the largest ratio $|1+e^{i\lambda_j}|^2/|1+e^{i\lambda_d}|^2$ over the other eigenvectors. After $k$ iterations the probability $P_k$ of the minimum-energy eigenvectors satisfies $P_k\ge1/\bigl(1+\frac{1-A}{A}R^{2k}\bigr)$ (Proposition~\ref{prop:kbound}). Hence $P_k\ge1-\epsilon$, where $\epsilon$ is the tolerated failure probability, is guaranteed when
\begin{equation}
\label{eq:kbound}
k \geq \frac{\ln\frac{1-A}{A}+\ln\frac{1-\epsilon}{\epsilon}}{\ln(1/R^2)}.
\end{equation}
For the uniform start the first term is about $n\ln2$, so $k$ grows like $n\ln2/\ln(1/R^2)$. On the random instances of this paper, Eq.~\eqref{eq:kbound} with $\epsilon=1/2$ asks for about $2\times10^2$ iterations at $n=4$ and $1.6\times10^4$ at $n=16$, and the probability that all of these post-selections succeed is below $10^{-80}$ from $n=8$ on (Appendix~\ref{app:coherent}). The bound concerns coherent iteration. VQPM follows a different update (Sec.~\ref{sec:VQPM}), and its runs with locking stop within 100 iterations, so their convergence does not come from the power filter alone.
The ratio $R$ can be directly related to the eigengap $\gamma = \lambda_s - \lambda_d$: Since the eigenvalues are assumed to be in $[0, \pi/2]$, the ratio $R$ can be simply written as $R = \cos(\gamma/2 + \lambda_d/2)/\cos(\lambda_d/2)$. For a small $\gamma$ value, this implies $R \approx 1$ and the iteration converges to the superposition of \ket{s} and \ket{d} and other closer eigenstates.

As mentioned before, this is fundamentally different from classical power iteration. Since on quantum computers the superposition disappears after the measurement, if we measure the quantum state at that point, we would still be able to get the dominant eigenstate if its probability in the superposition is sufficiently large. This is determined by the number of eigenstates that are close to the dominant eigenvector.

A different matter is the complexity of a single iteration: Both classical and quantum methods exhibit $\mathcal{O}(1/\gamma)$ scaling when they are applied to the unitaries. However, when we have an efficient circuit $\mathcal{U}$, i.e., a circuit with $poly(n)$ number of gates, the quantum power iteration avoids explicit matrix processing to find a circuit mapping for the unitary. Therefore, in those cases, one iteration of the algorithm only requires $O(poly(n))$ operations.

When the value of $\gamma$ is on the order of $2^{-n}$, then the required number of iterations for convergence obviously becomes exponential in the number of qubits, $O(2^n)$. Even if $\gamma$ is not exponentially small but is on the order of $1/poly(n)$, problems with such small eigengaps may still require many iterations for convergence, which is a challenging task for current computing devices. This can be reduced if prior knowledge related to the eigenstate is available. Another approach is turning the original algorithm into a variational circuit as done in Ref.\cite{daskin2021combinatorial}, which is outlined in the next section.

\section{Variational quantum power method (VQPM)}
\label{sec:VQPM}
This section outlines VQPM~\cite{daskin2021combinatorial}, states the update it performs, and describes the locking of qubits, incorrect locks and dynamic locking thresholds. Fig.~\ref{fig:Circuit Summary of VQPM} summarizes the circuit, and Appendix~\ref{alg:1} gives the pseudocode and the parameters (Table~\ref{tab:params}).

\subsection{Parameterized input state and the VQPM update}
In VQPM, to reduce the number of iterations in the standard power iteration, the algorithm is converted into a variational circuit by adding rotational-$Y$ gates to the qubits representing eigenstates. 
This allows the state to be reinitialized by feeding the measurement outcomes (the approximate tomography) of the previous iterations. The algorithm at the $k$ iteration can simply be represented by the following circuit:
\begin{center}
\includegraphics{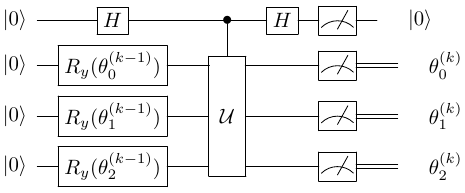}
\end{center}
For QUBO problems, we simply express \( \ket{\psi_0} \) in the eigenbasis \( \{\ket{j}\} \) and start with the vector:
\begin{equation}
    \ket{\psi_0} = \sum_j \alpha_j \ket{j}, \quad \alpha_j = \frac{1}{\sqrt{2^n}}.
\end{equation}
Therefore, for the initial state, the rotation-$y$ gates are considered as the Hadamard gates. Since any of the eigenvectors $\ket{j}$ is an unentangled tensor product state, this circuit is able to represent all the eigenstates by giving the right parameters to the single gates. However, the circuit at the end generates a final state which is the superposition of many eigenstates.

This circuit provides a way to feed the previous output state to the next iteration by individual qubit measurements. Thus, it does not necessitate a full quantum state tomography and can be considered as a very efficient circuit. For QUBO problems, however, the same feature makes it classically simulable: the input is a product state and $\mathcal{U}$ is diagonal, so every measurement shot can be reproduced by classical sampling (Remark~\ref{rem:classical} in Sec.~\ref{sec:sampling}). The simulability comes from the product state prepared in every iteration, not from the power method itself. A circuit that entangles the qubits before the diagonal operator, as the mixing layers of QAOA~\cite{farhi2014quantum} or the ansatz circuits of VQE~\cite{peruzzo2014variational} do, is in general hard to sample classically, but its parameters are then no longer fixed by the measured single-qubit probabilities; Remark~\ref{rem:classical} discusses this trade-off.

\paragraph{The VQPM update.} The feedback makes VQPM a different iteration from QPI. Write $w(x)=\cos^2(\lambda(x)/2)=|1+e^{i\lambda(x)}|^2/4$ for the \emph{post-selection weight} of the basis state $\ket{x}$, the probability that the control qubit is measured in $\ket{0}$ when the register is in $\ket{x}$; $\pi_i$ for the \emph{marginal} of qubit $i$, the probability that it is measured as $1$; and $p_\pi(x)=\prod_i\pi_i^{x_i}(1-\pi_i)^{1-x_i}$ for the \emph{product distribution} prepared by the $R_y$ gates. With exact marginals, and before locking or rounding, one VQPM iteration maps the marginals $\pi_i$ to
\begin{equation}
\label{eq:vqpm-update}
\pi_i'=\frac{1}{W(\pi)}\sum_{x:\,x_i=1}p_\pi(x)\,w(x),\qquad W(\pi)=\sum_xp_\pi(x)\,w(x),
\end{equation}
where the \emph{control probability} $W(\pi)$ is the probability of measuring the control qubit in $\ket{0}$. The next iteration starts from the product state with the new marginals $\pi_i'$, not from the post-selected state itself. The re-preparation keeps the marginals and discards every correlation between the qubits, so Eq.~\eqref{eq:vqpm-update} agrees with power iteration, Eq.~\eqref{eq:qpi}, only in the first iteration from the uniform state. The two-qubit example of Appendix~\ref{app:vqpm-map} shows the difference: after two iterations the probability that the first qubit is $0$ is $12/17$ under VQPM and $5/7$ under power iteration. The bound of Eq.~\eqref{eq:kbound} therefore does not apply to VQPM. Two properties of Eq.~\eqref{eq:vqpm-update} are used throughout (proofs in Appendix~\ref{app:vqpm-map}).
\begin{itemize}
\item[(i)] \emph{The control probability never decreases}: $W(\pi')\ge W(\pi)$ for the control probability $W$, with equality only at fixed points (Proposition~\ref{prop:monotone}). Eq.~\eqref{eq:vqpm-update} is the growth transform of Baum and Eagon~\cite{baum1967inequality}, and in the language of estimation-of-distribution algorithms it is the univariate marginal distribution algorithm with the post-selection weights $w$ as selection weights~\cite{muhlenbein1996recombination}.
\item[(ii)] \emph{Local minima attract the iteration}: every bit string is a fixed point, and it is stable exactly when every single bit flip raises its energy (Proposition~\ref{prop:stability}). The iteration can therefore converge to a suboptimal string even with exact marginals and no locking.
\end{itemize}
Sec.~\ref{sec:update} shows what the iteration does between the uniform start and these fixed points.

\paragraph{The control qubit.}
\label{sec:controlqubit}
The control qubit plays the role it has in phase estimation. Its probability $W$ of being measured in $\ket{0}$ carries the eigenvalue information of the circuit: once the register has converged to a string $x$, it equals $w(x)=|1+e^{i\lambda(x)}|^2/4$, the squared magnitude of the eigenvalue of $(I+\mathcal{U})/2$. By property (i), with exact marginals $W$ never decreases, so it can fall only through rounding, locking or shot noise. A fall of $W$ would flag a lock that moved the state toward higher energies, but in the runs with locking of Table~\ref{tab:app-locking} it fell only once in 3\,623 iterations. Accordingly, a rule that locks only when $W$ increases changed nothing in our simulations with three decimals.

\subsection{Measurement and information loss}
\label{sec:infoloss}
\begin{figure*}[t]
    \centering
    \includegraphics[width=1\linewidth]{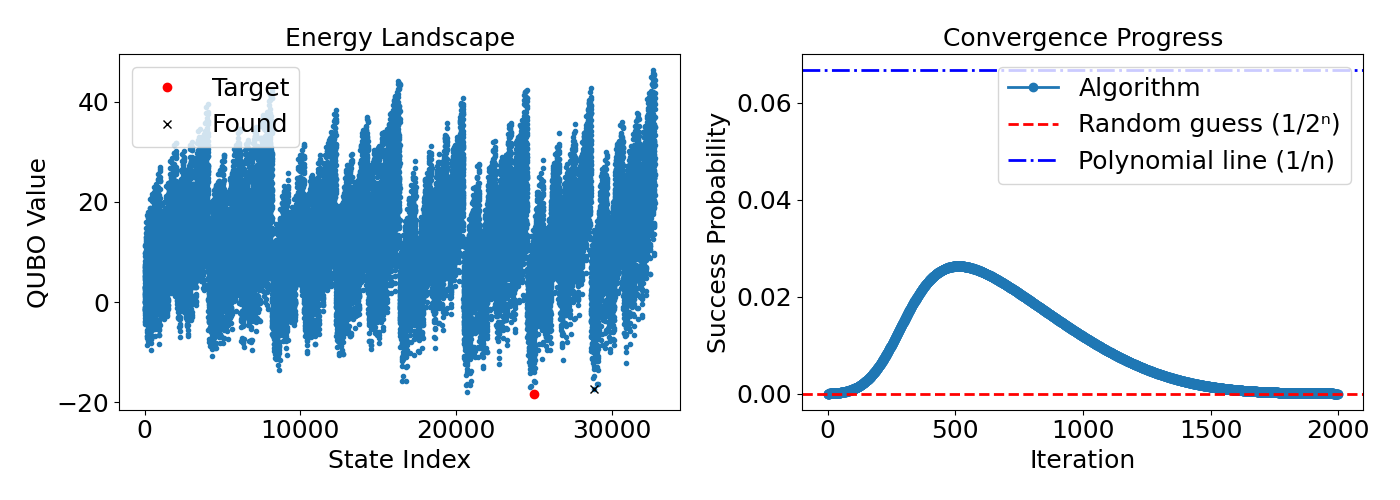}
    \caption{Convergence to a nearby local minimum: the optimization converges to a close energy value, but in another eigenstate (property (ii) of Eq.~\eqref{eq:vqpm-update}).}
    \label{fig-losing-eigenstate}
\end{figure*}

Because the re-preparation keeps only the marginals, a bias that the early iterations put on some qubits cannot always be reversed later. By property (ii) the iteration is pulled toward every 1-flip local minimum of the energy, and when such a minimum lies close in energy to the optimum, the marginals can move toward it and stay there. This is depicted for a random instance of QUBO in Fig.\ref{fig-losing-eigenstate}, where, after some iterations, the further iterations take the algorithm further away from the dominant eigenstate to another eigenstate whose energy value is very close to the dominant eigenvalue. The same happens with exact marginals: iterated without locking or rounding until it settles, Eq.~\eqref{eq:vqpm-update} ends at a suboptimal local minimum in 10 to 14\% of the instances for $n=6$ to $12$ (Table~\ref{tab:app-exact}) and in 27 to 76\% for $n=20$ to $100$ (Sec.~\ref{sec:update}). It is a property of the update, not of the measurement precision. Measuring with low precision in the early iterations does not prevent it; it only slows the iteration or stalls it, because rounded marginals can repeat unchanged (Appendix~\ref{app:locking}).

\begin{figure*}[t]
    \centering
    \begin{subfigure}[t]{0.95\linewidth}
        \centering
        \includegraphics[width=1\linewidth]{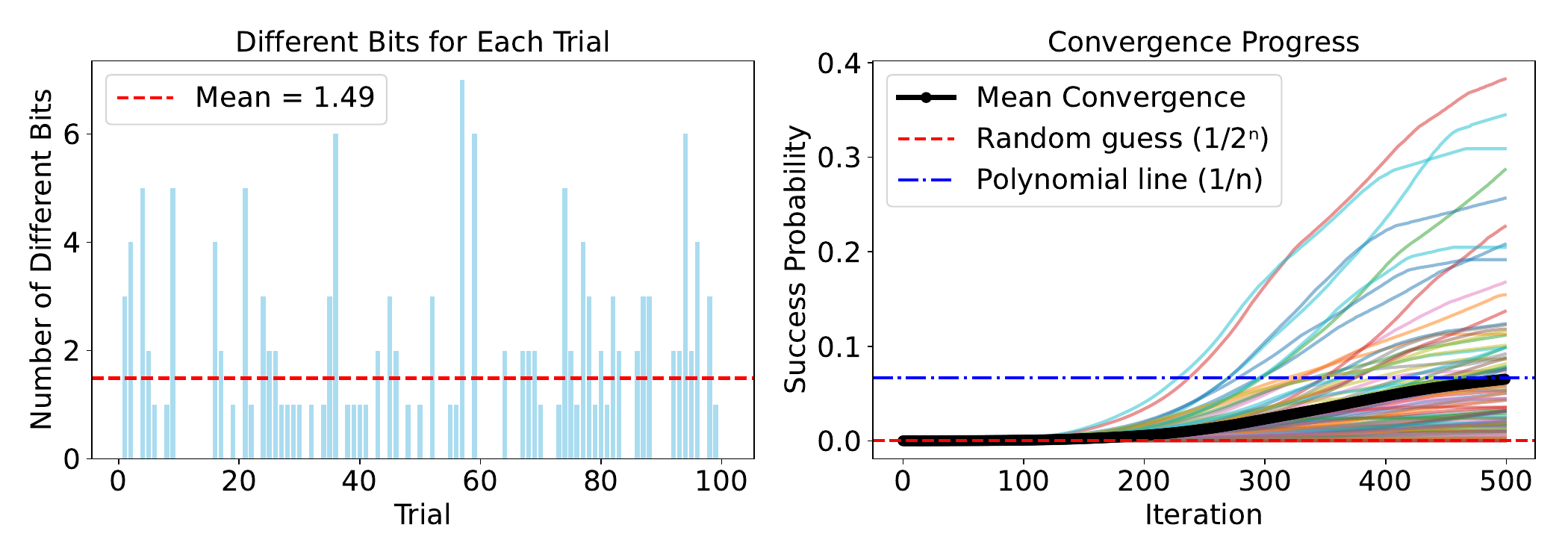}
        \caption{Convergence for 100 random trials of QUBO problems for \( n = 15 \) qubits. The convergence is slow, especially when the eigengap is small. Runs that miss the optimum are drawn toward a nearby 1-flip local minimum (property (ii) of Eq.~\eqref{eq:vqpm-update}).}
        \label{fig:runs-no-locking}
    \end{subfigure}
    \begin{subfigure}[t]{0.95\linewidth}
        \centering
        \includegraphics[width=1\linewidth]{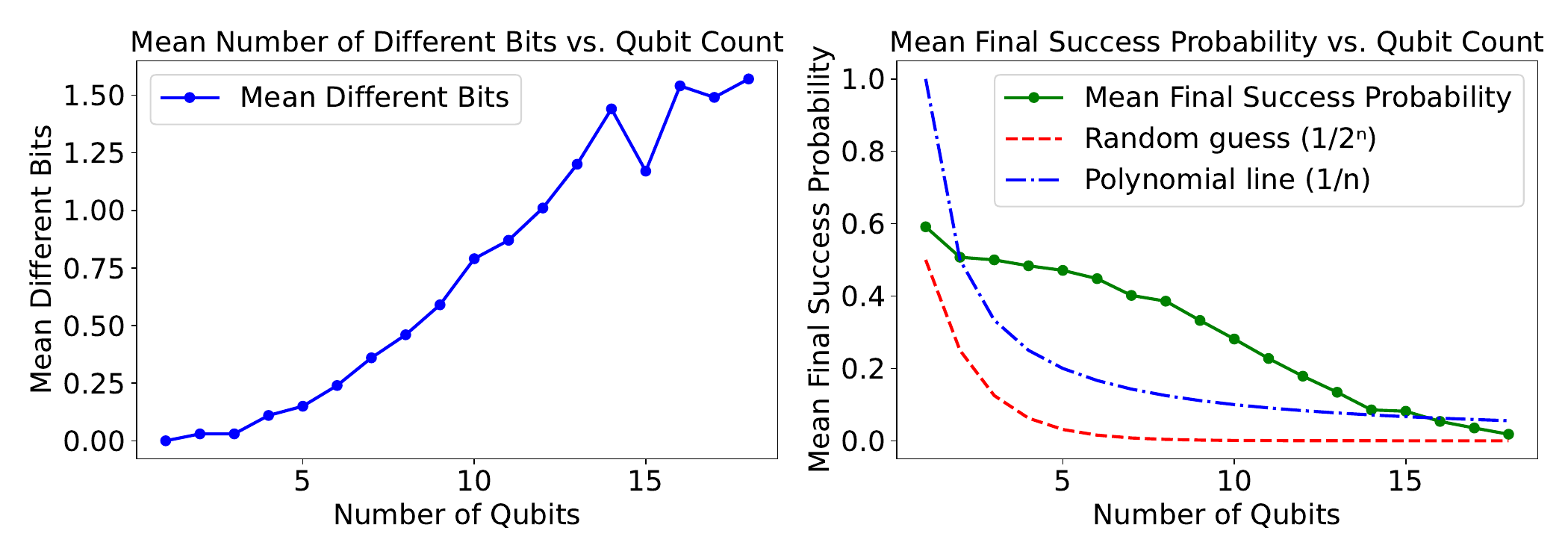}
        \caption{Mean convergence for 100 random trials of QUBO problems for \( n = 1 \dots 18 \) qubits. The maximum number of iterations is set to 500. The mean final success probability decreases roughly exponentially with $n$; part of this decline is the iteration budget (Sec.~\ref{sec:infoloss}).}
        \label{fig:means-no-locking}
    \end{subfigure}
    \caption{VQPM without locking the qubits. The algorithm can converge to a suboptimal local minimum, and 500 iterations are too few for many runs to settle.}
    \label{fig:both-no-locking}
\end{figure*}

Fig.\ref{fig:runs-no-locking} shows, for 100 random QUBO trials of \( n=15 \) parameters, how the success probability of the target (optimum) state increases and whether in these trials the optimization converges to different values. To indicate whether it has converged to a different value, we count the number of parameter values (0 or 1) that are different from the optimum solution. Similarly, Fig.\ref{fig:means-no-locking} shows the mean success probability and the mean number of bit difference values for \( n=1\dots 18 \) using 500 iterations and 100 trials for each \( n \) value. As can be seen from the lower right side of the figure, the mean success probability falls roughly exponentially with $n$. Part of this decline is the iteration budget: without locking, the exact update needs a median of $2.7\times10^3$ to $1.2\times10^4$ iterations to settle for $n=6$ to $12$ (Table~\ref{tab:app-exact}), far more than 500, and the number grows like $n^2$ (Sec.~\ref{sec:update}).

Therefore, for the above complexity dilemma, to reduce the required number of iterations and to increase the success probability at the same time, a locking mechanism is introduced to the algorithm: In each iteration, we check to see whether the probability difference of measuring 0 or 1 for a qubit is above a certain threshold. If yes (say, if the probability for 0 is much higher), then we lock that qubit to 0 (or 1 based on the difference). Below, this is described and analyzed in further detail.

\subsection{Locking qubits}
\label{sec:locking}
\begin{figure*}[t]
\centering
\begin{subfigure}[t]{0.95\linewidth}
\includegraphics[width=1\textwidth]{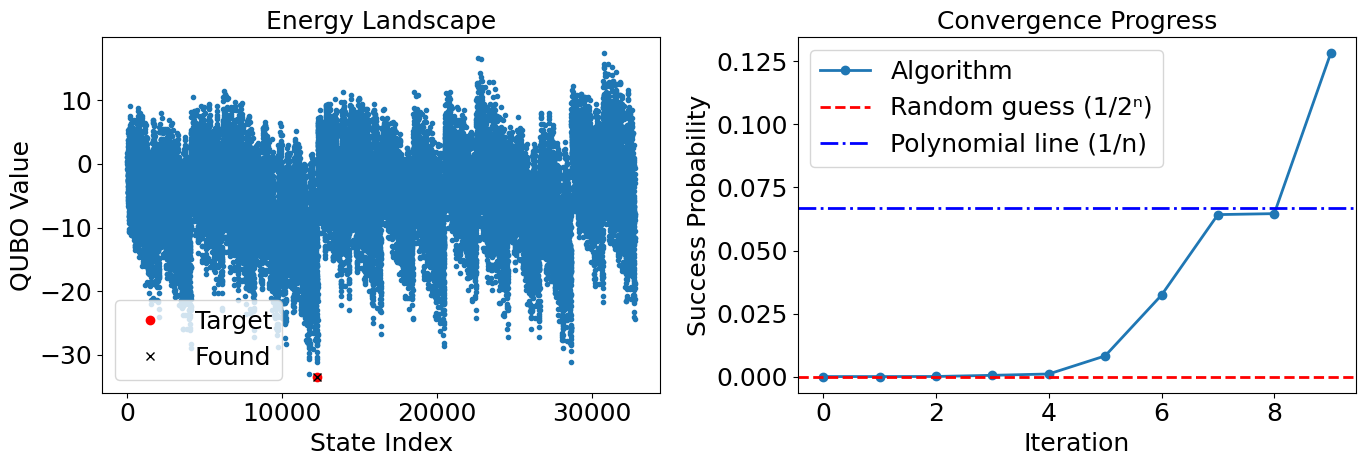}
\caption{Jumps in the iteration by locking qubits to the values early in the iterations: The optimization converges to the correct value in only a few iterations.}
\label{fig:jumps-correct}
\end{subfigure}
\begin{subfigure}[t]{0.95\linewidth}
\includegraphics[width=1\textwidth]{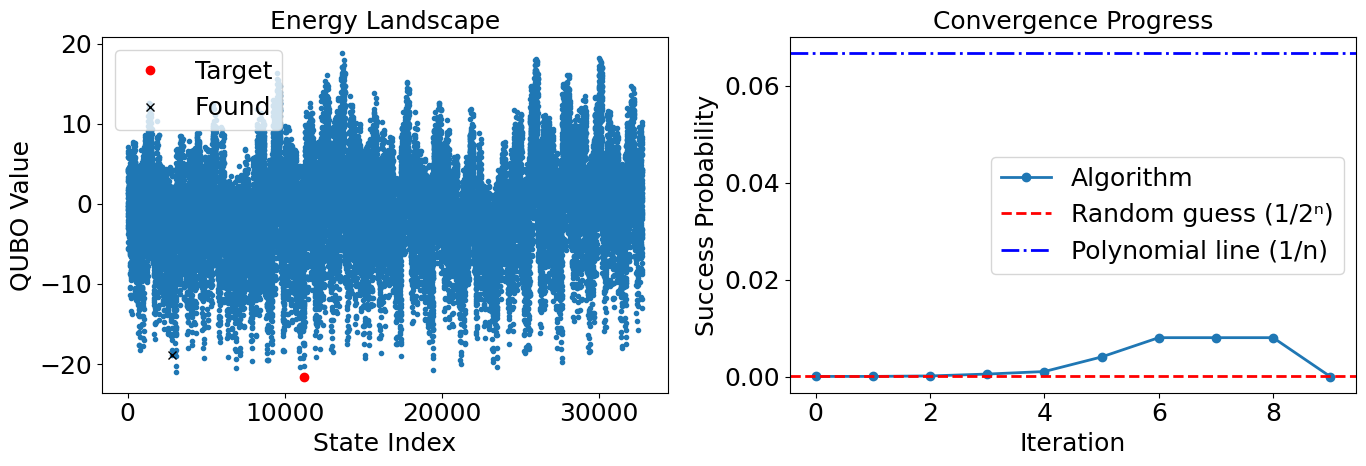}
\caption{Jumps in the iteration by locking qubits to the values early in the iterations: The optimization converges to a suboptimal value because of an early incorrect locking.}
\label{fig:jumps-incorrect}
\end{subfigure}
\caption{The correct and incorrect locking in VQPM for two random 15-parameter QUBO trials.}
\end{figure*}

\begin{figure*}[t]
    \centering
    \begin{subfigure}[t]{0.95\linewidth}
        \centering
        \includegraphics[width=1\linewidth]{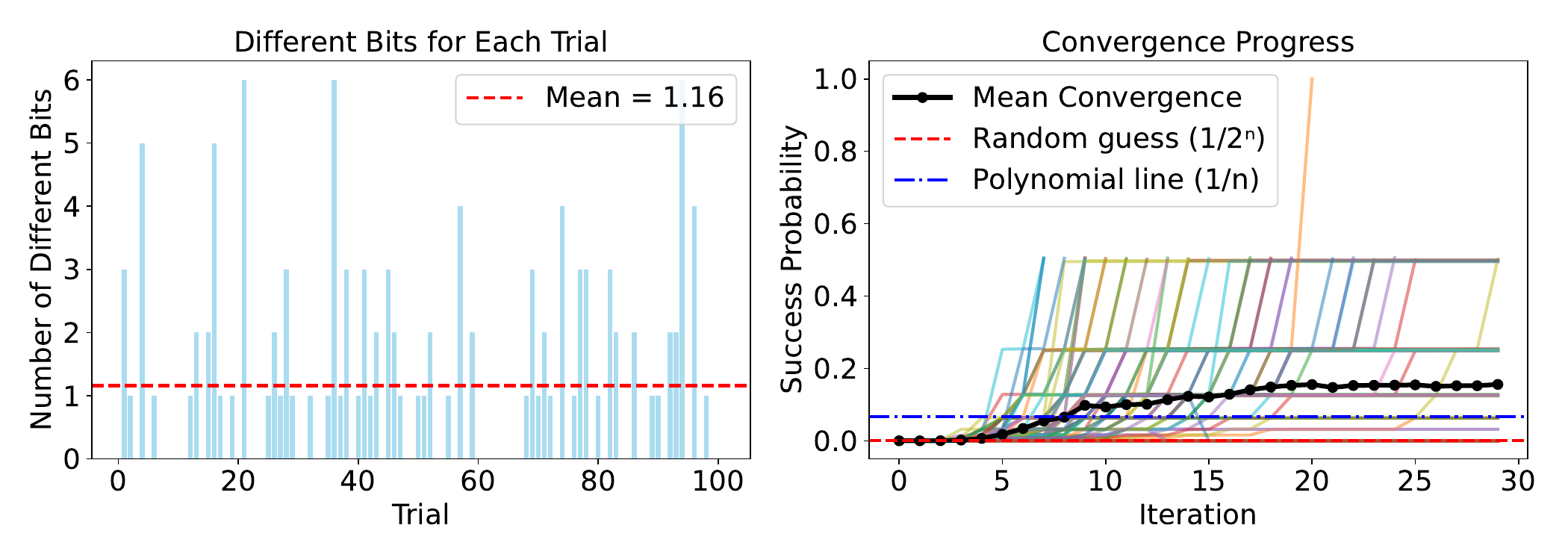}
        \caption{100 random QUBO trials for \(n=15\) with fixed \(p_{diff} = 0.01\) and three decimals.}
        \label{fig:run-pdiff-fixed}
    \end{subfigure}
    \begin{subfigure}[t]{0.95\linewidth}
        \centering
        \includegraphics[width=1\linewidth]{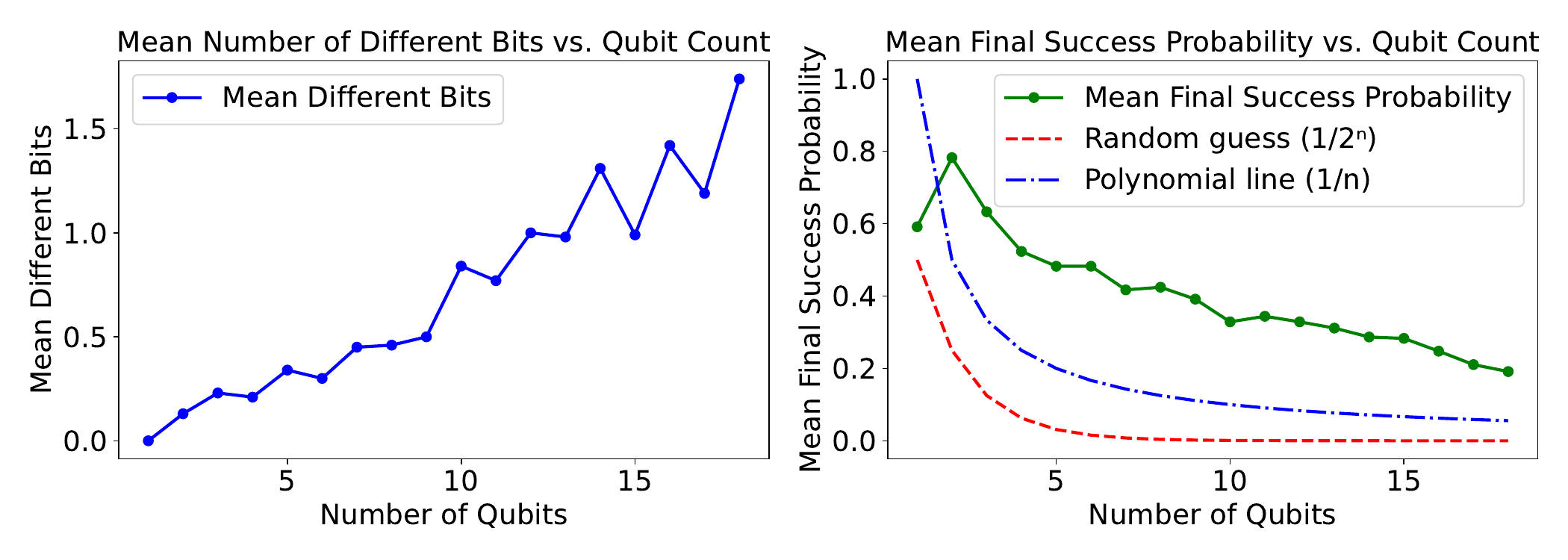}
        \caption{The convergence for the mean values of 100 random trials of QUBO problems for \(n = 1 \dots 18\) qubits. The maximum number of iterations is set to 30.}
        \label{fig:mean-pdiff-fixed}
    \end{subfigure}
    \caption{VQPM with locking of the qubits by using fixed \(p_{diff} = 0.01\) and three decimals. The algorithm converges to a suboptimal value if there is an incorrect locking or if the suboptimal values have higher probabilities.}
    \label{fig:both-pdiff-fixed}
\end{figure*}

As explained earlier, in addition to parameterizing the circuit, the second setting for VQPM is: If any qubit value can be determined in the earlier iterations, then these qubit states are locked to that determined value (set to 0 or 1 for the rest of the iterations). That means if the probability difference between 0 and 1 for any qubit is larger than an optimization parameter \(p_{diff}\), then the values of these qubits are locked to \(\{0, 1\}\) depending on the outcome. We call \(p_{diff}\) the \emph{locking threshold}. This can be considered as jumps in the iterations. An example of such jumps is depicted in Fig.\ref{fig:jumps-correct} for \(n = 15\) qubits. As can be seen from the figure, the algorithm converges to the solution in just 15 iterations. Fig.\ref{fig:both-pdiff-fixed} shows the same running test (the random seed is set to 42 while generating QUBO problem groups) as used for the case without locking. However, in this case, we only use a maximum of 30 iterations for \(n = 1 \dots 18\) and obtain higher success probabilities than without locking in 500 iterations. Without an iteration limit, the update without locking finds the optimum more often than with locking (Secs.~\ref{sec:locksigns} and~\ref{sec:largen}); locking makes it much faster.

In the simulations of this section we set \(p_{diff}=0.01\) and round the measured probabilities to 3 decimal places with the NumPy round function (e.g., 0.512 and 0.488), which we call \emph{three decimals}. These were the best values we observed for \(n \leq 18\) parameters. A second set of runs on 100 other instances per $n$ for $n=4$ to 22 (the instances of Sec.~\ref{sec:smalln}) shows that with this rounding the iteration budget does not matter: the found-optimum rate is the same for budgets of 30 and 300 iterations, 0.59 at $n=12$, 0.38 at $n=16$ and 0.28 at $n=20$, because the rounded marginals stop changing; with 300 iterations, 17\% (at $n=12$) to 46\% (at $n=20$) of the runs end stalled. With six decimals or exact marginals, which give the same results, the budget matters: with 300 iterations the rate stays between 0.51 and 0.66 for $n=12$ to 22, against 0.35 to 0.62 with 30 iterations (Fig.~\ref{fig:scaling}). Appendix~\ref{app:large} runs the same loop classically for $n=20$ to $100$: with three-decimal rounding the first-iteration differences fall below the rounding unit, and every run stalls from $n=60$ on (Table~\ref{tab:app-large}). Sec.~\ref{sec:locksigns} explains why no choice of $p_{diff}$ restores the success rate at larger $n$.

From the figures, it can also be seen that it is possible to lock some qubits to the wrong values, which would cause the algorithm to converge to a wrong solution. Sec.~\ref{sec:incorrectlocking} below analyzes these incorrect locks, Sec.~\ref{sec:locksigns} shows which signs a lock reads, and Sec.~\ref{sec:strategies} discusses other locking strategies.

\subsection{Incorrect locking}
\label{sec:incorrectlocking}
As mentioned earlier, a downside of this approach is the early locking of some qubit values: For a qubit \(q\) with the probabilities \(P_0\) for \(q=0\) and \(P_1\) for \(q=1\), let \(P_{diff} = |P_0 - P_1|\) be its measured \emph{probability difference}, which locking compares with the locking threshold \(p_{diff}\). By letting \(S_0\) and \(S_1\) denote the sets of eigenstates (basis strings) where \(q = 0\) and \(q = 1\), respectively, and \(\lambda_j\) the eigenphase of eigenstate \(j\), the probabilities after \(k\) iterations of power iteration, Eq.~\eqref{eq:qpi}, from the uniform state are:
\begin{align}
P_0^{(k)} = \frac{\sum_{j \in S_0} |1 + e^{i\lambda_j}|^{2k}}{\sum_{j} |1 + e^{i\lambda_j}|^{2k}}, & \quad \label{eq:marginal0}\\
P_1^{(k)} = \frac{\sum_{j \in S_1} |1 + e^{i\lambda_j}|^{2k}}{\sum_{j} |1 + e^{i\lambda_j}|^{2k}}. \label{eq:marginal1}
\end{align}
Here, the cardinalities of the sets are equal, \( |S_0| = |S_1| = 2^{n-1} \), and the dominant eigenvalue can be a member of the same set as the group of eigenvalues with the lowest magnitudes (this can be considered a marginal case for the qubit). For VQPM these equations hold only in the first iteration; afterwards the marginals follow Eq.~\eqref{eq:vqpm-update}, and Sec.~\ref{sec:locksigns} describes which of them a lock reads.

Therefore, a ``mistake'' (e.g., incorrectly locking \(q = 1\) while the dominant eigenstate \(d\) is in \(S_0\)) can occur if the probability difference \(P_{diff}^{(k)}\) exceeds the locking threshold \(p_{diff}\) with \(P_1^{(k)}>P_0^{(k)}\). This leads the algorithm to automatically eliminate the solution from the search space. This is shown in Fig.\ref{fig:jumps-incorrect}, where the third locking of a qubit value causes the removal of the optimum eigenvalue from the remaining search space, and the algorithm converges to a different eigenvalue. This incorrect locking can occur in any iteration; however, while in the early iterations it can occur arbitrarily, in the later iterations it is more likely to occur between the dominant eigenvalue and values that are closer to it. This can be prevented to some extent by setting \(p_{diff}\) high for early iterations and lowering it toward the later iterations. Note that a lower value of \(p_{diff}\) requires higher precision in the measurements and lower error rates in the circuit.

On the instances of Sec.~\ref{sec:smalln}, with $p_{diff}=0.01$ and six-decimal marginals, 4 to 7\% of all locks fix a qubit to the wrong value for $n=8$ to 20. Since a run makes about $n$ locks, the share of runs with at least one wrong lock grows with $n$, from 0.26 at $n=8$ to 0.44 at $n=20$. Wrong locks come early, with the first one at iteration 1 to 5 (median), and their margins $|P_0-P_1|$ are almost the same as those of correct locks (medians 0.010 to 0.011 against 0.011 to 0.013, just above $p_{diff}$). A larger fixed threshold therefore delays the locks rather than separating right from wrong ones. Since the marginals are exact here, the wrong locks come from the update itself, not from measurement noise; Sec.~\ref{sec:locksigns} shows which signs a lock reads.

Squeezing the energies into the eigenphases in \([0, \pi/2]\) does not lose them to numerical precision: the gap between the two lowest phases is $2.1\times10^{-3}$~rad at $n=16$ (Table~\ref{tab:app-phase}), far above double precision. What limits the resolution is the rounding of the measured marginals and, on a device, the number of shots. Our Gaussian random instances have a unique optimum with probability one, so degenerate optima play no role in the results.

Also note that in our experiments we always use the equal superposition state as an initial state. To minimize the incorrect early locking effect, one intuitive method is to start the algorithm with a different initial vector. However, this would require more parameterized gates and would change the algorithm into a variational quantum eigensolver. Instead of this, one can try setting the critical bits to different values and run the algorithm with this setting. However, for larger problem instances, it is hard to find which qubits are critical.  
Sec.~\ref{sec:strategies} describes dynamic \(p_{diff}\) settings for these cases, where a different value is used for every iteration or for each qubit based on its influence.  

\subsection{Dynamic locking thresholds}
\label{sec:strategies}
The threshold can also vary over the iterations or between qubits. Applying early locking only to the least significant bits brought no benefit in our runs, because a wrong lock of any bit removes the optimum from the search space. Starting with a high threshold and lowering it over the iterations, for example \(p_{diff}^{(k)}=p_{diff}^{(k-1)}/f(k)\) with \(f(k)=2^k\), delays the first locks; a threshold that stays high turns VQPM into the version without locking (Fig.~\ref{fig:both-no-locking}), and the measurement precision sets how low it can go. Below we test two rules on the instances of Fig.~\ref{fig:both-pdiff-fixed}.

\paragraph{A threshold from Hoeffding's inequality.}
\label{sec:hoeffding}
Hoeffding's inequality~\cite{hoeffding1994probability} bounds the deviation of the mean of independent bounded samples from its expectation. For the estimate \(\hat P_0\) of the probability \(P_0\) of measuring a qubit in \(\ket{0}\) from \(M\) shots it gives~\cite{Phillips2012ChernoffHoeffdingIA}
\begin{equation}
P\left(|\hat{P}_0 - P_0| \geq \epsilon\right) \leq 2e^{-2M\epsilon^2},
\end{equation}
so with confidence \(1-\delta\) the error is at most \(\epsilon=\sqrt{\ln(2/\delta)/(2M)}\), where \(\delta\) is the failure probability. Our implementation distributes a total failure probability \(\delta_{\text{total}}\) over the iterations as \(\delta_k=\delta_{\text{total}}/(T-k)\), where \(T\) is the iteration budget and \(k\) the current iteration, and uses the radius \(\epsilon\) as the locking threshold. The plain radius is too large for useful locking, and the simulations round exact marginals instead of drawing shots, so we set \(M=100\) and \(\delta_{\text{total}}=0.5\) and multiplies the number of samples by \(10n\):
\begin{equation}
\label{eq:hoeffdingdynamic}
       p_{diff}^{(k)} = \sqrt{\frac{\ln(2/\delta_k)}{2\times (10nM)}}.
\end{equation}
This gives a higher threshold for smaller \(n\) and a lower one for larger \(n\): between about 0.006 and 0.05 for \(n=1\) to 18 with \(T=30\), and between 0.007 and 0.013 at \(n=15\). Eq.~\eqref{eq:hoeffdingdynamic} is a heuristic. The factor \(10n\) is an empirical choice, and the split \(\delta_k\) overspends the union bound. Appendix~\ref{app:hoeffding} derives a threshold with a valid union bound over qubits and iterations (Proposition~\ref{prop:threshold}), which Sec.~\ref{sec:shots} tests on sampled marginals.

\begin{figure*}[t]
    \centering
    \begin{subfigure}[t]{0.95\linewidth}
        \centering
    \includegraphics[width=1\linewidth]{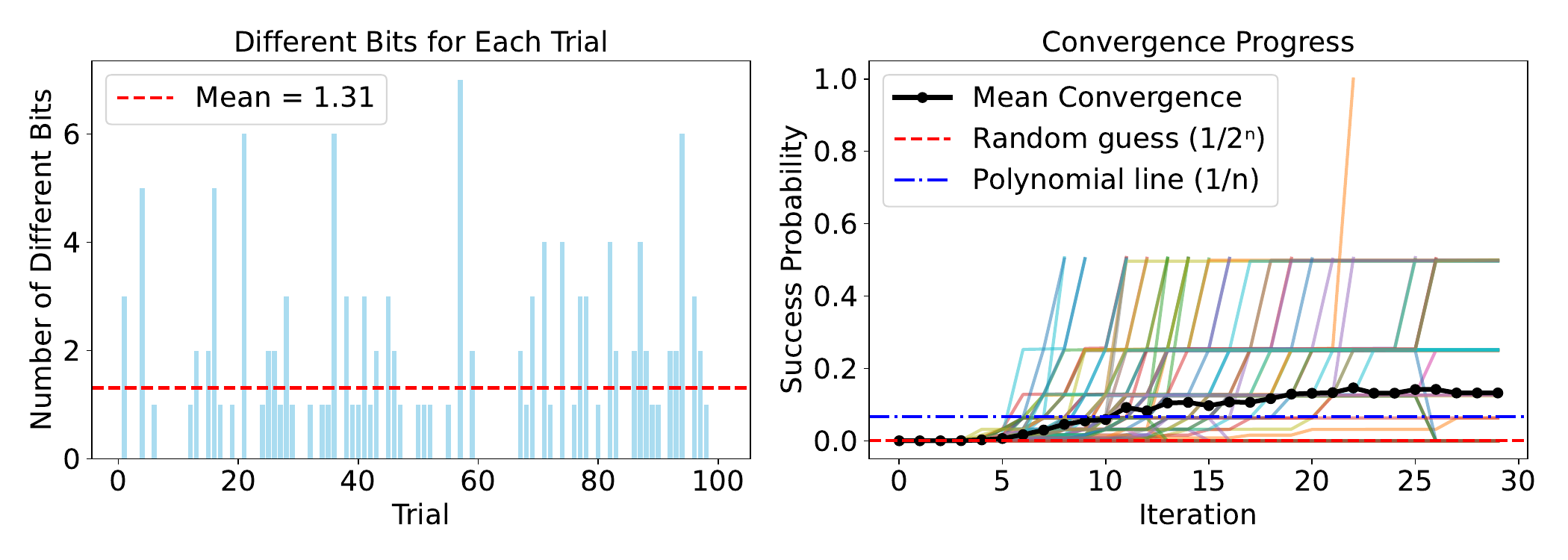}
    \caption{100 random QUBO trials for \(n=15\) with the threshold of Eq.~\eqref{eq:hoeffdingdynamic} (between 0.007 and 0.013 at \(n=15\)) and three decimals.}
    \label{fig:run-hoeffding}
    \end{subfigure}
    \begin{subfigure}[t]{0.95\linewidth}
        \centering
    \includegraphics[width=1\linewidth]{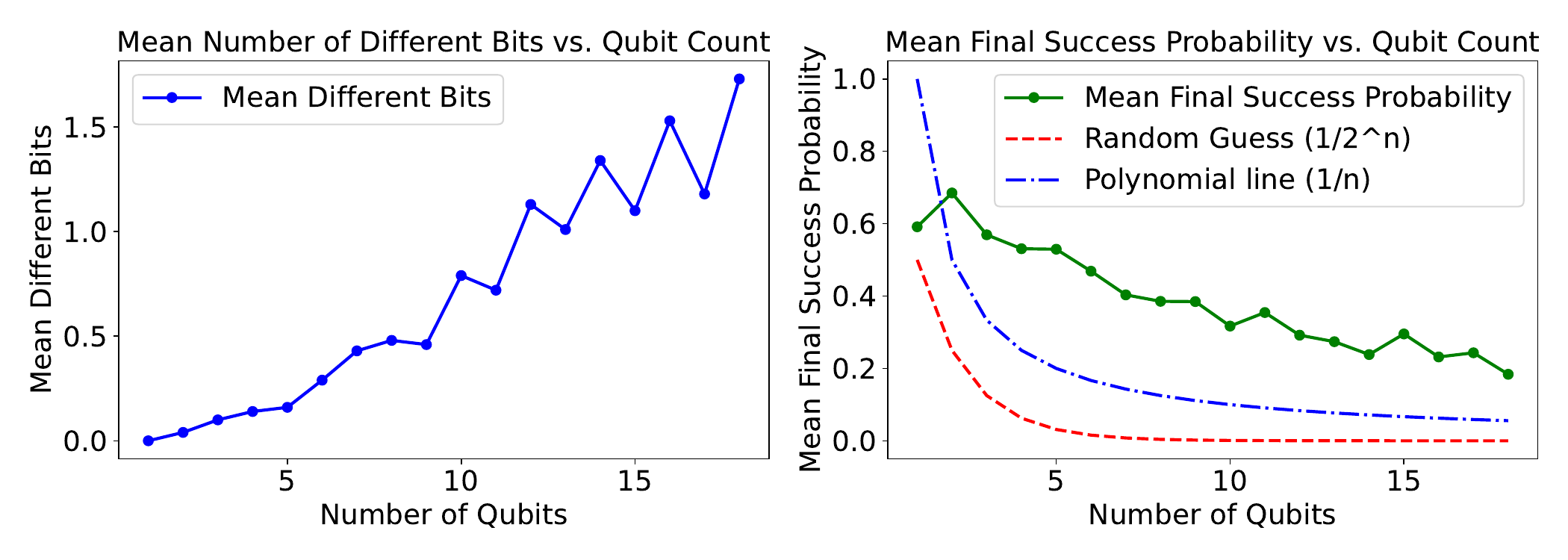}
    \caption{The convergence for the mean values of 100 random trials of QUBO problems for \(n = 1 \dots 18\) qubits. The maximum number of iterations is set to 30.}
    \label{fig:mean-hoeffding}
    \end{subfigure}
    \caption{VQPM with the dynamic threshold of Eq.~\eqref{eq:hoeffdingdynamic} (\(\delta_{total} = 0.5\), \(M=100\), \(10nM = 1000n\); between 0.006 and 0.05 for \(n=1\) to 18), at most 30 iterations and three decimals.}
    \label{fig:both-hoeffding}
\end{figure*}

Fig.~\ref{fig:both-hoeffding} shows the results with at most 30 iterations on the same instances. Overall the found-optimum rate is slightly lower than with the fixed threshold. In some instances in which the fixed threshold misses the optimum, however, the dynamic threshold finds it or a better string.

\begin{figure*}[t]
    \centering
    \begin{subfigure}[t]{0.95\linewidth}
        \centering
        \includegraphics[width=1\linewidth]{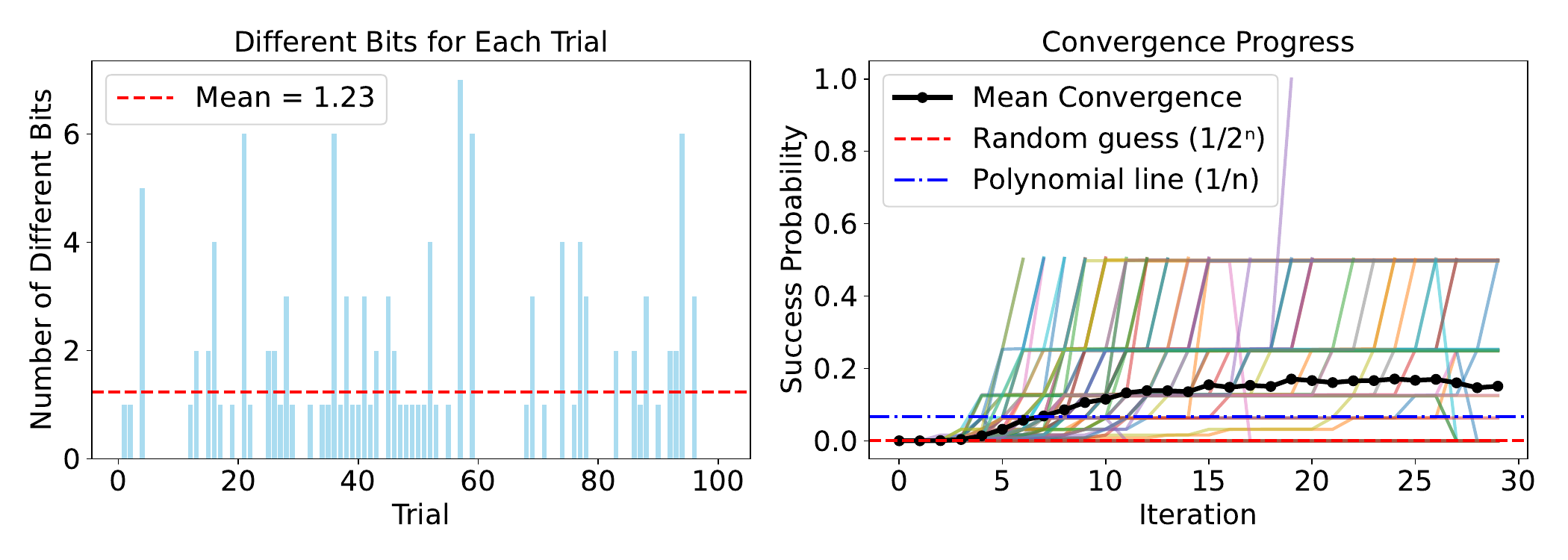}
        \caption{100 random QUBO trials for \(n=15\) with the threshold of Eq.~\eqref{eq:hoeffdingdynamic} multiplied by the influence of each qubit, and three decimals.}
        \label{fig:run-influence}
    \end{subfigure}
    \begin{subfigure}[t]{0.95\linewidth}
        \centering
        \includegraphics[width=1\linewidth]{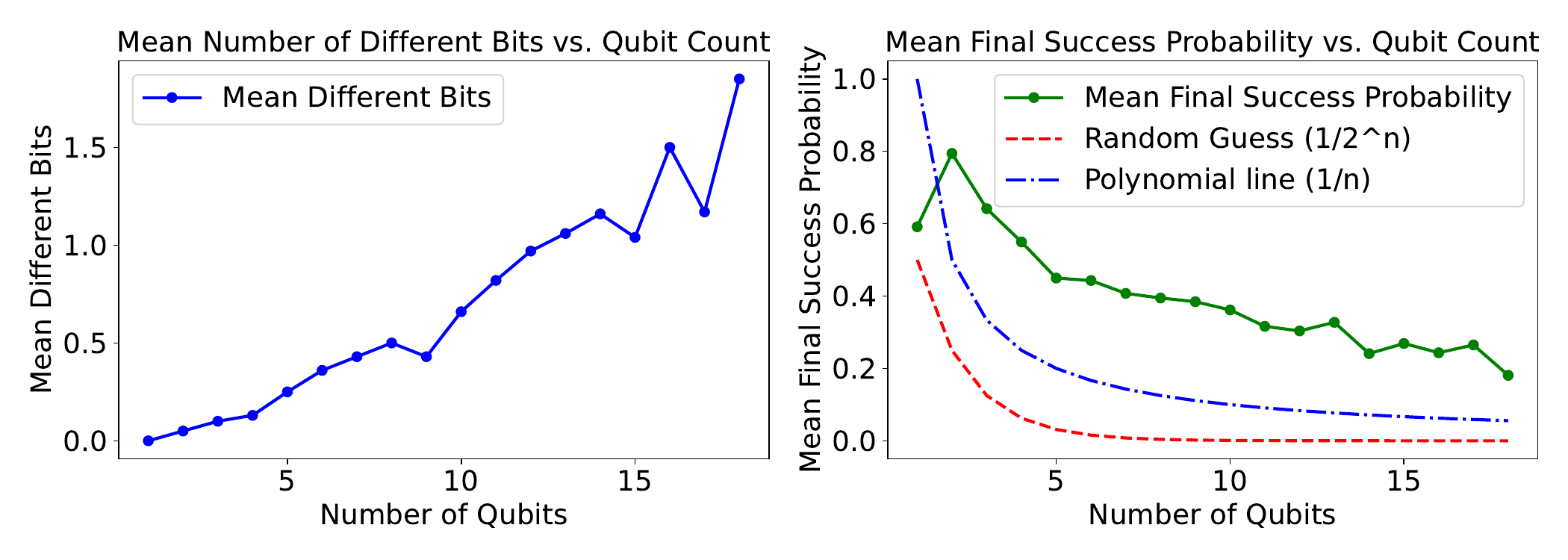}
        \caption{The convergence for the mean values of 100 random trials of QUBO problems for \(n = 1 \dots 18\) qubits. The maximum number of iterations is set to 30.}
        \label{fig:mean-influence}
    \end{subfigure}
    \caption{VQPM with the threshold of Eq.~\eqref{eq:hoeffdingdynamic} (\(\delta_{total} = 0.5\), \(M=100\)) multiplied for each qubit by its influence, Eq.~\eqref{eq:influence}; at most 30 iterations and three decimals.}
    \label{fig:both-influence}
\end{figure*}

\paragraph{Influence scores.}
\label{sec:influence}
The threshold can also differ between qubits, according to how much a change of the qubit can affect the energy. We define the influence of a parameter (a qubit) \(i\) as
\begin{equation}
\label{eq:influence}
    \text{Influence}(i) = \frac{\sum_j |Q_{ij}|}{\|Q\|_\infty},\qquad \|Q\|_\infty = \max_i\sum_j|Q_{ij}|,
\end{equation}
so that the most influential qubit has influence 1. Influence scores are also used in molecular similarity and network alignment problems, which are solved by obtaining the dominant eigenvector \cite{daskin2014multiple}. We multiply the threshold of Eq.~\eqref{eq:hoeffdingdynamic} by the influence of each qubit, so that the most influential qubits need the largest probability difference to lock. Lemma~\ref{lem:leakage} gives a reason for this choice: the influence bounds how far a single lock can move the mean energy of the product state. Fig.~\ref{fig:both-influence} shows that this gives slightly better results than the Hoeffding threshold alone. Like the fixed threshold, both rules read the signs of the update at some iteration; Sec.~\ref{sec:locksigns} shows what this implies.

\section{Classical equivalent of VQPM: sampling and a gradient method}
\label{sec:update}
For QUBO problems VQPM has an exact classical equivalent, shot-by-shot sampling (Sec.~\ref{sec:sampling}), and, to first order, a simple deterministic one, a gradient method on the mean energy (Sec.~\ref{sec:gradient}). The second shows what the update of Eq.~\eqref{eq:vqpm-update} does. We first study the update alone, with exact marginals and without locking or rounding (Secs.~\ref{sec:gradient}--\ref{sec:unlocked}), and then add locking (Sec.~\ref{sec:locksigns}). For $n\le20$ the update can be computed by enumerating all strings. For larger $n$ we use its first-order form below, which needs neither enumeration nor samples, and check it against sampled VQPM (Appendix~\ref{app:sampling}). The instances are random QUBOs from the same family as in the rest of the paper, 100 for each $n=20,30,\dots,100$. The reference string $x_{\mathrm{ref}}$ is the exhaustive optimum at $n=20$ and the best of 32 simulated-annealing runs for $n\ge30$; no run in this section found a string with lower energy.

\subsection{Shot-by-shot sampling}
\label{sec:sampling}
\begin{remark}[Classical simulability]
\label{rem:classical}
For QUBO problems, $\mathcal{U}$ is diagonal, the input of every VQPM iteration is a product state, and all qubits are measured in the computational basis. One measurement shot therefore returns the control bit $c$ and the bit string $x$ with probability $\Pr(c=0,x)=p(x)\cos^2(\lambda(x)/2)$, where $p$ is the product distribution prepared by the $R_y$ gates and $\cos^2(\lambda(x)/2)=w(x)$ is the post-selection weight of Sec.~\ref{sec:VQPM}. A classical computer can draw exactly the same samples: draw $x$ from $p$, evaluate its energy and hence $\lambda(x)$ in $O(n^2)$ operations, and accept $x$ with probability $\cos^2(\lambda(x)/2)$. The qubit probabilities used for locking can be estimated from these samples with the same number of shots, and with far fewer if each sample is weighted by $\cos^2(\lambda(x)/2)$ instead of being accepted or rejected (Appendix~\ref{app:classical}). VQPM with qubit locking is therefore a classically simulable heuristic for QUBO, and the analysis in this paper concerns its convergence and locking behaviour, not a quantum speedup.

The argument uses three features of the circuit: the input is a product state, $\mathcal U$ is diagonal, and the qubits are measured in the computational basis. If an entangling layer is placed between the $R_y$ gates and the controlled-$\mathcal U$, or the qubits are measured in another basis, the accepted shots no longer follow a product distribution reweighted by $w$, and the sampling argument fails; for QAOA circuits, already at depth one, exact classical sampling of the output is hard under standard complexity assumptions~\cite{farhi2016quantum}. Such a variant also loses the self-feeding update of Eq.~\eqref{eq:vqpm-update}. Its single-qubit marginals no longer determine the next input, so its additional parameters must be found by an outer classical optimizer, as in QAOA and VQE~\cite{peruzzo2014variational}, which brings back the training costs of those methods~\cite{bittel2021training}, or must be read from a measurement scheme that estimates more than the marginals, such as two-qubit correlations. We do not study such variants here. The proof is in Appendix~\ref{app:classical}, and the resulting limitation is discussed in Sec.~\ref{sec:conclusion}.
\end{remark}

\subsection{A gradient method on the mean energy}
\label{sec:gradient}
Let $\bar E(\pi)=\sum_xp_\pi(x)E(x)=\sum_iq_{ii}\pi_i+\sum_{i<j}q_{ij}\pi_i\pi_j$ be the \emph{mean energy} of the product distribution $p_\pi$, $g_i(\pi)=\partial\bar E/\partial\pi_i$ its \emph{gradient} with respect to the marginal $\pi_i$, and $\bar\lambda$ the \emph{mean phase}, the phase that Eq.~\eqref{eq:phasemap} gives to $\bar E(\pi)$. The phases of the strings that carry the probability lie close together; on our instances their spread is about $0.6/n$ rad at the uniform start. To first order in this spread, Eq.~\eqref{eq:vqpm-update} becomes (Proposition~\ref{prop:meanfield})
\begin{equation}
\label{eq:mf}
\pi_i'=\pi_i-\eta\,\pi_i(1-\pi_i)\,g_i(\pi),\qquad \eta=\frac{\pi\sin\bar\lambda}{8S\cos^2(\bar\lambda/2)}\approx\frac{0.33}{S}.
\end{equation}
Here $\eta$ is the \emph{step size} and $S$ the scale of the phase map, Eq.~\eqref{eq:phasemap}. In the \emph{logits} $\theta_i=\ln[\pi_i/(1-\pi_i)]$ of the marginals this reads $\theta_i'\approx\theta_i-\eta\,g_i$: every iteration moves each logit $\theta_i$ against the gradient $g_i$ of the mean energy, by the step size $\eta$. Without locking, VQPM for QUBO is therefore a continuous relaxation of the problem. It minimizes the mean energy $\bar E$, the multilinear extension of the energy, over the cube $[0,1]^n$ of marginals, starting from the centre of the cube, $\pi_i=\frac12$. In optimization terms this is mirror descent with the binary entropy~\cite{beck2003mirror}, and in continuous time a Hopfield--Tank network at zero temperature~\cite{hopfield1985neural}. These gradient methods are well known, and we do not propose Eq.~\eqref{eq:mf} as a new algorithm. We point out the equivalence because it makes VQPM analysable: it explains the behaviour of the marginals in the rest of this section, and it gives a starting point for further work, for instance for carrying step-size or locking rules from one form to the other. Its stable end points are the strings at which every single bit flip raises the energy, as for the exact update (property (ii) in Sec.~\ref{sec:VQPM}). Iterated until the marginals settle, Eq.~\eqref{eq:mf} ends at the same string as the exact update in 59 of 60 instances at $n=8$ and $12$.

Eq.~\eqref{eq:mf} also gives the size of the first step. From the uniform start, the \emph{first-iteration difference} $D_i=P_1-P_0$ of qubit $i$ is $-\eta g_i/2$, the step size times the gradient at the centre. Every qubit therefore first moves against its mean-energy gradient $g_i$, by an amount of order $\sqrt n/S$. On our instances the phase-map scale is $S\approx0.56\,n^2$, so the first-iteration difference $|D_i|$ shrinks like $n^{-3/2}$: its median falls from $1.6\times10^{-3}$ at $n=20$ to $1.4\times10^{-4}$ at $n=100$ (Appendix~\ref{app:large}).

\subsection{From the first-iteration differences to the couplings}
\label{sec:couplings}
Near the uniform start, write the marginals as $\pi=\frac12+u$, with $u$ the \emph{deviation from the centre}, and let $J$ be the \emph{coupling matrix}, $J_{ij}=J_{ji}=q_{ij}$ for $i<j$ and $J_{ii}=0$. To first order in the deviation $u$, Eq.~\eqref{eq:mf} is linear, and from $u=0$ it gives, after $t$ iterations,
\begin{equation}
\label{eq:linear}
u(t)=-\bigl(I-B^t\bigr)J^{-1}g^0,\qquad B=I-\frac{\eta_0}{4}J,
\end{equation}
where $g^0$ and $\eta_0$ are the gradient and the step size at the centre $\pi=\frac12$, and $B$ is the \emph{linearized iteration matrix}. Let $\mu_{\min}<0$ be the lowest eigenvalue of the coupling matrix $J$, and let $\tau=4/(\eta_0|\mu_{\min}|)$ be the \emph{turning time}. For $t\ll\tau$, the deviation is $u(t)\approx\frac t2D$: every qubit keeps moving in the direction of its first-iteration difference $D_i$. For $t\gtrsim\tau$, the components of the deviation $u$ along the eigenvectors of $J$ with negative eigenvalues grow geometrically, while the others level off, and the marginals turn toward the lowest eigenvectors of the couplings (Appendix~\ref{app:flow}).

The product state thus runs a power iteration of its own. The filter $(I+\mathcal{U})^k$ amplifies the eigenvector of $\mathcal{U}$ with the dominant eigenvalue, which is the ground state, in the space of all $2^n$ strings. The product state keeps only the $n$ marginals, and on them the filter acts, to first order, as power iteration with the $n\times n$ linearized iteration matrix $B$. The dominant eigenvectors of $B$ are the lowest eigenvectors of the coupling matrix $J$, the directions of the spectral relaxation $\min_{\|y\|=1}y^TJy$ of the problem, not the ground state. On our instances the turning time $\tau$ grows from 275 iterations at $n=20$ to 2\,610 at $n=100$, about $2.6\,n^{3/2}$.

\subsection{Where the unlocked update goes}
\label{sec:unlocked}
\begin{figure*}[!tp]
\centering
\includegraphics[width=0.92\linewidth]{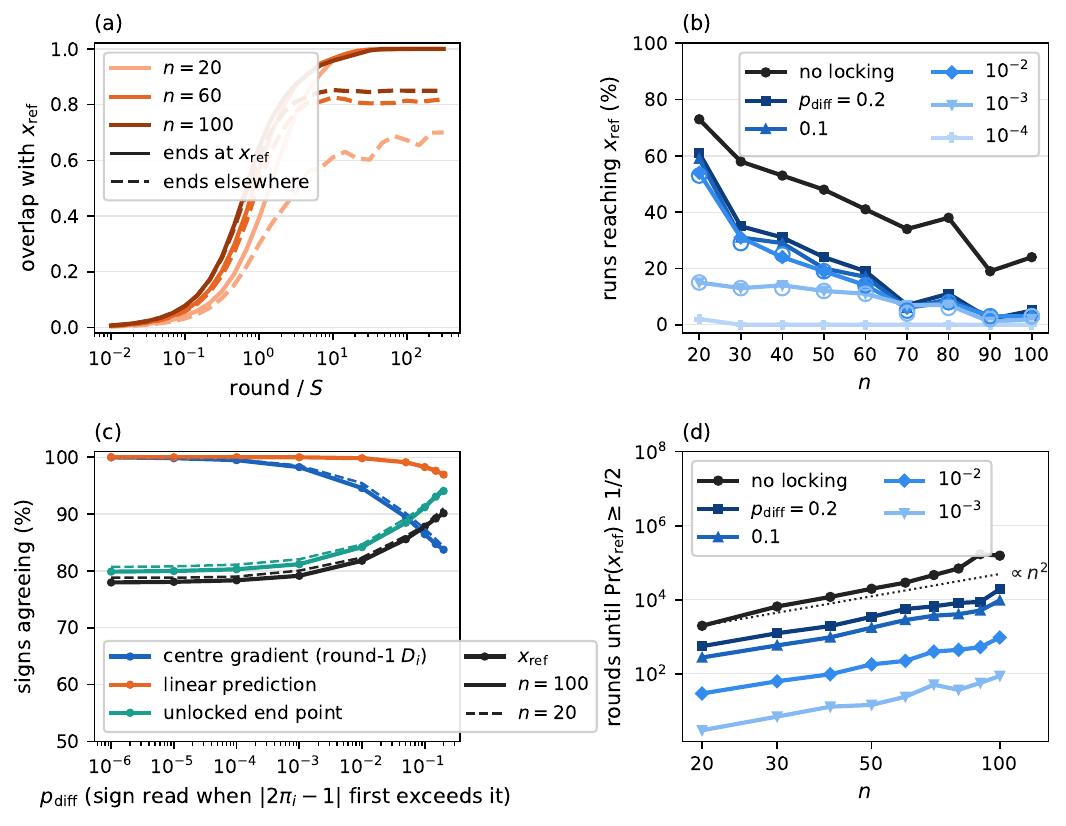}
\caption{The product-state map without and with locking, in the first-order form of Eq.~\eqref{eq:mf}, 100 instances per $n$. (a) Overlap $m(t)$ with $x_{\mathrm{ref}}$, Eq.~\eqref{eq:overlap}, along the unlocked map, against rounds in units of $S$: median over the runs that end at $x_{\mathrm{ref}}$ (solid) and over the others (dashed). (b) Share of runs that reach $x_{\mathrm{ref}}$, without locking and with locking at several thresholds (Table~\ref{tab:app-lockflow}); open circles: sampled VQPM at $10^{-2}$ and $10^{-3}$ (Table~\ref{tab:app-sweep}). (c) Signs read on the unlocked map in the round in which $|2\pi_i-1|$ first exceeds $p_{\mathrm{diff}}$, compared with the round-1 $D_i$, with the linear prediction of Eq.~\eqref{eq:linear}, with the unlocked end point and with $x_{\mathrm{ref}}$ (Table~\ref{tab:app-read}); solid $n=100$, dashed $n=20$. (d) Median round in which $x_{\mathrm{ref}}$ first reaches probability $1/2$, over the runs in which it does, with an $n^2$ reference line.}
\label{fig:app-flow}
\end{figure*}

To follow a run iteration by iteration we use its \emph{overlap} with the reference string $x_{\mathrm{ref}}$,
\begin{equation}
\label{eq:overlap}
m(t)=\frac1n\sum_is_i\bigl(2\pi_i(t)-1\bigr),\qquad s_i=2x_{\mathrm{ref},i}-1,
\end{equation}
where $s_i=\pm1$ are the \emph{reference signs}. The overlap $m(t)$ is 0 at the uniform start and 1 at $x_{\mathrm{ref}}$. Fig.~\ref{fig:app-flow}(a) shows its median over the runs that end at $x_{\mathrm{ref}}$ and over those that end elsewhere. Both rise together for about the first $S$ iterations (one to two turning times $\tau$), while the signs are being set. The runs that miss then level off at an overlap of $1-2h/n$, where $h$ is their distance in bits from $x_{\mathrm{ref}}$. The share of instances in which the unlocked update ends at $x_{\mathrm{ref}}$ falls from about three quarters at $n=20$ to a quarter at $n=100$ (Table~\ref{tab:app-lockflow}). When it misses, it ends at a 1-flip local minimum 3 to 8 bits from $x_{\mathrm{ref}}$, with a median energy gap of 0.6 to 1.3\% (Table~\ref{tab:app-flow}). It is also slow: the number of iterations until the median run settles grows like $n^{2.1}$ (Table~\ref{tab:app-lockflow}, bottom). With exact marginals, VQPM without locking is thus a local heuristic whose chance of ending at the optimum falls as $n$ grows and the energy landscape gains more local minima (Proposition~\ref{prop:stability}).

\subsection{The signs that locking reads}
\label{sec:locksigns}
Locking fixes qubit $i$ in the first iteration in which $|P_0-P_1|=|2\pi_i-1|$ exceeds the locking threshold $p_{diff}$, that is, when its logit $\theta_i$ passes $\pm\ln\frac{1+p_{diff}}{1-p_{diff}}\approx\pm2p_{diff}$. On the update of Sec.~\ref{sec:update} this has a simple meaning: locking reads the sign of the qubit at a time set by the threshold $p_{diff}$ and freezes it. In the first regime of Eq.~\eqref{eq:linear}, $|2\pi_i-1|$ grows like $t|D_i|$, so qubit $i$ reaches the threshold after about $p_{diff}/|D_i|$ iterations, the threshold divided by its first-iteration difference. A lock also shifts the other qubits, whose mean-energy gradients $g_j$ now see $\pi_i=0$ or $1$, and it ends the run sooner.

We ran Eq.~\eqref{eq:mf} with the locking rule of Sec.~\ref{sec:locking}, without rounding and without the stop rule that uses the known optimum, for $p_{diff}=0.2$, $0.15$, $0.1$, $0.05$ and $10^{-2},\dots,10^{-6}$ on the instances of Sec.~\ref{sec:update}. A run counts as a success if $x_{\mathrm{ref}}$ reaches probability $1/2$ before any lock contradicts it, the stop test of our simulations. Table~\ref{tab:app-lockflow} and Fig.~\ref{fig:app-flow}(b--d) give the results.
\begin{itemize}
\item \emph{Small thresholds read the first-iteration differences.} Thresholds up to $10^{-3}$ are passed within about ten iterations, and the sign read is that of the first-iteration difference $D_i$ for 98 to 100\% of the qubits. These signs agree with $x_{\mathrm{ref}}$ for only 78 to 80\% of the qubits, so with $n$ locks per run at least one is almost always wrong: from $10^{-4}$ down, no run reaches $x_{\mathrm{ref}}$ for $n\ge30$. On a device such thresholds are also costly, because resolving the median first-iteration difference $|D_i|$ at three standard deviations takes a number of shots per iteration that grows like $n^3$, about $5\times10^8$ at $n=100$ (Appendix~\ref{app:large}).
\item \emph{Larger thresholds read the couplings.} At $p_{diff}=0.2$ the sign is read after about half of the turning time $\tau$, a median of 1\,390 iterations at $n=100$. By then the couplings have turned 16\% of the signs away from the first-iteration differences $D_i$. Eq.~\eqref{eq:linear} predicts 94\% of these turns, and the signs read agree with $x_{\mathrm{ref}}$ for 90\% of the qubits (Fig.~\ref{fig:app-flow}(c), Table~\ref{tab:app-read}).
\item \emph{No threshold beats the unlocked update.} At every $n$ the share of runs that reach $x_{\mathrm{ref}}$ is highest without locking (Table~\ref{tab:app-lockflow}). For $p_{diff}\ge10^{-2}$, 95 to 97\% of the locks agree with the string at which the unlocked update ends, and locking turns a miss of the unlocked update into a success in at most 5 instances in 100. A run fails at its first wrong lock, and with $n$ locks the few percent of signs that would still turn later are enough: at $0.2$ a run makes on average 1.1 wrong locks at $n=20$ and 7.8 at $n=100$.
\item \emph{Locking is a speed-up.} At $n=100$ it cuts the median number of iterations by a factor of about 25 at $p_{diff}=0.2$ and 450 at $10^{-2}$ (Table~\ref{tab:app-lockflow}, bottom). For $p_{diff}\ge10^{-2}$ the count grows like $n^2$, and the first lock comes after about $p_{diff}/|D_i|$ iterations (threshold over first-iteration difference), which grows like $n^{1.4}$.
\end{itemize}
Locking therefore does not steer the iteration toward the optimum. It follows the update, makes it faster, and freezes its signs early; the threshold $p_{diff}$ decides how far along the update the signs are read. Sampled VQPM, with marginals estimated as on a device, gives the same outcome as Eq.~\eqref{eq:mf} on 98 to 100\% of the instances (Appendix~\ref{app:lockflow}). The dynamic thresholds of Sec.~\ref{sec:strategies} also read the signs of the update, only at times that differ between iterations or qubits; we did not run them on the first-order update.

\begin{table}[ht]
\centering
\small
\setlength{\tabcolsep}{4pt}
\caption{Eq.~\eqref{eq:mf} with the locking rule of Sec.~\ref{sec:locking} and without rounding, 100 instances per $n$. Each run continues until all qubits are locked or the free ones have settled at a vertex that is stable for them (no oracle stop); the unlocked runs are those of Table~\ref{tab:app-flow}. Top: runs that pass the stop test, $x_{\mathrm{ref}}$ reaching probability $1/2$ before any lock contradicts it (\%). Bottom: median rounds until the run ends.}
\label{tab:app-lockflow}
\begin{tabular}{lrrrrrrrrr}
\toprule
$p_{\mathrm{diff}}$ & $n=20$ & 30 & 40 & 50 & 60 & 70 & 80 & 90 & 100 \\
\midrule
\multicolumn{10}{l}{\emph{runs that reach $x_{\mathrm{ref}}$ (\%)}} \\
no locking & 73 & 58 & 53 & 48 & 41 & 34 & 38 & 19 & 24 \\
$0.2$ & 61 & 35 & 31 & 24 & 19 & 7 & 11 & 2 & 5 \\
$0.15$ & 59 & 34 & 31 & 22 & 17 & 7 & 11 & 2 & 4 \\
$0.1$ & 59 & 31 & 29 & 20 & 17 & 6 & 9 & 3 & 3 \\
$0.05$ & 59 & 31 & 27 & 20 & 16 & 6 & 9 & 3 & 3 \\
$10^{-2}$ & 54 & 31 & 24 & 19 & 14 & 6 & 8 & 3 & 3 \\
$10^{-3}$ & 15 & 13 & 14 & 12 & 11 & 7 & 7 & 1 & 2 \\
$10^{-4}$ & 2 & 0 & 0 & 0 & 0 & 0 & 0 & 0 & 0 \\
$\le10^{-5}$ & 0 & 0 & 0 & 0 & 0 & 0 & 0 & 0 & 0 \\
\midrule
\multicolumn{10}{l}{\emph{rounds until the run ends (median)}} \\
no locking & 13\,714 & 44\,762 & 60\,400 & 106\,826 & 165\,967 & 182\,482 & 275\,060 & 359\,390 & 563\,404 \\
$0.2$ & 978 & 2\,631 & 3\,752 & 6\,337 & 9\,532 & 11\,630 & 16\,129 & 21\,744 & 23\,148 \\
$0.15$ & 704 & 2\,074 & 3\,010 & 4\,538 & 6\,992 & 8\,244 & 12\,908 & 16\,116 & 18\,091 \\
$0.1$ & 476 & 1\,508 & 1\,924 & 3\,256 & 4\,712 & 5\,966 & 8\,200 & 10\,759 & 11\,668 \\
$0.05$ & 236 & 760 & 1\,072 & 1\,577 & 2\,184 & 2\,833 & 4\,230 & 5\,605 & 6\,202 \\
$10^{-2}$ & 53 & 132 & 200 & 295 & 456 & 542 & 786 & 1\,034 & 1\,271 \\
$10^{-3}$ & 4 & 11 & 20 & 32 & 50 & 52 & 78 & 108 & 116 \\
$10^{-4}$ & 1 & 2 & 2 & 3 & 4 & 5 & 6 & 8 & 10 \\
$10^{-5}$ & 1 & 1 & 1 & 1 & 2 & 2 & 2 & 2 & 2 \\
$10^{-6}$ & 1 & 1 & 1 & 1 & 1 & 1 & 1 & 1 & 1 \\
\bottomrule
\end{tabular}
\end{table}

\section{Comparison with QAOA and simulated annealing}
\label{sec:comparison}
We compare VQPM with QAOA~\cite{farhi2014quantum} and with simulated annealing~\cite{kirkpatrick1983optimization} in two steps. Sec.~\ref{sec:smalln} runs all three methods on the same instances up to $n=16$, where QAOA can still be simulated with full state vectors. Sec.~\ref{sec:largen} then follows VQPM to $n=100$ with its classical equivalent (Sec.~\ref{sec:update}), which a QAOA simulation cannot reach. The second step is needed because the first one alone says little about how the methods behave as $n$ grows: up to $n=16$ every method finds a good string, and the differences between them are of the same size as the differences between QAOA optimizers. Since these differences are of the order of 0.1 in the found-optimum rate, we use 50 to 100 instances per $n$ and several starts per QAOA run.

\subsection{Instances up to \texorpdfstring{$n=16$}{n=16}}
\label{sec:smalln}
\paragraph{Instances.} The instances are random QUBOs from the family used throughout the paper, $Q=G+G^T$ with $G_{ij}\sim\mathcal N(0,1)$, drawn with seed 2026: 100 instances for each $n=4,6,8,10$ and 50 for each $n=12,14,16$. The exact ground state of every instance is found by enumerating its $2^n$ energies.

\paragraph{QAOA.} We simulate QAOA with exact state vectors on the scaled QUBO of Eq.~\eqref{eq:phasemap}, at depths $p=2,4,6,8$. How well QAOA does depends on the classical optimizer of its $2p$ angles, on how the angles are initialized, and on the depth $p$. We therefore use four optimizers,
\begin{itemize}[noitemsep]
\item L-BFGS-B~\cite{byrd1995limited}, a quasi-Newton method, with exact gradients;
\item Adam~\cite{kingma2015adam}, gradient descent with adaptive step sizes, with learning rate 0.3;
\item COBYLA~\cite{powell1994direct}, a derivative-free method;
\item plain gradient descent with learning rate 1;
\end{itemize}
and two kinds of start:
\begin{itemize}[noitemsep]
\item five random restarts, with angles drawn uniformly from $[0,\pi)$ and the same angles for every optimizer, keeping the restart with the lowest final energy;
\item the INTERP strategy of Zhou et al.~\cite{zhou2020quantum}, which interpolates the optimized angles at depth $p$ into a start for depth $p+1$, beginning from five random starts at depth 1.
\end{itemize}
The learning rates were chosen on 20 separate tuning instances at $n=8$. On these instances a pilot at $p=8$ with random starts gave the optimum a probability of 0.61 in the final state of L-BFGS-B and 0.13 in that of COBYLA. Each optimizer runs for at most 500 iterations. We measure the cost of QAOA as the number of evaluations of its expected energy, counting one gradient as $2p+1$ evaluations, the cost of finite differences. QAOA returns the most probable string of its final state.

\paragraph{VQPM and simulated annealing.} VQPM uses the setting of Sec.~\ref{sec:locking}: marginals rounded to three decimals, locking threshold $p_{diff}=0.01$, and at most 30 or 100 iterations. We use two stop rules. The \emph{oracle rule}, used in Sec.~\ref{sec:VQPM}, stops when the known optimum reaches probability $1/2$. The \emph{max-prob rule} stops when any string reaches probability $1/2$ or all qubits are locked, and needs no knowledge of the optimum. VQPM returns the most probable string of its final state. Simulated annealing runs one chain of single-bit flips from a random string, with a geometric temperature schedule from half the typical energy change of a flip down to a thousandth of it. It makes 10, 100 or 1000 sweeps of $n$ proposals each, and returns the best string it has visited. Each proposal costs $O(n)$ operations.

\paragraph{Metrics.} For each run we record whether the returned string is the ground state, the probability $P(\mathrm{opt})$ of the ground state in the final state, the \emph{bit difference} (Hamming distance) between the returned string and the ground state, the \emph{normalized energy gap} $(E_{\mathrm{found}}-E_{\min})/(E_{\max}-E_{\min})$ of the returned string, and the normalized gap of the expected energy of the final state. The \emph{found-optimum rate} is the share of instances in which the returned string is the ground state. The bit difference treats all bit flips alike: for the target \texttt{1111}, the strings \texttt{1110} and \texttt{0111} both differ by one bit, though their energies can differ a lot. The energy gaps measure what the methods optimize: QAOA minimizes its expected energy directly, and by Sec.~\ref{sec:gradient} the VQPM update is, to first order, a gradient method on the mean energy of a product distribution. Intervals are 95\% Wilson intervals~\cite{wilson1927probable}; with 50 instances their half width is up to 0.14.

\begin{table}[ht]
\centering
\small
\setlength{\tabcolsep}{5pt}
\caption{Found-optimum rate on the instances of Sec.~\ref{sec:smalln} (100 instances per $n$ for $n\le10$, 50 for $n\ge12$), with QAOA at depth $p=8$, and the cost per instance at $n=16$: evaluations of the expected energy for QAOA, iterations (mean) for VQPM and single-flip proposals for simulated annealing. VQPM: three decimals, $p_{diff}=0.01$, at most 100 iterations.}
\label{tab:qaoa}
\begin{tabular}{l cccc r}
\toprule
 & $n=4$ & $n=8$ & $n=12$ & $n=16$ & cost at $n=16$ \\
\midrule
QAOA, L-BFGS-B, random restarts & 0.96 & 0.90 & 0.86 & 0.74 & $4.6\times10^4$ \\
QAOA, L-BFGS-B, INTERP & 0.97 & 0.90 & 0.86 & 0.78 & $3.2\times10^4$ \\
QAOA, Adam, random restarts & 0.93 & 0.92 & 0.86 & 0.68 & $4.3\times10^4$ \\
QAOA, Adam, INTERP & 0.93 & 0.91 & 0.86 & 0.76 & $1.3\times10^5$ \\
QAOA, COBYLA, random restarts & 0.89 & 0.59 & 0.62 & 0.56 & $2.5\times10^3$ \\
QAOA, COBYLA, INTERP & 0.95 & 0.91 & 0.78 & 0.62 & $2.0\times10^4$ \\
QAOA, gradient descent, random restarts & 0.86 & 0.61 & 0.64 & 0.58 & $4.3\times10^4$ \\
QAOA, gradient descent, INTERP & 0.94 & 0.83 & 0.66 & 0.40 & $2.0\times10^5$ \\
\midrule
VQPM, oracle rule & 0.80 & 0.69 & 0.62 & 0.46 & 33 \\
VQPM, max-prob rule & 0.77 & 0.59 & 0.52 & 0.42 & 50 \\
\midrule
Simulated annealing, 10 sweeps & 0.96 & 0.91 & 0.92 & 0.94 & 160 \\
Simulated annealing, 100 sweeps & 1.00 & 1.00 & 0.98 & 0.98 & 1\,600 \\
\bottomrule
\end{tabular}
\end{table}

\begin{figure*}[t]
\centering
\includegraphics[width=0.92\textwidth]{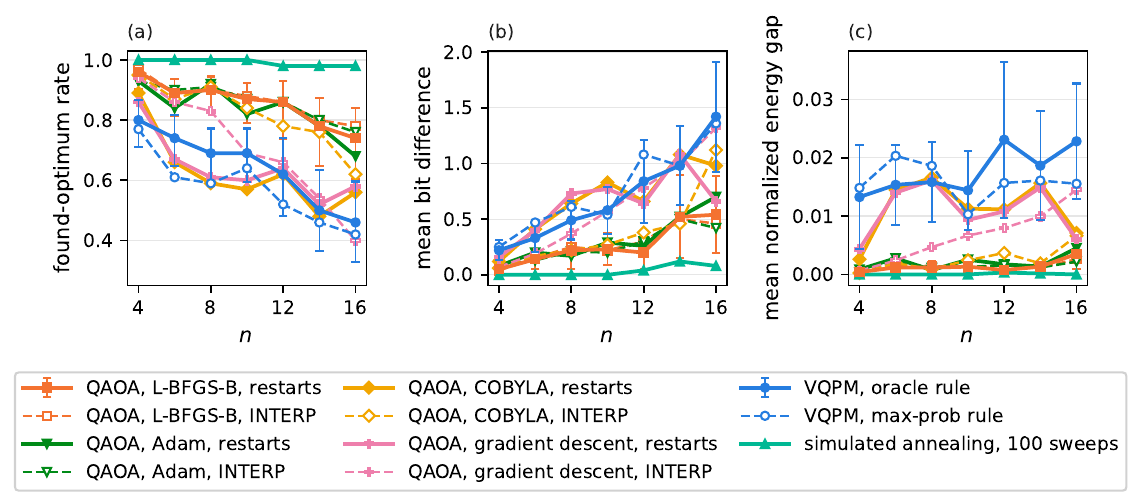}
\caption{QAOA at depth $p=8$ with four optimizers (solid: five random restarts; dashed: INTERP), VQPM (three decimals, $p_{diff}=0.01$, at most 100 iterations; oracle and max-prob rules) and simulated annealing with 100 sweeps, on the instances of Sec.~\ref{sec:smalln}: (a) found-optimum rate, (b) mean bit difference to the ground state, (c) mean normalized energy gap of the returned string. Error bars (95\%) are shown for QAOA with L-BFGS-B and restarts and for VQPM with the oracle rule; the other series have intervals of similar width. Other depths and the gap of the expected energy are given with the data in the repository.}
\label{fig:qaoa}
\end{figure*}

\paragraph{Results.} Table~\ref{tab:qaoa} and Fig.~\ref{fig:qaoa} give the results at depth $p=8$.
\begin{itemize}
\item \emph{QAOA depends on its optimizer and its start.} With a well-chosen optimizer and start, QAOA finds the optimum more often than VQPM with locking and three decimals at every $n$, and the 95\% intervals of the two overlap at most at their ends: at $n=16$ its found-optimum rate is 0.74 to 0.78 for the best settings, against 0.46 for VQPM. With a weaker choice, such as a derivative-free optimizer or plain gradient descent from random starts, QAOA falls to the level of VQPM, and the INTERP start helps some optimizers and hurts others. In this setting VQPM with locking is therefore comparable only to a weakly optimized QAOA.
\item \emph{The accuracy of QAOA can be raised with its depth.} The decline of QAOA with $n$ in Table~\ref{tab:qaoa} is at a fixed depth. QAOA has a guarantee that VQPM lacks: as $p\to\infty$, suitable angles let it follow the adiabatic path to the ground state~\cite{farhi2014quantum}, so its accuracy at a given $n$ can be raised by adding layers, at the price of deeper circuits and more angles to optimize. In our runs with the best optimizer, the found-optimum rate at $n=16$ rises from 0.30 at $p=2$ to 0.74--0.78 at $p=8$. VQPM has no such parameter: its runs end at the stable points of its update (Sec.~\ref{sec:unlocked}), however many iterations they use.
\item \emph{All methods return strings close in energy to the optimum} (Fig.~\ref{fig:qaoa}c). When VQPM misses the optimum it usually ends at a nearby local minimum, as Sec.~\ref{sec:unlocked} predicts. The probability of the optimum in the final state declines with $n$ for both QAOA and VQPM, to about 0.3 at $n=16$.
\item \emph{The costs are of different kinds} (last column of Table~\ref{tab:qaoa}). QAOA uses tens of thousands of evaluations of its expected energy per instance, VQPM tens of iterations. Our QAOA energies are exact, so these counts do not include shots. Sec.~\ref{sec:cost} compares the costs of all methods.
\item \emph{Simulated annealing solves almost every instance}, with at most 1\,600 single-flip proposals at $n=16$, and all of them with 1000 sweeps. Since VQPM for QUBO is itself a classical sampling method (Remark~\ref{rem:classical}), a classical heuristic of this kind is the baseline it has to be measured against.
\end{itemize}
These runs have limits. All instances are dense random QUBOs; QAOA is run at depths up to 8 and without shot noise, with learning rates tuned at $n=8$ only; and VQPM uses three-decimal rounding and at most 100 iterations, which by Sec.~\ref{sec:locking} lowers its found-optimum rate for $n\ge9$. With exact marginals and without locking, VQPM is on a par with QAOA (Sec.~\ref{sec:largen}).

\subsection{Larger instances with classical VQPM}
\label{sec:largen}
\begin{figure*}[t]
\centering
\includegraphics[width=0.92\textwidth]{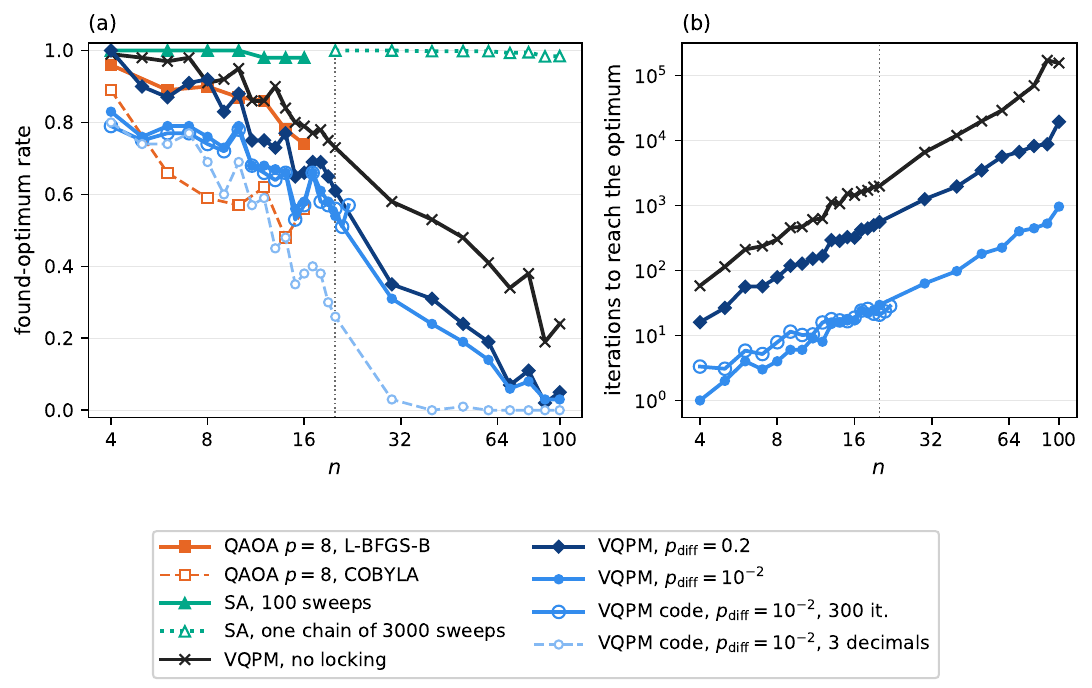}
\caption{(a) Found-optimum rate from $n=4$ to 100. QAOA ($p=8$, random restarts) and simulated annealing with 100 sweeps: instances of Sec.~\ref{sec:smalln}. VQPM without locking and with $p_{diff}=0.2$ or $10^{-2}$ (solid lines, filled markers): first-order update without rounding and without an iteration limit, on the instances of Sec.~\ref{sec:smalln} for $n<20$ and on those of Sec.~\ref{sec:update} for $n\ge20$ (Table~\ref{tab:app-lockflow}). VQPM code (open markers): the experiment code of Sec.~\ref{sec:smalln} with exact marginals and at most 300 iterations, or with three decimals and at most 100 iterations, for $n\le22$; the three-decimal loop with at most 30 iterations on the instances of Sec.~\ref{sec:update} for $n\ge20$ (Appendix~\ref{app:large}). A run counts as a success if the string it returns is the optimum. Simulated annealing for $n\ge20$: share of single chains of 3000 sweeps that reach $x_{\mathrm{ref}}$, which is itself the best of 32 such chains for $n\ge30$. The dotted line at $n=20$ marks the change of instance set. (b) VQPM iterations until $x_{\mathrm{ref}}$ reaches probability $1/2$: median over the runs that reach it for the first-order update (lines), and the mean number of iterations of the VQPM code with exact marginals and $p_{diff}=10^{-2}$ (open circles).}
\label{fig:scaling}
\end{figure*}

For QUBO problems VQPM can be simulated classically (Remark~\ref{rem:classical}), and to first order its update is the gradient method of Sec.~\ref{sec:gradient}. Neither needs the $2^n$-dimensional state vector, so we can follow VQPM to $n=100$, far beyond the reach of a state-vector QAOA simulation. Fig.~\ref{fig:scaling} combines three sets of runs: those of Sec.~\ref{sec:smalln}; VQPM with exact marginals for $n=4$ to 22 on the instances of Sec.~\ref{sec:smalln} (100 per $n$, at most 300 iterations); and the first-order update without locking and with the locking thresholds $p_{diff}=0.2$ and $10^{-2}$, on the same instances for $n=4$ to 22 and on the instances of Sec.~\ref{sec:update} for $n=20$ to 100 (100 per $n$, no rounding, run until all qubits are locked or the free ones have settled). For $n\le14$ we also iterated the exact update, Eq.~\eqref{eq:vqpm-update}, with the same thresholds: it returns the same string as the first-order update in 98 to 99.5\% of the runs. For $n=20$ to 100 the first-order update and sampled VQPM agree on the outcome in 98 to 100\% of the instances (Appendix~\ref{app:lockflow}). The two instance sets overlap at $n=20$, where their found-optimum rates differ by at most 0.03.

\begin{itemize}
\item \emph{Without locking, VQPM is on a par with QAOA up to $n=16$, but slow.} With exact marginals and no locking, the update finds the optimum as often as QAOA with its best optimizer (0.79 against 0.74 at $n=16$, Table~\ref{tab:cost}). It needs about 100 times more iterations for this than with locking (Fig.~\ref{fig:scaling}b): a median of $1.4\times10^3$ at $n=16$ until the optimum reaches probability $1/2$, against 17 with $p_{diff}=10^{-2}$. Locking lowers the found-optimum rate, and no threshold does better than the update without locking by more than 0.01 at any $n$ (Sec.~\ref{sec:locksigns}).
\item \emph{VQPM keeps declining beyond $n=16$} (Fig.~\ref{fig:scaling}a). At $n=100$ it reaches $x_{\mathrm{ref}}$ in 24\% of the instances without locking and in at most 5\% with locking (Table~\ref{tab:cost}). With three-decimal rounding and 30 iterations the decline is faster, because rounding erases the first-iteration differences and the run stalls (Appendix~\ref{app:large}).
\item \emph{The number of iterations grows like $n^2$ or faster} (Fig.~\ref{fig:scaling}b; fitted exponents 2.0 to 2.7 for $n\ge20$), with and without locking. A fixed budget of 30 iterations, as in Sec.~\ref{sec:locking}, is therefore too small from about $n=13$ on.
\item \emph{At large $n$, simulated annealing is measured only against itself.} For $n\ge30$ the optimum is unknown, and $x_{\mathrm{ref}}$ is the best string found by 32 chains of simulated annealing. A single chain reaches $x_{\mathrm{ref}}$ in 98 to 100\% of its runs at every $n$ up to 100, but this shows how often the chains agree with each other, not how often they find the optimum. Its rate in Fig.~\ref{fig:scaling}a for $n\ge30$ is therefore an upper bound that carries its accuracy at small $n$ forward, and the true rate may well decline with $n$. The rates of VQPM for $n\ge30$ use the same reference; no VQPM run found a string with lower energy.
\end{itemize}

Taken together, QAOA with a well-chosen optimizer is ahead of VQPM with locking wherever we can simulate both, and level with VQPM without locking, which however needs about 100 times more iterations than with locking. At larger $n$ VQPM declines further with and without locking. VQPM therefore offers no advantage in solution quality on these instances. What it offers is a transparent mechanism: by Secs.~\ref{sec:update} and~\ref{sec:locksigns}, each of its runs is a gradient method on the mean energy whose signs are read and frozen by locking, and its failures are the local minima of that method.

\section{Finite shots, circuit size, noise and cost}
\label{sec:noise}
The results so far use exact marginals or the first-order update. On a device, each iteration estimates the marginals from a finite number of shots, runs a deep controlled-$\mathcal U$ circuit, and suffers gate and readout errors. We study the three effects separately and then compare the cost per instance of all methods, on a classical computer and on a device.

\subsection{Finite shots}
\label{sec:shots}
We sampled VQPM shot by shot as in Remark~\ref{rem:classical}. In each iteration $M$ shots are drawn; the number $M'$ that pass the post-selection is binomial with the control probability $W$, and the marginals are the frequencies among the accepted shots, rounded to three decimals. We used 100 instances for each $n=8,10,\dots,16$ (those of Sec.~\ref{sec:smalln} and their extension to 100 instances), $M=10^3$ to $10^6$ shots per iteration, at most 100 iterations and the oracle rule, and returned the most frequent string of the last iteration. Table~\ref{tab:shots} gives the results for two locking rules.

\begin{table}[ht]
\centering
\small
\setlength{\tabcolsep}{5pt}
\caption{VQPM with marginals estimated from $M$ shots per iteration (100 instances per $n$, three decimals, oracle rule, at most 100 iterations). Top: found-optimum rate. Bottom, for $M=10^6$: share of runs with at least one wrong lock, mean number of iterations and mean number of shots per run. Fixed: $p_{diff}=0.01$. Hoeffding: lock only if $|P_0-P_1|>2\epsilon+10^{-3}$ with $\epsilon=\sqrt{\ln(2nT/\delta)/(2M')}$, $\delta=0.05$, $T=100$.}
\label{tab:shots}
\begin{tabular}{l ccc ccc}
\toprule
 & \multicolumn{3}{c}{fixed $p_{diff}$} & \multicolumn{3}{c}{Hoeffding} \\
 \cmidrule(lr){2-4}\cmidrule(lr){5-7}
$M$ & $n=8$ & $n=12$ & $n=16$ & $n=8$ & $n=12$ & $n=16$ \\
\midrule
$10^3$ & 0.02 & 0.00 & 0.00 & 0.20 & 0.03 & 0.01 \\
$10^4$ & 0.12 & 0.01 & 0.00 & 0.47 & 0.16 & 0.08 \\
$10^5$ & 0.54 & 0.34 & 0.24 & 0.65 & 0.43 & 0.22 \\
$10^6$ & 0.75 & 0.53 & 0.49 & 0.55 & 0.49 & 0.44 \\
\midrule
runs with a wrong lock ($10^6$) & 0.22 & 0.39 & 0.48 & 0.44 & 0.49 & 0.53 \\
iterations ($10^6$) & 6.7 & 10.0 & 15.6 & 3.7 & 5.5 & 8.3 \\
shots per run ($10^6$) & $6.7\times10^6$ & $1.0\times10^7$ & $1.6\times10^7$ & $3.7\times10^6$ & $5.5\times10^6$ & $8.3\times10^6$ \\
\bottomrule
\end{tabular}
\end{table}

With the fixed threshold $p_{diff}=0.01$, VQPM needs $10^5$ to $10^6$ shots per iteration. With $10^3$ or $10^4$ shots, 86 to 100\% of the runs lock a qubit to a wrong value within the first two or three iterations: the sampling error of the estimated difference $P_1-P_0$, about $1/\sqrt{M'}$, exceeds both the threshold and the first-iteration differences $D_i$ of Sec.~\ref{sec:gradient}, so the locks follow the noise. With $10^6$ shots the found-optimum rate comes close to that with exact marginals rounded to three decimals on the same instances, at about $10^7$ shots per run (Table~\ref{tab:shots}). This agrees with the estimate of Appendix~\ref{app:shots}.

The second rule is the threshold of Proposition~\ref{prop:threshold}, a version of the Hoeffding rule of Sec.~\ref{sec:strategies} with a valid union bound over all qubits and iterations, so that with probability at least $1-\delta$ no lock reads a sign that sampling noise has flipped. It helps when shots are scarce, with $10^3$ and $10^4$ shots per iteration (Table~\ref{tab:shots}). It does not remove wrong locks, however: with $10^6$ shots about half of the runs at $n=12$ to 16 still make one. The bound guarantees the sign of the current marginal, not its agreement with the optimum, and by Sec.~\ref{sec:locksigns} the sign of the current marginal is the sign of the update, which differs from the optimum for a few percent of the qubits. The same section also explains why the Hoeffding rule does worse than the fixed threshold at $n=8$ with $10^6$ shots: there its threshold $2\epsilon+10^{-3}$ is about 0.006, below 0.01, so it locks earlier and reads signs that are closer to the first-iteration differences. With $10^5$ shots its threshold is about 0.017, above 0.01, and it does better.

\subsection{Circuit size}
\label{sec:circuitcost}
One VQPM iteration applies the controlled-$\mathcal U$, which contains one doubly controlled phase gate for each of the $n(n-1)/2$ pairs of variables and one controlled phase gate for each diagonal term. Table~\ref{tab:gates} gives its size after transpilation with Qiskit~\cite{javadi2024quantum} (optimization level 3) for all-to-all connectivity and for the heavy-hex lattices of two IBM devices, simulated with the FakeTorino and FakeSherbrooke backends. The number of two-qubit gates grows like $n^2$, and routing onto the lattices multiplies it by 2.4 to 3. The product of the calibrated gate fidelities of FakeTorino, which ignores decoherence, gives an estimated fidelity of one iteration of 4\% at $n=8$, 0.7\% at $n=10$ and 0.1\% at $n=12$; for the newer FakeFez backend it gives 19\%, 6\% and 1\%.

\begin{table}[ht]
\centering
\small
\setlength{\tabcolsep}{5pt}
\caption{Size of the controlled-$\mathcal U$ of one VQPM iteration for a dense random QUBO after transpilation (Qiskit, optimization level 3): number of two-qubit gates and circuit depth.}
\label{tab:gates}
\begin{tabular}{r cc cc cc}
\toprule
 & \multicolumn{2}{c}{all-to-all (CX)} & \multicolumn{2}{c}{FakeTorino (CZ)} & \multicolumn{2}{c}{FakeSherbrooke (ECR)} \\
\cmidrule(lr){2-3}\cmidrule(lr){4-5}\cmidrule(lr){6-7}
$n$ & gates & depth & gates & depth & gates & depth \\
\midrule
8 & 172 & 292 & 407 & 995 & 399 & 1\,483 \\
12 & 400 & 694 & 1\,137 & 2\,321 & 1\,150 & 3\,339 \\
16 & 718 & 1\,261 & 2\,132 & 4\,123 & 2\,185 & 5\,712 \\
\bottomrule
\end{tabular}
\end{table}

\subsection{Gate and readout noise}
\label{sec:gatenoise}
We simulated the VQPM circuit with Qiskit Aer (density matrices). Each iteration prepares the product state with $R_y$ gates, applies the Hadamard gate, the controlled-$\mathcal U$ and the Hadamard gate to the control qubit, and measures all qubits; the marginals are taken from the shots in which the control qubit is $0$. We used at most 100 iterations and two locking settings: the threshold $p_{diff}=0.01$ with three decimals, the setting of Secs.~\ref{sec:VQPM}--\ref{sec:comparison}, and $p_{diff}=0.1$ with four decimals. A run returns its locked bits and the more likely value of each free qubit. The noise models are depolarizing errors with probability $p_1$ on single-qubit gates and $p_2$ on CX gates together with a symmetric readout error $r$, with all-to-all connectivity, and the full FakeTorino model with routing onto its lattice. We used $10^5$ shots per iteration, and $10^6$ for the median error rates of FakeTorino. We ran 5 instances with 3 simulator seeds each at $n=8$, and 5 instances with one seed at $n=10$. Larger $n$ was out of reach: with depolarizing noise one iteration takes about 25 seconds at $n=10$ and 13 minutes at $n=12$, and the routed FakeTorino circuit needs 17\,GB at $n=10$.

\begin{table}[ht]
\centering
\small
\setlength{\tabcolsep}{5pt}
\caption{Circuit VQPM under noise: found-optimum rate (at most 100 iterations) and, for $p_{diff}=0.01$, the share of runs with at least one wrong lock and the mean number of iterations. At $n=8$: 5 instances with 3 simulator seeds each; the 95\% intervals of a rate over 15 runs are about $\pm0.25$, and they treat the runs as independent and are therefore too narrow. At $n=10$: 5 instances with one seed for $p_{diff}=0.01$, and 3 seeds for $p_{diff}=0.1$. Exact: state-vector VQPM with the same settings, one run per instance. A dash marks settings that were not run.}
\label{tab:noise}
\begin{tabular}{l r c ccc}
\toprule
 & & $p_{diff}=0.1$ & \multicolumn{3}{c}{$p_{diff}=0.01$} \\
\cmidrule(lr){3-3}\cmidrule(lr){4-6}
noise ($p_1$, $p_2$, $r$) & shots & found & found & wrong lock & iterations \\
\midrule
\multicolumn{6}{l}{$n=8$} \\
exact marginals, no noise & -- & 5/5 & 5/5 & 0/5 & 8 \\
sampling, no noise & $10^5$ & 0.67 & 0.60 & 0.27 & 44 \\
($10^{-5}$, $10^{-4}$, $10^{-3}$) & $10^5$ & 0.73 & 0.60 & 0.33 & 39 \\
($10^{-4}$, $10^{-3}$, $10^{-2}$) & $10^5$ & 0.47 & 0.47 & 0.47 & 39 \\
($2.7\times10^{-4}$, $4.1\times10^{-3}$, $2.3\times10^{-2}$) & $10^5$ & 0.27 & 0.40 & 0.53 & 70 \\
($2.7\times10^{-4}$, $4.1\times10^{-3}$, $2.3\times10^{-2}$) & $10^6$ & -- & 0.80 & 0.07 & 82 \\
FakeTorino, routed & $10^5$ & 0.00 & -- & -- & -- \\
\midrule
\multicolumn{6}{l}{$n=10$} \\
exact marginals, no noise & -- & 3/5 & 2/5 & 2/5 & 45 \\
sampling, no noise & $10^5$ & -- & 1/5 & 3/5 & 65 \\
($2.7\times10^{-4}$, $4.1\times10^{-3}$, $2.3\times10^{-2}$) & $10^5$ & 0.00 & 0/5 & 4/5 & 100 \\
\bottomrule
\end{tabular}
\end{table}

\begin{figure*}[t]
\centering
\includegraphics[width=0.92\textwidth]{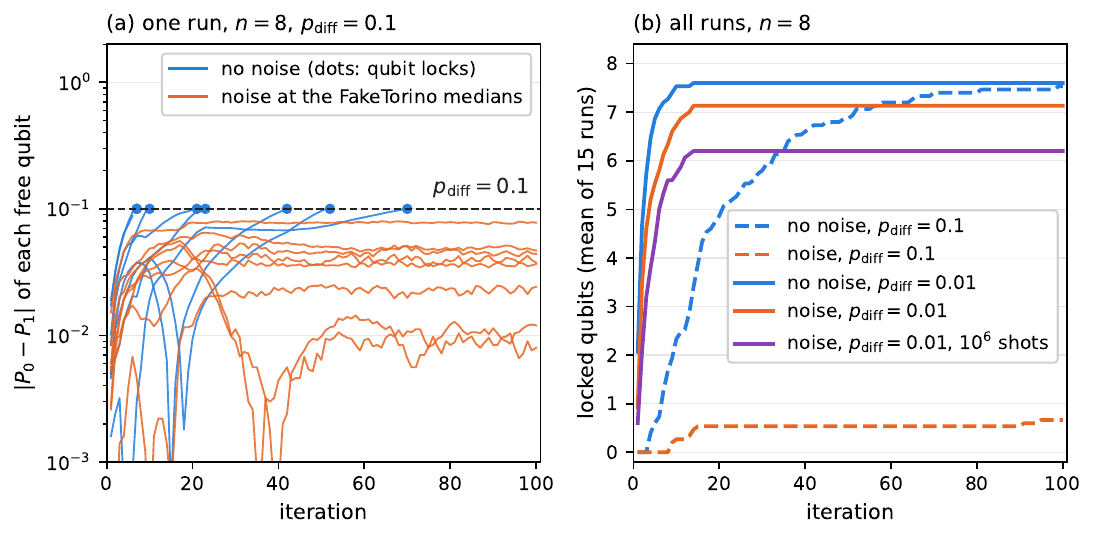}
\caption{Circuit VQPM at $n=8$ with and without depolarizing and readout noise at the median error rates of FakeTorino. (a) The difference $|P_0-P_1|$ of each free qubit in one run (instance 0, simulator seed 0, $10^5$ shots, $p_{diff}=0.1$); a dot marks the iteration in which a qubit locks. Without noise the differences grow until they pass the threshold; with noise they settle between 0.01 and 0.08 and no qubit locks. (b) Mean number of locked qubits over all 15 runs of each setting.}
\label{fig:noisy}
\end{figure*}

Table~\ref{tab:noise} and Fig.~\ref{fig:noisy} give the results. With the threshold $p_{diff}=0.1$, VQPM fails under noise for a reason that the analysis of Sec.~\ref{sec:update} predicts. Depolarizing noise mixes the state with the maximally mixed state and readout errors flip bits, so in every iteration the noise shrinks the measured differences $|P_0-P_1|$ by a fixed factor, while the update adds only a small step to them: the first-iteration differences $D_i$ have a median of 0.005 at $n=8$ (Sec.~\ref{sec:gradient}). The differences therefore stop growing at a level set by the noise, 0.03 in the median at the error rates of FakeTorino (Fig.~\ref{fig:noisy}a). A threshold of 0.1 lies above this level, so on average only 0.7 of the 8 qubits lock in 100 iterations, and the found-optimum rate falls well below its value without noise (Table~\ref{tab:noise}).

The threshold $p_{diff}=0.01$ of the rest of the paper lies below this level, and the qubits lock within about ten iterations with and without noise (Fig.~\ref{fig:noisy}b). The noise then acts through the signs that the locks read: with $10^5$ shots, noise at the error rates of FakeTorino doubles the share of runs with a wrong lock and lowers the found-optimum rate (Table~\ref{tab:noise}). With $10^6$ shots per iteration only 1 of the 15 noisy runs makes a wrong lock, and the noisy circuit finds the optimum as often as noiseless sampling with $10^6$ shots (Table~\ref{tab:shots}). At $n=8$, VQPM therefore tolerates depolarizing noise at the median error rates of current IBM hardware if it uses a threshold below the level at which the noise holds the differences and enough shots to read their signs. At $n=10$ five instances are too few for rates, but they point the same way (Table~\ref{tab:noise}). The full FakeTorino model, with routing onto its lattice, was run only with $p_{diff}=0.1$, where no run found the optimum; we did not repeat it with $p_{diff}=0.01$.

In short, at $n=8$ VQPM keeps its noiseless performance up to two-qubit error rates of about $4\times10^{-3}$ when it uses $10^6$ shots per iteration and a small threshold. The requirement grows with $n$, because the controlled-$\mathcal U$ has $O(n^2)$ two-qubit gates (Table~\ref{tab:gates}) and the first-iteration differences shrink like $n^{-3/2}$. Since VQPM for QUBO is classically simulable, a run on a device would test the device, not offer a speedup.

\subsection{Cost per instance}
\label{sec:cost}
Table~\ref{tab:cost} lists the cost of each method in its own basic step, together with its found-optimum rate. The steps differ in price. A proposal of simulated annealing and an iteration of the first-order VQPM update are cheap classical steps, with $O(n)$ and $O(n^2)$ operations. An energy evaluation of QAOA is a state-vector simulation with $2^n$ amplitudes, or on a device a batch of shots of a circuit with $2p$ layers. A VQPM iteration with exact marginals needs all $2^n$ energies classically, and on a device it is replaced by $M$ shots of a circuit with $O(n^2)$ two-qubit gates (Sec.~\ref{sec:circuitcost}).

\begin{table}[ht]
\centering
\small
\setlength{\tabcolsep}{4pt}
\caption{Cost per instance and found-optimum rate at $n=16$ (instances of Sec.~\ref{sec:smalln}) and $n=100$ (instances of Sec.~\ref{sec:update}). Costs are means for QAOA, simulated annealing and the VQPM code, and medians for the first-order update, whose runs end when all qubits are locked or the free marginals have settled. The VQPM code at $n=100$ is the three-decimal loop of Appendix~\ref{app:large}, which stalls in its first iteration because rounding erases the first-iteration differences. Simulated annealing at $n=100$ is one chain of 3000 sweeps, measured against the best of 32 chains. A dash marks runs that are out of reach.}
\label{tab:cost}
\begin{tabular}{l l rr rr}
\toprule
 & & \multicolumn{2}{c}{$n=16$} & \multicolumn{2}{c}{$n=100$} \\
\cmidrule(lr){3-4}\cmidrule(lr){5-6}
method & cost counted in & cost & found & cost & found \\
\midrule
Simulated annealing & proposals, $O(n)$ each & $1.6\times10^3$ & 0.98 & $3.0\times10^5$ & 0.98 \\
QAOA $p=8$, L-BFGS-B, restarts & energy evaluations & $4.6\times10^4$ & 0.74 & -- & -- \\
QAOA $p=8$, L-BFGS-B, INTERP & energy evaluations & $3.2\times10^4$ & 0.78 & -- & -- \\
\midrule
VQPM, first-order, no locking & iterations, $O(n^2)$ each & $8.9\times10^3$ & 0.79 & $5.6\times10^5$ & 0.24 \\
VQPM, first-order, $p_{diff}=0.2$ & iterations, $O(n^2)$ each & 596 & 0.66 & $2.3\times10^4$ & 0.05 \\
VQPM, first-order, $p_{diff}=10^{-2}$ & iterations, $O(n^2)$ each & 30 & 0.58 & $1.3\times10^3$ & 0.03 \\
VQPM code, three decimals, $p_{diff}=10^{-2}$ & iterations & 33 & 0.46 & 1 & 0.00 \\
\midrule
VQPM circuit, $10^6$ shots, fixed $p_{diff}$ & shots & $1.6\times10^7$ & 0.49 & -- & -- \\
VQPM circuit, $10^6$ shots, Hoeffding rule & shots & $8.3\times10^6$ & 0.44 & -- & -- \\
\bottomrule
\end{tabular}
\end{table}

At $n=16$, simulated annealing and the first-order VQPM update with locking are the cheapest methods, at about $10^4$ operations each, but simulated annealing finds the optimum far more often. Without locking, the first-order update matches QAOA at about 300 times the iterations of the locked update, about $2\times10^6$ operations, while QAOA needs tens of thousands of energy evaluations, each a simulation of $2^{16}$ amplitudes through 16 layers. On a device, VQPM needs about $10^7$ shots of a circuit with hundreds of two-qubit gates (Table~\ref{tab:gates}), and QAOA would need its energy evaluations times the shots per evaluation. At $n=100$ only the classical methods remain, and simulated annealing is measured only against its own best runs (Sec.~\ref{sec:largen}). A device version of VQPM at $n=100$ would need about $5\times10^8$ shots per iteration only to resolve the median first-iteration difference (Fig.~\ref{fig:app-large}).

\FloatBarrier
\section{Conclusion}
\label{sec:conclusion}
We have analyzed VQPM for QUBO problems and compared it with QAOA and simulated annealing. Because the circuit re-prepares a product state in every iteration, VQPM is not power iteration on the $2^n$-dimensional state. Its update of the marginals is the growth transform of Baum and Eagon: it never lowers the control probability, and its stable points are the strings at which every single bit flip raises the energy (Sec.~\ref{sec:VQPM}). For QUBO, VQPM has an exact classical equivalent, shot-by-shot sampling, and to first order its update is a gradient method on the mean energy of the product distribution (Sec.~\ref{sec:update}). This classical form explains the behaviour of the algorithm. The marginals first follow their first-iteration differences and then turn toward the lowest eigenvectors of the coupling matrix. Without locking, the update ends at the optimum or, more and more often as $n$ grows, at a nearby local minimum. Locking reads the signs of this update at a time set by the threshold $p_{diff}$ and freezes them; it makes VQPM faster but does not steer it toward the optimum, and no fixed threshold restores the success rate at larger $n$.

The classical form also lets us follow VQPM to $n=100$ (Sec.~\ref{sec:comparison}). Up to $n=16$, QAOA with a well-chosen optimizer and start finds the optimum more often than VQPM with locking at every $n$, while QAOA with a weak optimizer is at the level of VQPM with locking. At a fixed depth the accuracy of QAOA also declines with $n$, but unlike that of VQPM it can be raised by adding layers. Without locking and with exact marginals, VQPM is level with the best QAOA, at about 100 times more iterations than with locking. Beyond $n=16$ the found-optimum rate of VQPM keeps falling, to at most 0.05 at $n=100$ with locking, and the number of iterations grows like $n^2$. Simulated annealing finds the optimum in 98 to 100\% of the instances up to $n=20$, where the optimum is known; at larger $n$ we could only check that its chains agree with their own best string.

On a device, VQPM needs $10^5$ to $10^6$ shots per iteration to approach its performance with exact marginals at $n=8$ to 16. Gate and readout noise shrink the differences between the marginals toward a level set by the noise, so a fixed threshold must lie below that level; with $p_{diff}=0.01$ and $10^6$ shots, VQPM at $n=8$ keeps its noiseless performance under depolarizing noise at the median error rates of current IBM hardware (Sec.~\ref{sec:gatenoise}). The requirement grows with $n$, because the controlled-$\mathcal U$ needs $O(n^2)$ two-qubit gates and the first-iteration differences shrink like $n^{-3/2}$.

The analysis suggests directions for further work: locking rules that read the signs later in the update, at the turning time of Sec.~\ref{sec:couplings}; re-initialization from richer measured data, such as the two-point correlations of Appendix~\ref{app:sos}, or after a block of coherent rounds (Appendix~\ref{app:coherent}); and VQPM on problems whose Hamiltonians are not diagonal, where the classical equivalent of Sec.~\ref{sec:update} does not apply. Structured QUBO problems are natural test cases, for instance graph cuts in labeling problems \cite{komodakis2007approximate}, where a cut minimizes a quadratic energy function \cite{kolmogorov2004energy}; locking a qubit splits the candidate strings into two sets, as a cut splits the vertices of a graph.

\textbf{Limitations.} The core limitation of VQPM is that it carries the state from one iteration to the next as a product state: it keeps the $n$ single-qubit marginals of the post-selected state and discards its correlations. VQPM is therefore not power iteration on the $2^n$-dimensional state, and the most it can return is the product state at a stable point of its own update. For a QUBO problem the ground state is a basis string and hence itself a product state, yet the update can settle at another string that no single bit flip improves, a 1-flip local minimum (Sec.~\ref{sec:unlocked}), or lock a qubit to a wrong value. For a Hamiltonian whose ground state is entangled, even the best product state can be a poor approximation of the ground state. The accuracy can be raised by adding entangling layers to the input state. Their angles are then no longer fixed by the measured marginals and must be found by a classical optimizer, which turns VQPM into a standard variational quantum eigensolver~\cite{peruzzo2014variational}, with the training costs of such methods~\cite{bittel2021training} (Remark~\ref{rem:classical}).

The product-state input is also what makes VQPM for QUBO classically simulable (Remark~\ref{rem:classical}): with a diagonal $\mathcal{U}$, every measurement shot, and therefore every locking decision, can be generated by classical sampling at $O(n^2)$ cost per shot (Appendix~\ref{app:classical}). A quantum advantage would require non-diagonal Hamiltonians, input states that cannot be sampled classically, or coherent amplitude amplification of the post-selection, which can reduce the number of repetitions at most quadratically. Coherent QPI keeps the full state, but it keeps a run only if all of its post-selections succeed, which on our instances happens with probability below $10^{-80}$ from $n=8$ on; VQPM trades these repetitions for iterations that each succeed with probability at least $1/2$ (Appendix~\ref{app:coherent}).

The numerical results have further limits. All instances are dense random QUBOs with Gaussian coefficients, and structured problems may behave differently. For $n\ge30$ the reference string is the best of 32 simulated-annealing runs, not a certified optimum, so the rates at large $n$, in particular that of simulated annealing, are measured against this reference only. QAOA was run at depths up to 8, without shot noise and with learning rates tuned at $n=8$ only. The noise simulations cover $n=8$ and 10 with 5 instances each, depolarizing and readout errors with all-to-all connectivity, and the full FakeTorino model only with $p_{diff}=0.1$.

\section*{Data availability}
The code, the data and the scripts that generate every figure and table of this paper are publicly available at \url{https://github.com/adaskin/vqpm}. The README of the repository lists the script and the data behind each figure and table, and Appendix~\ref{alg:1} gives the pseudocode and the parameters.

\section*{Funding}
This project is not funded by any funding agency. 

\section*{Acknowledgment}
When writing the simulation code and the paper, we have benefited from suggestions provided by various AI tools. The initial version of the paper and most of the equations were written by the author and have been proofread, without changing the structure of the paragraphs and sentences. The later version of the paper was drafted and edited with the help of Claude (Anthropic), which was also used to run the simulations and to generate the new figures.

\bibliographystyle{unsrtdoi}
\bibliography{main}

\onecolumn
\appendix

\section{Pseudocode and parameters of VQPM}
\label{alg:1}
Below is the VQPM loop as we implemented it, with exact marginals or marginals estimated from shots. Table~\ref{tab:params} lists its parameters and the values used in this paper.
\begin{algorithmic}[1]
\State \textbf{Input:} QUBO matrix \(Q\), locking threshold \(p_{diff}\), precision (decimals or none), iteration budget \(T\), stop rule (oracle or max-prob); optionally a dynamic threshold and influence weights
\State Map the energies to phases, \(\lambda(x)=\frac{\pi}{4}+\frac{\pi}{4}E(x)/S\) (Eq.~\eqref{eq:phasemap}), and set \(\mathcal{U}=\mathrm{diag}\bigl(e^{i\lambda(x)}\bigr)\)
\State \(\ket{\psi_0}\gets\) uniform superposition; no qubit is locked
\For{\(k=0\) \textbf{to} \(T-1\)}
    \State \(\ket{\phi}\gets(I+\mathcal{U})\ket{\psi_k}/\|(I+\mathcal{U})\ket{\psi_k}\|\) \Comment{control qubit measured in \(\ket{0}\)}
    \If{oracle rule and \(|\braket{x^*}{\phi}|^2\ge\frac12\)}
        \State \textbf{stop} (success)
    \ElsIf{oracle rule and \(|\braket{x^*}{\phi}|^2=0\)}
        \State \textbf{stop} (a lock has removed the optimum)
    \ElsIf{max-prob rule and \(\max_x|\braket{x}{\phi}|^2\ge\frac12\)}
        \State \textbf{stop}
    \EndIf
    \State If a dynamic threshold is used, set \(p_{diff}\) by Eq.~\eqref{eq:hoeffdingdynamic} with \(\delta_k=\delta_{\text{total}}/(T-k)\)
    \For{each unlocked qubit \(q\)}
        \State \((P_0,P_1)\gets\) marginals of qubit \(q\) in \(\ket{\phi}\) (exact or from shots), rounded to the precision
        \State \(t_q\gets p_{diff}\,\mathrm{Influence}(q)\) if influence weights are used, else \(t_q\gets p_{diff}\)
        \If{\(|P_0-P_1|>t_q\)}
            \State lock qubit \(q\) to the more probable value for all later iterations
        \Else
            \State set the \(R_y\) angle of qubit \(q\) so that it is measured as 1 with probability \(P_1\)
        \EndIf
    \EndFor
    \State \(\ket{\psi_{k+1}}\gets\) product state of the locked and unlocked qubits
\EndFor
\State \textbf{Output:} the most probable string of \(\ket{\phi}\), its probability, and the probability of \(x^*\) if \(x^*\) is known
\end{algorithmic}

\newcolumntype{P}{>{\raggedright\arraybackslash\hsize=.7\hsize}X}
\newcolumntype{s}{>{\raggedright\arraybackslash\hsize=1.1\hsize}X}

\begin{table}[ht]
\centering
\small
\caption{Parameters of VQPM and the values used in this paper.}
\label{tab:params}
\begin{tabularx}{\textwidth}{Psss}
\toprule
\textbf{Parameter} & \textbf{Purpose} & \textbf{Values in this paper} & \textbf{Effect} \\
\midrule
\(n\) & number of qubits (variables) & 1--18 (Sec.~\ref{sec:VQPM}), 4--22 (Sec.~\ref{sec:smalln}), 20--100 (classical equivalent), 8--10 (noise) & the state vector has \(2^n\) entries; the first-iteration differences shrink like \(n^{-3/2}\) \\
\midrule
\(T\) & iteration budget & 30 (Sec.~\ref{sec:VQPM}), 100 or 300 (Sec.~\ref{sec:smalln}), none (first-order runs) & a larger budget helps only if the marginals are not rounded to three decimals \\
\midrule
\(p_{diff}\) & locking threshold & 0.01 (fixed), 0.006--0.05 (Eq.~\eqref{eq:hoeffdingdynamic}), \(10^{-6}\)--0.2 (Sec.~\ref{sec:locksigns}), 0.1 and 0.01 (noise) & a small threshold locks early and reads the first-iteration differences; a larger one locks later, reads better signs and needs more iterations \\
\midrule
precision & decimals of the marginals & 3, 6 or none (exact) & three decimals stall runs as \(n\) grows \\
\midrule
\(M\) & shots per iteration & \(10^3\)--\(10^6\) (Sec.~\ref{sec:noise}) & sampling error about \(1/\sqrt{M}\) \\
\midrule
\(\delta_{\text{total}}\), \(\delta\) & failure probability of the Hoeffding rules & 0.5 (Eq.~\eqref{eq:hoeffdingdynamic}), 0.05 (Proposition~\ref{prop:threshold}) & a lower value raises the threshold \\
\midrule
Influence\((i)\) & qubit weights, Eq.~\eqref{eq:influence} & \((0,1]\) & scale the threshold of each qubit \\
\bottomrule
\end{tabularx}
\end{table}


\section{Theoretical analysis}
\setcounter{equation}{0}
\renewcommand{\theequation}{\thesection\arabic{equation}}
\label{app:theory}

This appendix gives the proofs for Secs.~\ref{sec:qpi} and~\ref{sec:VQPM} and keeps two procedures apart: \emph{coherent} quantum power iteration (QPI), which applies $(I+\mathcal{U})^k$ to one state (Eq.~\ref{eq:qpi}), and VQPM, which measures the single-qubit marginals after every round and prepares a new \emph{product} state from them. It also gives a moment-matrix (degree-2 sum-of-squares) view of VQPM. Classical simulability, shot counts and runs at larger $n$ are in Appendix~\ref{app:sampling}.

All numbers below are computed with the same code as the main text. The instances are random QUBOs $Q=G+G^T$ with $G_{ij}\sim\mathcal N(0,1)$, drawn with seed $2026+n$, 100 instances per $n$ unless stated. Probabilities are exact (statevector level).

\subsection{Setting}
\label{app:setting}
A computational-basis string $x\in\{0,1\}^n$ labels an eigenstate $\ket{j}$ of $\mathcal{U}$ in the main text. We evaluate
\begin{equation}
E(x)=\sum_{i\le j}Q_{ij}\,x_ix_j ,
\label{eq:app-energy}
\end{equation}
that is, we use the upper triangle of $Q$ including the diagonal, and assign to $x$ an eigenphase $\lambda(x)$. For an input state whose computational-basis distribution is $p(x)$, one run of the circuit of Fig.~\ref{fig:Circuit Summary of VQPM} gives the control outcome $0$ together with the register outcome $x$ with probability
\begin{equation}
\Pr(c=0,\,x)=p(x)\,w(x),\qquad w(x)=\frac{|1+e^{i\lambda(x)}|^2}{4}=\cos^2\!\frac{\lambda(x)}{2}.
\label{eq:app-filter}
\end{equation}
The control is therefore measured in $\ket{0}$ with probability $W(p)=\sum_x p(x)w(x)$, and the post-selected register has distribution $q(x)=p(x)w(x)/W(p)$. Relative phases of the input amplitudes play no role, because $\mathcal{U}$ is diagonal and every qubit is measured in the computational basis.

\subsection{The phase map}
\label{app:phasemap}
\begin{lemma}[Phase window]
\label{lem:window}
Let $S=\sum_{i\le j}|Q_{ij}|$. The map of Eq.~\eqref{eq:phasemap},
\begin{equation}
\lambda(x)=\frac{\pi}{4}+\frac{\pi}{4}\,\frac{E(x)}{S},
\label{eq:app-map}
\end{equation}
places all $2^n$ eigenphases in $[0,\pi/2]$, preserves the order of the energies, and costs $O(n^2)$ operations with no spectral information.
\end{lemma}
\begin{proof}
For $x\in\{0,1\}^n$, $|E(x)|\le\sum_{i\le j}|Q_{ij}|x_ix_j\le S$. The map is affine with positive slope.
\end{proof}

\begin{remark}[Choice of the window]
\label{rem:window}
\leavevmode
\begin{itemize}
\item[(i)] Any window $[a,b]\subset[0,\pi)$ works: $w$ is strictly decreasing on $[0,\pi)$, so the minimum-energy string always receives the largest weight. The choice $b=\pi/2$ guarantees $W(p)\ge\cos^2(\pi/4)=1/2$ for every input, so post-selection costs at most a factor of two in shots.
\item[(ii)] With $L=\sum_{i\le j}\min(Q_{ij},0)$ and $U=\sum_{i\le j}\max(Q_{ij},0)$ one has $L\le E(x)\le U$ and $U-L=S$, so the map $\lambda=a+(b-a)(E-L)/S$ uses twice the resolution of Eq.~\eqref{eq:app-map}. For a spin formulation $\sum_ih_iz_i+\sum_{i<j}J_{ij}z_iz_j$ with $z_i=\pm1$, the analogous bound is $\sum_i|h_i|+\sum_{i<j}|J_{ij}|$.
\item[(iii)] \emph{Effect on convergence.} For phases $\lambda_d<\lambda_s$ the per-round weight ratio is $R^2=w(\lambda_s)/w(\lambda_d)$, with $R$ as in Eq.~\eqref{eq:ratio}. For a small gap $\gamma=\lambda_s-\lambda_d$,
\begin{equation}
1-R^2=\gamma\tan\frac{\lambda_d}{2}+O(\gamma^2).
\label{eq:app-contrast}
\end{equation}
An affine map with slope $c$ gives $\gamma=c\,\Delta E$. Stretching the window raises the per-round contrast linearly in $c$, and moving it toward $\pi$ raises it through $\tan(\lambda_d/2)$, but moving it toward $\pi$ also lowers the acceptance bound $\cos^2(b/2)$.
\end{itemize}
\end{remark}

Table~\ref{tab:app-phase} shows what the map does on the random instances of this paper. Because $S$ bounds the worst case over all strings, the energies occupy only 17\% to 29\% of $[0,\pi/2]$, and the gap between the two lowest phases falls from $4.2\times10^{-2}$~rad at $n=4$ to $2.1\times10^{-3}$~rad at $n=16$. Table~\ref{tab:app-windows} compares windows at $n=10$: the first-order formula~\eqref{eq:app-contrast} matches the exact $1-R^2$ to within 2\%, and the windows with larger contrast pay for it in acceptance.
Sec.~\ref{sec:qpi} compares four maps in VQPM runs with locking.

\begin{table}[ht]
\centering
\caption{Paper phase map, Eq.~\eqref{eq:app-map}, on random instances (medians over 100 instances per $n$). $\rho^*$ is the local VQPM rate at the optimum (Proposition~\ref{prop:stability}). The two iteration counts are for coherent iteration with $\epsilon=1/2$: the simpler bound $\ln(1/\epsilon)/\ln(1/R^2)$, which omits $A$, and Eq.~\eqref{eq:kbound}. The last column is the probability that all $k$ post-selections of coherent iteration succeed, Eq.~\eqref{eq:app-survival}, at the corrected $k$.}
\label{tab:app-phase}
\begin{tabular}{rcccccrr}
\toprule
$n$ & used window (rad) & gap $\lambda_s-\lambda_d$ (rad) & $1-R^2$ & $1-\rho^*$ & $k$ without $A$ & $k$, Eq.~\eqref{eq:kbound} & $\log_{10}Z_k$ \\
\midrule
4  & $[0.58,\,1.05]$ & $4.2\times10^{-2}$ & $1.2\times10^{-2}$ & $1.4\times10^{-2}$ & 56   & 219    & $-8.7$ \\
8  & $[0.62,\,0.98]$ & $8.8\times10^{-3}$ & $2.9\times10^{-3}$ & $3.8\times10^{-3}$ & 237  & 1\,896 & $-82$ \\
12 & $[0.62,\,0.94]$ & $3.5\times10^{-3}$ & $1.1\times10^{-3}$ & $1.4\times10^{-3}$ & 614  & 7\,373 & $-325$ \\
16 & $[0.65,\,0.92]$ & $2.1\times10^{-3}$ & $6.8\times10^{-4}$ & $9.3\times10^{-4}$ & 1\,015 & 16\,247 & $-781$ \\
\bottomrule
\end{tabular}
\end{table}

\begin{table}[ht]
\centering
\caption{Per-round contrast and acceptance for four phase windows at $n=10$ (medians over 100 instances). Acceptance is $W$ for the uniform input.}
\label{tab:app-windows}
\begin{tabular}{lccc}
\toprule
Map & $1-R^2$ & $\gamma\tan(\lambda_d/2)$, Eq.~\eqref{eq:app-contrast} & Acceptance $\Pr(c=0)$ \\
\midrule
Paper: $[-S,S]\to[0,\pi/2]$        & $2.4\times10^{-3}$ & $2.4\times10^{-3}$ & 0.85 \\
Tight: $[L,U]\to[0,\pi/2]$         & $3.2\times10^{-3}$ & $3.2\times10^{-3}$ & 0.85 \\
Tight: $[L,U]\to[0,0.9\pi]$        & $1.1\times10^{-2}$ & $1.1\times10^{-2}$ & 0.58 \\
Paper, shifted: $\to[\pi/2,\pi]$   & $1.5\times10^{-2}$ & $1.5\times10^{-2}$ & 0.15 \\
\bottomrule
\end{tabular}
\end{table}

\subsection{Coherent iteration and its iteration bound}
\label{app:coherent}
After $k$ coherent rounds in which every control outcome is $0$ (Eq.~\ref{eq:qpi}), the register distribution is
\begin{equation}
q^{\mathrm{coh}}_k(x)=\frac{p_0(x)\,w(x)^k}{Z_k},\qquad Z_k=\sum_x p_0(x)\,w(x)^k ,
\label{eq:app-coherent}
\end{equation}
and $Z_k$ is the probability that all $k$ post-selections succeed. Eqs.~\eqref{eq:marginal0}--\eqref{eq:marginal1} are the single-qubit marginals of Eq.~\eqref{eq:app-coherent} for the uniform $p_0$.

\begin{proposition}[Iteration bound]
\label{prop:kbound}
Let $X^*$ be the set of optimal strings, $A=\sum_{x\in X^*}p_0(x)>0$ the initial probability on it, $w^*$ the weight of an optimal string and $R^2=\max_{x\notin X^*}w(x)/w^*$. Then
\begin{equation}
P_k:=q^{\mathrm{coh}}_k(X^*)\;\ge\;\frac{1}{1+\frac{1-A}{A}R^{2k}},
\end{equation}
so $P_k\ge1-\epsilon$ is guaranteed when
\begin{equation}
k\;\ge\;\frac{\ln\frac{1-A}{A}+\ln\frac{1-\epsilon}{\epsilon}}{\ln(1/R^2)} .
\label{eq:app-kbound}
\end{equation}
For the uniform start $A=|X^*|/2^n$, so $k$ grows as $n\ln2/\ln(1/R^2)$.
\end{proposition}
\begin{proof}
$P_k=Aw^{*k}/(Aw^{*k}+\sum_{x\notin X^*}p_0(x)w(x)^k)$ and $\sum_{x\notin X^*}p_0(x)w(x)^k\le(1-A)R^{2k}w^{*k}$. Rearranging $(1-A)R^{2k}/A\le\epsilon/(1-\epsilon)$ gives Eq.~\eqref{eq:app-kbound}.
\end{proof}

Eq.~\eqref{eq:app-kbound} is Eq.~\eqref{eq:kbound} of the main text. The simpler bound $k\ge\ln(1/\epsilon)/\ln(1/R^2)$ omits the initial mass $A$; $\epsilon$ is the tolerated failure probability. The bound is tight when all suboptimal strings have the same weight, as in the two-qubit example $\lambda(x)=\frac{\pi}{2}(x_1+x_2-x_1x_2)$: there $A=1/4$, $R^2=1/2$ and $P_k=1/(1+3\cdot2^{-k})=0.40,\,0.57,\,0.73,\,0.84,\,0.91$ for $k=1,\dots,5$. For $\epsilon=1/4$, the bound without $A$ gives $k=2$, while Eq.~\eqref{eq:app-kbound} gives $k\ge3.17$, that is $k=4$, the first $k$ with $P_k\ge3/4$.

\paragraph{Cost of coherent iteration.} A run of coherent iteration is kept only if all $k$ control outcomes are $0$, which happens with probability
\begin{equation}
Z_k=\sum_xp_0(x)w(x)^k\;\ge\;A\,w^{*k}.
\label{eq:app-survival}
\end{equation}
This probability is tied to the eigenvalues of $I+\mathcal{U}$. Round $t$ succeeds with probability $Z_t/Z_{t-1}=\sum_xq^{\mathrm{coh}}_{t-1}(x)\,w(x)$, which tends to $w^*=\cos^2(\lambda_d/2)=|1+e^{i\lambda_d}|^2/4$, the squared magnitude of the dominant eigenvalue of $(I+\mathcal{U})/2$, as the state converges to the optimum; this is the eigenvalue information carried by the control qubit (Sec.~\ref{sec:controlqubit}). Once the state has converged, $Z_k$ therefore decays like $w^{*k}$. The map of Lemma~\ref{lem:window} puts $\lambda_d$ near $0.6$~rad on our instances (Table~\ref{tab:app-phase}), so $w^*\approx0.9$. Moving $\lambda_d$ toward $0$ would make post-selection nearly free, but by Eq.~\eqref{eq:app-contrast} the per-round contrast $1-R^2\approx\gamma\tan(\lambda_d/2)$ would vanish with it.

\paragraph{Restarting versus reloading.} With mid-circuit measurement, each round is a repeat-until-success step whose success probability $p_t=Z_t/Z_{t-1}$ is set by the eigenvalues as above. What makes the total cost exponential is that a failed round destroys the state: the control outcome $1$ leaves $(I-\mathcal{U})\ket{\psi_{t-1}}$, which carries the opposite filter $\sin^2(\lambda(x)/2)$, so the run has to restart from $\ket{\psi_0}$. The $k$ successes are then needed in a row, the probability of that is the product $p_1\cdots p_k=Z_k$, and the expected number of rounds to finish is $\sum_{s=1}^{k}Z_{s-1}/Z_k$, between $1/Z_k$ and $k/Z_k$. If instead the post-selected state $\ket{\psi_{t-1}}$ could be reloaded after every failure, each round could be retried on its own, and the expected cost would drop to $\sum_{t=1}^{k}1/p_t$ rounds. This is at most $2k$ under the map of Lemma~\ref{lem:window}, where every $w(x)\ge1/2$, and about $k/w^*\approx1.1k$ once the state has converged. Reloading, however, needs a fresh copy of $\ket{\psi_{t-1}}$ for every attempt. A quantum memory holding one copy does not supply this: the failed attempt consumes that copy, and an unknown state cannot be copied \cite{wootters1982single}. A new copy must be prepared either by re-running the $t-1$ successful rounds, which is the restart cost above, or from a classical description of the state, for example loaded through a quantum RAM \cite{giovannetti2008quantum}. For coherent iteration that description is the amplitude vector $\psi_{t-1}(x)\propto\psi_0(x)(1+e^{i\lambda(x)})^{t-1}$, which needs $\lambda(x)$ for all $2^n$ strings, so building it is as costly as exhaustive search. VQPM is the cheapest version of reloading: it keeps only the $n$ measured marginals and reloads the product state with those marginals, so its rounds are independent and each succeeds with probability at least $1/2$. The price is that the correlations of $\ket{\psi_{t-1}}$ are not reloaded (Appendix~\ref{app:vqpm-map}).

\paragraph{Between QPI and VQPM.} Two kinds of variants lie between the two ends. Storing richer classical data, such as the two-point correlations of Appendix~\ref{app:sos}, would reload part of the correlations, but requires correlated input states. One can also measure and re-prepare after every block of $m$ coherent rounds: $m=1$ is VQPM and $m=k$ is QPI. For an input distribution $p$ a block succeeds with probability $\sum_xp(x)w(x)^m$, and under the map of Lemma~\ref{lem:window} every $w(x)\ge1/2$, so a block needs at most $2^m$ attempts on average. Inside a block the filter $w^m$ acts with all correlations kept; between blocks only the marginals survive. Larger $m$ therefore keeps more of the coherent filter and needs fewer re-preparations, at a cost in repetitions that grows exponentially with $m$; $m=1$ keeps the repetitions lowest. For QUBO, every member of this family remains classically samplable (Proposition~\ref{prop:classical}).

On the paper's instances (Table~\ref{tab:app-phase}), Eq.~\eqref{eq:app-kbound} asks for 219 rounds at $n=4$ and about $1.6\times10^4$ at $n=16$, four to sixteen times the counts of the bound without $A$, and at those $k$ the survival probability $Z_k$ is $10^{-9}$ at $n=4$ and below $10^{-80}$ from $n=8$ on. Coherent iteration with these $k$ is therefore not practical, and the iteration counts of Sec.~\ref{sec:qpi} do not describe the VQPM runs with locking of Secs.~\ref{sec:VQPM} and~\ref{sec:comparison}, which stop within 30 to 100 iterations. The VQPM update without locking is not faster either: from the uniform start it needs a median of $2.7\times10^3$ to $1.2\times10^4$ rounds to settle for $n=6$ to $12$ (Table~\ref{tab:app-exact}). In the runs of the main text, the locking mechanism, not the power filter, does most of the work (Appendix~\ref{app:locking}).

\subsection{The VQPM update}
\label{app:vqpm-map}
VQPM does not apply Eq.~\eqref{eq:app-coherent}. After each round it estimates the single-qubit marginals of $q$ and prepares the product state with those marginals. Write $\pi_i=\Pr(x_i=1)$ and $p_\pi(x)=\prod_i\pi_i^{x_i}(1-\pi_i)^{1-x_i}$. With exact marginals and without rounding or locking, one VQPM round is the map
\begin{equation}
\pi_i'=\frac{1}{W(\pi)}\sum_{x:\,x_i=1}p_\pi(x)\,w(x),\qquad W(\pi)=\sum_xp_\pi(x)\,w(x).
\label{eq:app-vqpm}
\end{equation}
Locking replaces $\pi_i'$ by $0$ or $1$ for all later rounds, and rounding replaces it by its value rounded to the measurement precision.

From the uniform start, the first round of Eq.~\eqref{eq:app-vqpm} agrees with Eqs.~\eqref{eq:marginal0}--\eqref{eq:marginal1} at $k=1$. From the second round on they differ, because the product state $p_{\pi'}$ is not the filtered distribution $q$. In the two-qubit example of Appendix~\ref{app:coherent} the first round gives $q_1=(2,1,1,1)/5$ in the order $00,01,10,11$, so each bit is $0$ with probability $3/5$ and VQPM prepares $(9,6,6,4)/25$. The second round then gives $(9,3,3,2)/17$ for VQPM and $(4,1,1,1)/7$ for coherent iteration, so $\Pr(x_1=0)$ is $12/17$ and $5/7$, respectively. Our implementation of the VQPM update reproduces these numbers exactly. Eqs.~\eqref{eq:marginal0}--\eqref{eq:marginal1} describe coherent iteration; VQPM follows Eq.~\eqref{eq:app-vqpm}.

The product state with the marginals of $q$ is the product distribution closest to $q$ in relative entropy $D(q\,\|\,p_\pi)$, since $D(q\|p_\pi)=\mathrm{const}-\sum_i[q_i\ln\pi_i+(1-q_i)\ln(1-\pi_i)]$ is minimized at $\pi_i=q_i$. The re-preparation therefore keeps the marginals and discards every correlation of $q$; Appendix~\ref{app:sos} quantifies what is discarded. In the language of estimation-of-distribution algorithms, Eq.~\eqref{eq:app-vqpm} is the univariate marginal distribution algorithm with selection weights $w$~\cite{muhlenbein1996recombination}. This connection lets us use two classical results.

\begin{proposition}[The acceptance never decreases]
\label{prop:monotone}
Under Eq.~\eqref{eq:app-vqpm}, $W(\pi')\ge W(\pi)$, with equality only at fixed points. Equivalently,
\begin{equation}
\pi_i'-\pi_i=\frac{\pi_i(1-\pi_i)}{W(\pi)}\,\frac{\partial W}{\partial\pi_i}.
\label{eq:app-wright}
\end{equation}
\end{proposition}
\begin{proof}
Write $y_{i1}=\pi_i$, $y_{i0}=1-\pi_i$. Then $W=\sum_xw(x)\prod_iy_{i,x_i}$ is a polynomial with nonnegative coefficients, homogeneous of degree $n$ and of degree one in each pair $(y_{i0},y_{i1})$. Hence $y_{i1}\,\partial W/\partial y_{i1}=\sum_{x:x_i=1}p_\pi(x)w(x)$ and $\sum_b y_{ib}\,\partial W/\partial y_{ib}=W$, so Eq.~\eqref{eq:app-vqpm} reads $\pi_i'=y_{i1}\,\partial_{y_{i1}}W/\sum_by_{ib}\,\partial_{y_{ib}}W$. This is the growth transform of Baum and Eagon, which increases any such polynomial strictly unless it is at a fixed point~\cite{baum1967inequality,baum1968growth}. Eq.~\eqref{eq:app-wright} follows from $W=\pi_iW_{i1}+(1-\pi_i)W_{i0}$ with $W_{ib}=\mathbb{E}_{p_\pi}[w\mid x_i=b]$.
\end{proof}

$W$ is the probability of measuring the control qubit in $\ket{0}$. Proposition~\ref{prop:monotone} therefore makes the observation of Sec.~\ref{sec:controlqubit} precise: with exact marginals, VQPM is an ascent method for $W$, and the control probability can decrease only through rounding, locking or shot noise. Even with rounding and locking it almost never does: in the 3\,623 rounds of the paper-setting runs of Table~\ref{tab:app-locking} it fell once. This is why the policy ``lock only when the probability of $\ket{0}$ increases'' of Sec.~\ref{sec:controlqubit} changed nothing.

\begin{proposition}[Fixed points and their stability]
\label{prop:stability}
Every vertex $\pi=x\in\{0,1\}^n$ is a fixed point of Eq.~\eqref{eq:app-vqpm}. Near $x$, a small deviation $\delta_i$ of coordinate $i$ toward the flipped value evolves as
\begin{equation}
\delta_i'=r_i\,\delta_i+O(\|\delta\|^2),\qquad r_i=\frac{w(x\oplus e_i)}{w(x)} .
\end{equation}
Hence $x$ is asymptotically stable if every single bit flip raises the energy ($x$ is a strict 1-flip local minimum), and unstable if some single flip lowers it.
\end{proposition}
\begin{proof}
Near $x$, $p_\pi$ puts mass $\delta_i+O(\|\delta\|^2)$ on $x\oplus e_i$ and $O(\|\delta\|^2)$ on strings at Hamming distance two or more, and $W=w(x)+O(\|\delta\|)$. The new deviation is the post-selected mass on strings with bit $i$ flipped, $\delta_iw(x\oplus e_i)/w(x)+O(\|\delta\|^2)$. Since $w$ decreases with $\lambda$ and $\lambda$ increases with $E$, $r_i<1$ exactly when $E(x\oplus e_i)>E(x)$.
\end{proof}

Three consequences follow.
\begin{itemize}
\item[(a)] VQPM can converge to a suboptimal string even with exact marginals and no locking: every strict 1-flip local minimum of $E$ attracts the iteration. This accounts for the behaviour in Fig.~\ref{fig-losing-eigenstate} without any measurement error: the information loss of Sec.~\ref{sec:infoloss} is a property of the update itself. In Table~\ref{tab:app-exact}, 10\% to 14\% of exact runs end at a suboptimal local minimum.
\item[(b)] Near the optimum $x^*$ the contraction rate is $\rho^*=\max_iw(x^*\oplus e_i)/w(x^*)\le R^2$. It is set by the single-flip neighbours of $x^*$, not by the global gap. On our instances $1-\rho^*$ is about 15\% to 35\% larger than $1-R^2$ (Table~\ref{tab:app-phase}).
\item[(c)] The number of attracting vertices grows with $n$: on our instances the mean number of strict 1-flip local minima is 1.2 at $n=4$, 2.0 at $n=8$, 2.7 at $n=12$ and 4.3 at $n=16$, with up to 19 at $n=16$. For the Sherrington--Kirkpatrick spin glass this number grows as $e^{0.199n}$~\cite{tanaka1980analytic}.
\end{itemize}

Table~\ref{tab:app-exact} checks Propositions~\ref{prop:monotone} and~\ref{prop:stability} by iterating Eq.~\eqref{eq:app-vqpm} from the uniform start until every marginal is within $10^{-6}$ of $0$ or $1$ (at most $3\times10^5$ rounds). $W$ never decreased, and every run that reached a vertex ended at a strict 1-flip local minimum.

\begin{table}[ht]
\centering
\small
\caption{Exact VQPM map, Eq.~\eqref{eq:app-vqpm}, without rounding or locking, from the uniform start (100 instances per $n$). Across all rounds and runs the smallest change of $W$ was $+3\times10^{-15}$, so $W$ never decreased. A few runs (1 to 3 per $n$) were still moving slowly after $3\times10^5$ rounds; for them the limit is taken as the nearest vertex.}
\label{tab:app-exact}
\begin{tabularx}{\textwidth}{r>{\centering\arraybackslash}X>{\centering\arraybackslash}X>{\centering\arraybackslash}X>{\centering\arraybackslash}X>{\centering\arraybackslash}X}
\toprule
$n$ & Reached a vertex & Limit is a strict 1-flip local min. & Limit is the optimum & Rounds to settle (median) & Strict 1-flip local minima per instance (median, max) \\
\midrule
6  & 99\% & 100\% & 89\% & 2\,672  & 1, 4 \\
8  & 98\% & 100\% & 90\% & 4\,265  & 2, 6 \\
10 & 99\% & 100\% & 86\% & 5\,547  & 2, 7 \\
12 & 97\% & 100\% & 88\% & 11\,739 & 2, 7 \\
\bottomrule
\end{tabularx}
\end{table}

\subsection{Locking, rounding and wrong locks}
\label{app:locking}
A qubit is locked to $s\in\{0,1\}$ when the measured difference $D_i=P_1-P_0=2\pi_i'-1$ satisfies $|D_i|>p_{\mathrm{diff}}$, and $s$ is the more probable value. A lock is \emph{wrong} if $s\ne x^*_i$; it removes the optimum from the remaining search space permanently. There are two mechanisms.

\paragraph{Structural wrong locks.} The exact marginal itself favours the wrong value. By Proposition~\ref{prop:stability} the update is pulled toward every 1-flip local minimum, and with $p_{\mathrm{diff}}=0.01$ a qubit locks as soon as its marginal moves $0.005$ away from $1/2$. The first lock comes in round 1 or 2 (Table~\ref{tab:app-locking}), long before the filter has separated the optimum from nearby strings. More shots do not remove these locks.

\paragraph{Statistical wrong locks.} With $M$ post-selected shots in a round, $\hat D_i-D_i=2(\hat P_1-P_1)$. If $x^*_i=0$ and $D_i<p_{\mathrm{diff}}$, Hoeffding's inequality gives
\begin{equation}
\Pr\bigl(\text{lock to }1\bigr)=\Pr\bigl(\hat D_i>p_{\mathrm{diff}}\bigr)\le\exp\!\Bigl(-\tfrac{M}{2}\,(p_{\mathrm{diff}}-D_i)^2\Bigr),
\label{eq:app-statlock}
\end{equation}
and symmetrically for $x^*_i=1$. For an undecided qubit ($D_i\approx0$) and $p_{\mathrm{diff}}=0.01$ the bound is $0.95$, $0.61$, $6.7\times10^{-3}$ and $2\times10^{-22}$ for $M=10^3,10^4,10^5,10^6$. A threshold of $0.01$ thus needs on the order of $10^5$ post-selected shots per round before statistical wrong locks become rare; Appendix~\ref{app:hoeffding} makes this precise for all qubits and rounds.

\paragraph{Rounding.} The simulations round the exact marginals to $d$ decimals instead of sampling. The rounding error, at most $\frac12 10^{-d}$, equals the shot-noise standard error at $P=1/2$ when $M\approx10^{2d}$, that is $10^6$ shots for $d=3$. Rounding is deterministic and therefore creates fixed points: if every unlocked marginal moves by less than $\frac12 10^{-d}$ in a round (Eq.~\eqref{eq:app-wright}), the rounded input repeats and the iteration stalls for good. This is why raising the iteration budget beyond 30 does not change the results at precision 3.

Table~\ref{tab:app-locking} reports these effects for the paper's settings. With exact marginals every wrong lock is structural. Wrong locks occur in 26\% to 36\% of runs and account for 57\% to 79\% of the failures; the other failures ran out of rounds. Stalls from rounding grow with $n$, from 8\% of runs at $n=8$ to 37\% at $n=16$.

\begin{table}[ht]
\centering
\small
\caption{Paper settings ($p_{\mathrm{diff}}=0.01$, precision 3, at most 30 rounds, stop rule of Appendix~\ref{alg:1} using the known optimum), exact marginals, 100 instances per $n$. ``Lowers $\bar E$'': the lock lowers the expected energy of the next input state (Lemma~\ref{lem:leakage}). ``Stalled'': the rounded input state repeated before all qubits were locked.}
\label{tab:app-locking}
\begin{tabularx}{\textwidth}{r*{8}{>{\centering\arraybackslash}X}}
\toprule
$n$ & Rounds used (mean) & First lock (round, median) & Locks that lower $\bar E$ & Runs with a wrong lock & Failures with a wrong lock & Runs stalled & Optimum found & Output is a 1-flip local min. \\
\midrule
8  & 7.8  & 1 & all 655 & 26\% & 26 of 33 & 8\%  & 67\% & 71\% \\
12 & 13.9 & 2 & all 943 & 33\% & 33 of 55 & 29\% & 45\% & 52\% \\
16 & 17.5 & 2 & all 1\,211 & 36\% & 36 of 63 & 37\% & 37\% & 45\% \\
\bottomrule
\end{tabularx}
\end{table}

\subsection{The Hoeffding threshold}
\label{app:hoeffding}
\paragraph{What Eq.~\eqref{eq:hoeffdingdynamic} is.} Hoeffding's inequality gives the radius $\epsilon=\sqrt{\ln(2/\delta)/(2N)}$ for the mean of $N$ samples in $[0,1]$. Eq.~\eqref{eq:hoeffdingdynamic} sets $N=10nM$, which is the radius for a marginal estimated from $10nM=1000n$ shots. The simulations do not draw shots (they round exact marginals), so $M$ in Eq.~\eqref{eq:hoeffdingdynamic} is not the shot count of any experiment, and the factor $10n$ is an empirical choice that keeps $p_{\mathrm{diff}}$ in a workable range; it is not derived. Two further corrections apply. First, the split $\delta_k=\delta_{\mathrm{total}}/(T-k)$ over $T$ rounds sums to $\delta_{\mathrm{total}}H_T$, where $H_T$ is the harmonic number; for $T=30$ this is $4.0\,\delta_{\mathrm{total}}=2.0$ for $\delta_{\mathrm{total}}=0.5$, so the union bound is overspent; $\delta_k=\delta_{\mathrm{total}}/T$ is the correct split. Second, the implemented rule gives $p_{\mathrm{diff}}\in[0.0062,0.049]$ over $n=1,\dots,18$ with $T=30$, and $[0.0068,0.0126]$ at $n=15$.

\begin{proposition}[A derived threshold]
\label{prop:threshold}
Suppose each round uses $M$ post-selected shots in which all $n$ qubits are measured, and the run has at most $T$ rounds. Let
\begin{equation}
\epsilon=\sqrt{\frac{\ln(2nT/\delta)}{2M}},
\label{eq:app-threshold}
\end{equation}
and lock qubit $i$ only if $|\hat P_0-\hat P_1|\ge p_{\mathrm{diff}}=2\epsilon$. Then, with probability at least $1-\delta$, every lock in the run agrees with the sign of the exact marginal difference $D_i$ of its round.
\end{proposition}
\begin{proof}
For each qubit and round, $\Pr(|\hat P_0-P_0|\ge\epsilon)\le2e^{-2M\epsilon^2}=\delta/(nT)$. A union bound over the $nT$ estimates gives probability at least $1-\delta$ that all of them are within $\epsilon$. On that event $|\hat D_i-D_i|\le2\epsilon$, so $|\hat D_i|\ge2\epsilon$ implies that $D_i$ has the same sign as $\hat D_i$ (or is zero).
\end{proof}

The rule depends on $n$ only through $\ln n$, from the union bound over qubits. Solving Eq.~\eqref{eq:app-threshold} for $M$ gives $M=2\ln(2nT/\delta)/p_{\mathrm{diff}}^2$: for $p_{\mathrm{diff}}=0.01$, $T=30$ and $\delta=0.05$ this is $1.8\times10^5$ post-selected shots per round at $n=8$ and $2.0\times10^5$ at $n=20$. Dividing by the acceptance $W\ge1/2$ (about $0.85$ in practice) gives the total. Proposition~\ref{prop:threshold} removes statistical wrong locks only; structural wrong locks (Appendix~\ref{app:locking}) do not depend on $M$.

\subsection{Moment-matrix (degree-2 sum-of-squares) view}
\label{app:sos}
For a distribution $q$ on $\{0,1\}^n$ let $Y(q)=\mathbb{E}_q[(1,x)(1,x)^{T}]$, the $(n+1)\times(n+1)$ matrix with $Y_{00}=1$, $Y_{0i}=q(x_i=1)$ and $Y_{ij}=q(x_i=x_j=1)$, so that $Y_{ii}=Y_{0i}$. $Y(q)$ is a convex combination of the rank-one matrices $(1,x)(1,x)^T$ and lies in the boolean quadric (cut) polytope~\cite{deza1997geometry}; in particular it is feasible for the degree-2 sum-of-squares (Shor) relaxation~\cite{shor1987quadratic,lasserre2001global}
\begin{equation}
\mathrm{SDP}^*=\min_Y\ \sum_iQ_{ii}Y_{0i}+\sum_{i<j}Q_{ij}Y_{ij}\quad\text{s.t.}\quad Y\succeq0,\ Y_{00}=1,\ Y_{ii}=Y_{0i},
\label{eq:app-shor}
\end{equation}
whose objective at $Y(q)$ is the expected energy $\mathbb{E}_q[E(x)]$. This view gives three statements about VQPM.

\paragraph{What the re-preparation discards.} The product state keeps the first row $Y_{0i}$ of $Y(q)$ and replaces each $Y_{ij}$, $i\ne j$, by $Y_{0i}Y_{0j}$. The discarded covariance $C_{ij}=Y_{ij}-Y_{0i}Y_{0j}$ changes the expected energy of the next input by $-\sum_{i<j}Q_{ij}C_{ij}$. In the two-qubit example, $q_1=(2,1,1,1)/5$ has $Y_{01}=Y_{02}=2/5$ and $Y_{12}=1/5$, so the covariance $C_{12}=1/25$ is lost; this is exactly why the coherent and the VQPM second rounds differ.

\paragraph{Locking.} For a product input with marginals $\pi$, fixing $x_i=s$ sets $Y_{0i}=s$ and $Y_{ij}=sY_{0j}$, the same affine constraint that fixing a variable imposes in Eq.~\eqref{eq:app-shor}. Its effect on the expected energy $\bar E(\pi)=\mathbb{E}_{p_\pi}[E(x)]=\sum_iQ_{ii}\pi_i+\sum_{i<j}Q_{ij}\pi_i\pi_j$ is exact and simple.

\begin{lemma}[Energy change under locking]
\label{lem:leakage}
Setting $\pi_i$ to $s\in\{0,1\}$ changes the expected energy of the product state by
\begin{equation}
\Delta\bar E=(s-\pi_i)\,g_i,\qquad g_i=\frac{\partial\bar E}{\partial\pi_i}=Q_{ii}+\sum_{j\ne i}Q_{ij}\pi_j ,
\end{equation}
where $Q_{ij}$ with $i>j$ stands for $Q_{ji}$. If the lock passes the threshold, $\min(\pi_i,1-\pi_i)\le(1-p_{\mathrm{diff}})/2$, and
\begin{equation}
|\Delta\bar E|\le\frac{1-p_{\mathrm{diff}}}{2}\Bigl(|Q_{ii}|+\sum_{j\ne i}|Q_{ij}|\Bigr).
\label{eq:app-leakage}
\end{equation}
\end{lemma}
\begin{proof}
$\bar E$ is affine in $\pi_i$ with slope $g_i$. Locking to the more probable value moves $\pi_i$ by $\min(\pi_i,1-\pi_i)$, and $|g_i|\le|Q_{ii}|+\sum_{j\ne i}|Q_{ij}|$ because $\pi_j\in[0,1]$.
\end{proof}

The right-hand side of Eq.~\eqref{eq:app-leakage} is proportional to the influence score of Eq.~\eqref{eq:influence}. This gives a reason for scaling $p_{\mathrm{diff}}$ by influence (Sec.~\ref{sec:influence}): the influence bounds how far a single lock can move the expected energy. In the runs of Table~\ref{tab:app-locking}, $\Delta\bar E<0$ for all 2\,809 locks: in the paper's settings locking always moved the input toward lower expected energy, so a wrong lock is a greedy step into the wrong basin rather than an increase in energy.

\paragraph{An a posteriori optimality gap.} Because every $Y(q)$ is feasible for Eq.~\eqref{eq:app-shor},
\begin{equation}
\mathrm{SDP}^*\le\min_xE(x)\le E(x_{\mathrm{VQPM}})
\end{equation}
for any string $x_{\mathrm{VQPM}}$ returned by VQPM. The semidefinite program has size $n+1$ and is solvable in polynomial time, so $E(x_{\mathrm{VQPM}})-\mathrm{SDP}^*$ certifies the quality of a VQPM output for sizes at which the optimum cannot be enumerated. Table~\ref{tab:app-sdp} shows how tight this certificate is on our instances: the relaxation is exact for 7\% to 33\% of them, and the certified gap of a VQPM output is about twice its true error (median $0.027$ against $0.013$ of the energy range at $n=16$).

\begin{table}[ht]
\centering
\small
\caption{Shor (degree-2 SoS) lower bound, Eq.~\eqref{eq:app-shor}, against the optimum $E^*$ and the energy $E_{\mathrm{VQPM}}$ of the VQPM output (paper settings), as fractions of the energy range $E_{\max}-E_{\min}$. Medians over 30 instances per $n$; the maximum is in brackets.}
\label{tab:app-sdp}
\begin{tabularx}{\textwidth}{r*{4}{>{\centering\arraybackslash}X}}
\toprule
$n$ & Relaxation gap $E^*-\mathrm{SDP}^*$ & Relaxation exact ($<10^{-4}$) & VQPM error $E_{\mathrm{VQPM}}-E^*$ & Certified gap $E_{\mathrm{VQPM}}-\mathrm{SDP}^*$ \\
\midrule
8  & 0.005 [0.066] & 33\% & 0.000 & 0.009 \\
12 & 0.009 [0.037] & 13\% & 0.012 & 0.019 \\
16 & 0.015 [0.043] & 7\%  & 0.013 & 0.027 \\
\bottomrule
\end{tabularx}
\end{table}

\begin{remark}[Hyperplane rounding]
For Max-Cut instances with nonnegative weights one could apply Goemans--Williamson hyperplane rounding~\cite{goemans1995improved} to $Y(q)$ and obtain an expected cut of at least $\alpha_{\mathrm{GW}}\approx0.878$ times $\mathbb{E}_q[\mathrm{cut}]$. Sampling $q$ directly already gives $\mathbb{E}_q[\mathrm{cut}]$, so this does not improve on measuring the state, and for QUBO instances with mixed signs, as here, no constant-factor guarantee of this kind holds. We therefore do not use rounding. Martinez, Fawzi and Stilck Fran\c{c}a apply Gaussian rounding of this kind to the two-qubit marginals of noisy quantum circuits for Max-Cut~\cite{martinez2026sampling}.
\end{remark}


\section{Classical sampling of VQPM}
\setcounter{equation}{0}
\renewcommand{\theequation}{\thesection\arabic{equation}}
\label{app:sampling}
For QUBO problems, everything VQPM measures can be generated by classical sampling. This appendix states this precisely (Appendix~\ref{app:classical}), compares the number of device shots with the number of classical draws (Appendix~\ref{app:shots}), uses the classical sampler to run VQPM up to $n=100$ (Appendix~\ref{app:large}), and derives the first-order form of the update and gives the details of the runs of Sec.~\ref{sec:update} (Appendices~\ref{app:flow} and~\ref{app:lockflow}). The notation is that of Appendix~\ref{app:theory}.

\subsection{Shot-by-shot simulation}
\label{app:classical}
\begin{proposition}[Shot-by-shot classical simulation]
\label{prop:classical}
Let $\mathcal{U}$ be diagonal and let the input have a computational-basis distribution $p$ that can be sampled classically, such as the product states of VQPM. Then one VQPM shot, the control bit $c$ together with all register bits $x$, has the distribution $\Pr(c=0,x)=p(x)w(x)$ and $\Pr(c=1,x)=p(x)(1-w(x))$. A classical algorithm samples it exactly: draw $x\sim p$ in $O(n)$ time, compute $E(x)$ and $\lambda(x)$ in $O(n^2)$, and output $c=0$ with probability $w(x)$. The same holds for $k$ coherent rounds: given $x$, the $k$ control outcomes are independent, each equal to $0$ with probability $w(x)$.
\end{proposition}
\begin{proof}
Each round multiplies the amplitude of $\ket{x}$ by $(1\pm e^{i\lambda(x)})/2$ for control outcome $0$ or $1$, whose squared moduli are $w(x)$ and $1-w(x)$. No amplitudes of different $x$ interfere, because $\mathcal{U}$ is diagonal and $x$ is measured in the computational basis.
\end{proof}

Consequently, VQPM for QUBO as defined here can be reproduced classically shot by shot, at $O(n^2)$ cost per shot and with the same number of shots, including its finite-shot locking decisions. It cannot by itself give a quantum speedup. An advantage would require a non-diagonal Hamiltonian, inputs whose distribution cannot be sampled classically, or coherent amplitude amplification of the post-selection in the $k$-round version, which can at most reduce the $1/Z_k$ repetitions of Eq.~\eqref{eq:app-survival} quadratically. We therefore present VQPM here as a heuristic whose convergence and locking behaviour we analyse, not as evidence of quantum advantage (Remark~\ref{rem:classical} and Sec.~\ref{sec:conclusion}).

\subsection{Number of shots: device and classical sampler}
\label{app:shots}
The sampler of Proposition~\ref{prop:classical} reproduces the shot statistics of the device exactly, so for any locking rule it needs at most as many draws as the device needs shots. The work per shot is of the same order. A device shot prepares the product state with $n$ $R_y$ gates, applies controlled-$\mathcal{U}$, which for a dense $Q$ takes one controlled phase gate per term of $E(x)$ ($n$ two-qubit and $n(n-1)/2$ three-qubit gates), and measures $n+1$ qubits. A classical draw takes $n+1$ random numbers and the $n(n+1)/2$ multiply-adds of $E(x)$.

The classical sampler can, however, use far fewer draws than the device, for two reasons. First, it knows $w(x)$, while a shot reveals only one Bernoulli bit of it, so it can drop the control bit and weight every draw by $w(x)$. Second, the input marginals $\pi_i$ are known exactly, since they set the $R_y$ angles, so $\mathbb{E}_p[x_i]=\pi_i$ can serve as a control variate. Together these give the estimate
\begin{equation}
\hat\pi_i'=\pi_i+\frac{\frac1N\sum_{s=1}^{N}\bigl(x^{(s)}_i-\pi_i\bigr)\bigl(w(x^{(s)})-\bar w\bigr)}{\bar w},\qquad \bar w=\frac1N\sum_{s=1}^{N}w(x^{(s)}),\quad x^{(s)}\sim p .
\label{eq:app-weighted}
\end{equation}
Write $a_i=x_i-\pi_i$ and $\Delta_i=\pi_i'-\pi_i$. To first order in $1/N$, the device's estimate from its accepted shots, as in the main text, has variance $\mathbb{E}_p[w\,(x_i-\pi_i')^2]/(NW^2)$. The device can lower this to $\mathbb{E}[(a_i(c-W)-\Delta_ic)^2]/(NW^2)$ by keeping the rejected shots and using the same control variate. Eq.~\eqref{eq:app-weighted} has variance $\mathbb{E}_p[(a_i(w-W)-\Delta_iw)^2]/(NW^2)$. The last is never larger than the device's control-variate estimate, because $c$ is $w(x)$ plus Bernoulli noise of variance $w(1-w)$, which adds $\mathbb{E}_p[(a_i-\Delta_i)^2w(1-w)]$. The size of the gap is set by the spread of $w$ under $p$, which is small because the phase map compresses all energies into a window narrower than $\pi/2$. Table~\ref{tab:app-shots} evaluates the three variances in every round of the paper-setting runs, for the worst unlocked qubit. Keeping the rejected shots and using the control variate saves the device a factor of about $10$. Eq.~\eqref{eq:app-weighted} needs at most $2\times10^{-3}$ of the device's shots in any round, and typically $10^{-5}$. Most of this gain comes from the control variate: weighting alone changes the variance only by a factor within $2\times10^{-3}$ of $W\approx0.9$. For the locking guarantee of Proposition~\ref{prop:threshold} with $p_{\mathrm{diff}}=0.01$, $\delta=0.05$ and $T=30$, the device needs $2.1$ to $2.2\times10^5$ shots per round with the post-selected estimate, or about $2\times10^4$ with the control variate, while the classical sampler reaches the same variance with fewer than $500$ draws per round. For comparison, rounding the marginals to three decimals, as in the main-text simulations, corresponds to a standard error of about $5\times10^{-4}$, which near $\pi_i'=1/2$ needs about $10^6$ post-selected shots per round on the device.

\begin{table}[ht]
\centering
\small
\caption{Shots per round on the paper-setting runs of Table~\ref{tab:app-locking} (100 instances per $n$, all rounds). Terms: number of terms of $E(x)$, which is the number of controlled phase gates per device shot and of multiply-adds per classical draw. Post-selected: device shots $M/W$ for the guarantee of Proposition~\ref{prop:threshold} ($p_{\mathrm{diff}}=0.01$, $\delta=0.05$, $T=30$, $M=2\ln(2nT/\delta)/p_{\mathrm{diff}}^2$), with the median $W$. The last two columns give the number of draws that reaches the same variance for the worst unlocked qubit, as a fraction of the post-selected count, median over rounds [largest]: the device keeping all shots with the control variate $\mathbb{E}_p[x_i]=\pi_i$, and the classical estimate of Eq.~\eqref{eq:app-weighted}.}
\label{tab:app-shots}
\begin{tabular}{rrcrcc}
\toprule
$n$ & terms & $W$, median [range] & post-selected & device, control variate & classical, Eq.~\eqref{eq:app-weighted} \\
\midrule
8 & 36 & 0.897 [0.797, 0.972] & $2.05\times10^5$ & 0.10 [0.20] & $3\times10^{-5}$ [$2.1\times10^{-3}$] \\
12 & 78 & 0.901 [0.817, 0.946] & $2.13\times10^5$ & 0.10 [0.18] & $5\times10^{-6}$ [$7.1\times10^{-4}$] \\
16 & 136 & 0.892 [0.829, 0.934] & $2.21\times10^5$ & 0.11 [0.17] & $7\times10^{-6}$ [$3.6\times10^{-4}$] \\
\bottomrule
\end{tabular}
\end{table}

For $k$ coherent rounds, weighting each draw by $w(x)^k$ instead of post-selecting already changes the variance by a factor of at most $\max_xw(x)^k=\cos^{2k}(\lambda_d/2)$, since $w^{2k}\le w^k\max_xw^k$. Weighting thus removes the eigenvalue factor of Eq.~\eqref{eq:app-survival}, which the device pays in repetitions. It does not remove the cost of finding the optimum by sampling from $p_0$: the effective sample size $NZ_k^2/Z_{2k}$ of the weighted sampler tends to $NA$ for large $k$, so from the uniform start it still needs about $2^n/|X^*|$ draws.

\subsection{Larger instances}
\label{app:large}
The estimate of Eq.~\eqref{eq:app-weighted} needs no $2^n$ state vector, so the VQPM loop can be run classically far beyond the $n\le20$ of the main text. With $N=2\times10^4$ draws per round its marginals agree with exact enumeration at $n=20$ to within $1.1\times10^{-4}$, well below the rounding unit $5\times10^{-4}$ of the main-text simulations. We ran 100 instances for each $n=20,30,\dots,100$ from the same family as above ($Q=G+G^T$, $G_{ij}\sim\mathcal N(0,1)$, seed $2026+n$), with the phase map of Eq.~\eqref{eq:phasemap} and the following stop rule: stop when the reference string $x_{\mathrm{ref}}$ reaches probability $1/2$ after the filter, or when a lock contradicts it. At $n=20$, $x_{\mathrm{ref}}$ is the exhaustive optimum. For $n\ge30$ it is the best of 32 simulated-annealing chains, each followed by 1-flip descent. All 32 chains returned the same energy in 100\% of the instances for $n\le30$, in 93 to 99\% for $n=40$ to $80$, and in 83 and 88\% for $n=90$ and $100$; at $n=20$ annealing found the exhaustive optimum every time.

\begin{figure}[ht]
\centering
\includegraphics[width=\linewidth]{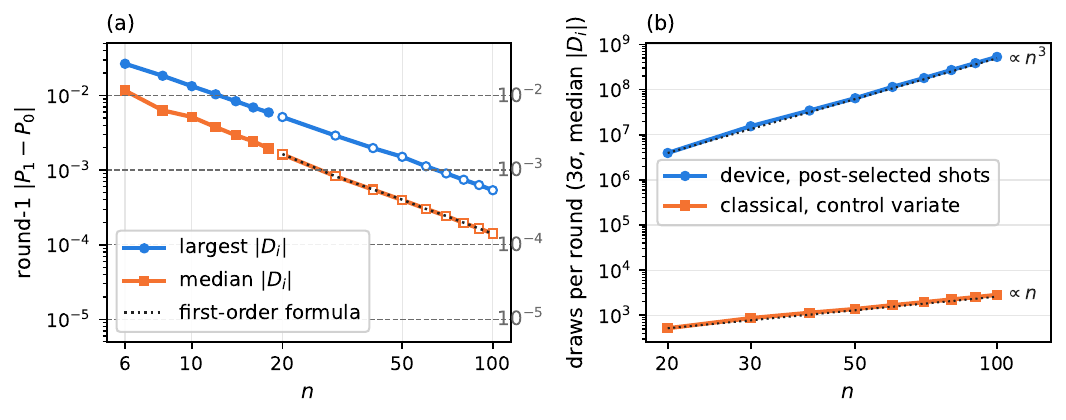}
\caption{(a) Round-1 $|P_1-P_0|$ from the uniform start: largest and median over the qubits, median over 100 instances per $n$. Filled markers: exact enumeration ($n\le18$); open markers: sampled with Eq.~\eqref{eq:app-weighted} ($n\ge20$); dotted: Eq.~\eqref{eq:app-round1}. Dashed lines: thresholds $10^{-2}$ to $10^{-5}$; $10^{-3}$ is also the smallest difference that survives rounding to three decimals. (b) Draws per round that resolve the median $|D_i|$ at three standard deviations, for the device's post-selected estimate and the classical estimate of Eq.~\eqref{eq:app-weighted}, with reference slopes $n^3$ and $n$. The threshold sweep without rounding is in Fig.~\ref{fig:app-flow}.}
\label{fig:app-large}
\end{figure}

\paragraph{The first round.} From the uniform start, expanding $w$ to first order around the phase $\bar\lambda$ of the mean energy, $w\approx w(\bar\lambda)-\tfrac12\sin\bar\lambda\,(\lambda-\bar\lambda)$, and using $\mathrm{Cov}(x_i,E)=g_i/4$ under the uniform distribution gives
\begin{equation}
D_i:=P_1-P_0\;\approx\;-\frac{\pi\sin\bar\lambda}{16\,S\cos^2(\bar\lambda/2)}\,g_i,\qquad g_i=Q_{ii}+\tfrac12\sum_{j\ne i}Q_{ij},
\label{eq:app-round1}
\end{equation}
where $g_i=\partial\bar E/\partial\pi_i$ at $\pi=1/2$ is the mean-field gradient of Appendix~\ref{app:sos}. Eq.~\eqref{eq:app-round1} is the first round of the first-order map of Appendix~\ref{app:flow}. The first round therefore moves every qubit against its mean-field gradient, by an amount proportional to $g_i/S$. For these instances $g_i$ has standard deviation $\sqrt{(n+7)/2}$ and $S\approx0.56\,n^2$, so $|D_i|$ shrinks like $n^{-3/2}$. Table~\ref{tab:app-round1} and Fig.~\ref{fig:app-large}(a) confirm this: the median $|D_i|$ falls from $1.6\times10^{-3}$ at $n=20$ to $1.4\times10^{-4}$ at $n=100$, with fitted exponents $-1.50$ for the median and $-1.40$ for the largest $|D_i|$. Eq.~\eqref{eq:app-round1} is within 3\% of the sampled values and has the same sign for 99\% of the qubits. At every $n$, about 78\% of the round-1 signs agree with $x_{\mathrm{ref}}$, while the qubit with the largest $|D_i|$ agrees in 98 to 100\% of the instances.

\paragraph{Shots needed to see the first round.} A device estimates $D_i$ from post-selected shots with standard deviation $1/\sqrt{NW}$, so resolving the median $|D_i|$ at three standard deviations takes $N=9/(WD^2)$ shots. This grows like $n^3$: from $3.9\times10^6$ at $n=20$ to $5.3\times10^8$ at $n=100$, with fitted exponent $3.01$ (Table~\ref{tab:app-round1}, Fig.~\ref{fig:app-large}(b)). The shots can be spread over several rounds, but not saved: over $t$ rounds the drift adds up to about $tD$ and the noise only to $\sqrt t$ times its one-round size, so the total is still about $9/(WD^2)$. The classical estimate of Eq.~\eqref{eq:app-weighted} has standard deviation about $\sigma_w/(W\sqrt N)$, where $\sigma_w$ is the spread of $w$ under the input distribution, and needs $9\sigma_w^2/(W^2D^2)$ draws. Since $\sigma_w$ falls like $1/n$ (fitted exponent $-0.99$), this grows only like $n$: from $5.2\times10^2$ to $2.9\times10^3$ (exponent $1.03$).

\begin{table}[ht]
\centering
\small
\caption{Round 1 from the uniform start, 100 instances per $n$, $2\times10^5$ draws each with Eq.~\eqref{eq:app-weighted}. Median and largest: median over instances of the per-instance median and largest $|D_i|$, $D_i=P_1-P_0$. Erased: qubits with $|D_i|<10^{-3}$, whose marginal rounds back to $0.500$ at three decimals; no qubit had $|D_i|\ge0.01$ at any $n\ge20$. Agrees: share of the signs of $D_i$ that agree with $x_{\mathrm{ref}}$. Device and classical: draws per round that resolve the median $|D_i|$ at three standard deviations, $9/(WD^2)$ post-selected shots and $9\sigma_w^2/(W^2D^2)$ draws of Eq.~\eqref{eq:app-weighted}.}
\label{tab:app-round1}
\begin{tabular}{rccccrr}
\toprule
$n$ & median $|D_i|$ & largest $|D_i|$ & erased & agrees & device & classical \\
\midrule
20 & $1.6\times10^{-3}$ & $5.1\times10^{-3}$ & 32\% & 78\% & $3.9\times10^6$ & 520 \\
30 & $8.3\times10^{-4}$ & $2.9\times10^{-3}$ & 57\% & 78\% & $1.5\times10^7$ & 870 \\
40 & $5.5\times10^{-4}$ & $2.0\times10^{-3}$ & 76\% & 79\% & $3.4\times10^7$ & 1140 \\
50 & $4.0\times10^{-4}$ & $1.5\times10^{-3}$ & 90\% & 78\% & $6.4\times10^7$ & 1380 \\
60 & $3.0\times10^{-4}$ & $1.1\times10^{-3}$ & 98\% & 78\% & $1.2\times10^8$ & 1680 \\
70 & $2.4\times10^{-4}$ & $9.0\times10^{-4}$ & 100\% & 79\% & $1.8\times10^8$ & 1960 \\
80 & $2.0\times10^{-4}$ & $7.4\times10^{-4}$ & 100\% & 78\% & $2.7\times10^8$ & 2250 \\
90 & $1.7\times10^{-4}$ & $6.3\times10^{-4}$ & 100\% & 78\% & $3.9\times10^8$ & 2530 \\
100 & $1.4\times10^{-4}$ & $5.4\times10^{-4}$ & 100\% & 78\% & $5.3\times10^8$ & 2850 \\
\bottomrule
\end{tabular}
\end{table}

\paragraph{Paper settings and a finite-shot device.} We first ran the loop with $p_{\mathrm{diff}}=0.01$ and at most 30 rounds, as in the main text, in two versions (Table~\ref{tab:app-large}).
\begin{itemize}
\item \emph{Marginals rounded to three decimals}, as in the main-text simulations. Rounding erases most of the round-1 signal, and the rounded marginals then repeat unchanged, so the run stalls. All runs stall for $n\ge60$. The share that stalls in round 1 with nothing locked rises from 27\% at $n=60$ to 74\% at $n=70$ and 99 to 100\% for $n\ge80$. Only 11\% of the runs reach $x_{\mathrm{ref}}$ at $n=20$, and none do for $n\ge30$. If a run that stalls with $x_{\mathrm{ref}}$ as its most probable string is also counted as a success, as in Sec.~\ref{sec:comparison}, the share is 26\% at $n=20$, 3\% at $n=30$ and at most 1\% for $n\ge40$.
\item \emph{A device with $2\times10^5$ shots per round} (post-selected estimate, three decimals). The shot noise in $\hat D_i$ has standard deviation $1/\sqrt{MW}\approx2.4\times10^{-3}$, which exceeds the largest round-1 difference for $n\ge40$. The locks then follow the noise: 75\% of the runs end in a wrong lock at $n=20$, 98\% at $n=30$ and all of them for $n\ge40$. In 33\% (at $n=20$) to 93\% (at $n=100$) of the runs, the same loop without rounding or shot noise made no wrong lock in the same 30 rounds.
\end{itemize}

\begin{table}[ht]
\centering
\small
\caption{VQPM with the main-text settings ($p_{\mathrm{diff}}=0.01$, at most 30 rounds), 100 instances per $n$. Reached: runs in which $x_{\mathrm{ref}}$ reaches probability $1/2$. Wrong lock: runs stopped by a lock that contradicts $x_{\mathrm{ref}}$. Stalled: rounded marginals repeat unchanged.}
\label{tab:app-large}
\begin{tabular}{r ccc cc}
\toprule
 & \multicolumn{3}{c}{three decimals} & \multicolumn{2}{c}{device, $2\times10^5$ shots} \\
\cmidrule(lr){2-4}\cmidrule(lr){5-6}
$n$ & reached & wrong lock & stalled & reached & wrong lock \\
\midrule
20 & 11\% & 39\% & 50\% & 18\% & 75\% \\
30 & 0\% & 53\% & 42\% & 1\% & 98\% \\
40 & 0\% & 20\% & 70\% & 0\% & 100\% \\
50 & 0\% & 2\% & 94\% & 0\% & 100\% \\
60 & 0\% & 0\% & 100\% & 0\% & 100\% \\
70 & 0\% & 0\% & 100\% & 0\% & 100\% \\
80 & 0\% & 0\% & 100\% & 0\% & 100\% \\
90 & 0\% & 0\% & 100\% & 0\% & 100\% \\
100 & 0\% & 0\% & 100\% & 0\% & 100\% \\
\bottomrule
\end{tabular}
\end{table}

\subsection{The product-state map without locking}
\label{app:flow}
Every VQPM run mixes two mechanisms: the product-state update of Eq.~\eqref{eq:app-vqpm} and locking. This section studies the update alone, with exact marginals and without locking or rounding; Appendix~\ref{app:lockflow} then adds locking. For $n\ge30$ the exact map cannot be iterated, but the phases of the strings that carry the probability lie close together, and to first order in their spread the map has a closed form.

\begin{proposition}[First-order form of the product-state map]
\label{prop:meanfield}
Let $J$ be the coupling matrix, $J_{ij}=J_{ji}=Q_{ij}$ for $i<j$ and $J_{ii}=0$. Let $\bar E(\pi)=\mathbb{E}_{p_\pi}[E]=\sum_iQ_{ii}\pi_i+\frac12\pi^TJ\pi$ be the mean energy of the product state, $g_i(\pi)=\partial\bar E/\partial\pi_i=Q_{ii}+\sum_jJ_{ij}\pi_j$ its gradient, and $\bar\lambda=\frac\pi4+\frac\pi4\bar E(\pi)/S$ the phase of the mean energy. To first order in the spread of the phases under $p_\pi$, Eq.~\eqref{eq:app-vqpm} is
\begin{equation}
\pi_i'=\pi_i-\eta(\pi)\,\pi_i(1-\pi_i)\,g_i(\pi),\qquad\eta(\pi)=\frac{\pi\sin\bar\lambda}{8S\cos^2(\bar\lambda/2)}.
\label{eq:app-mf}
\end{equation}
\end{proposition}
\begin{proof}
Let $\bar w=\cos^2(\bar\lambda/2)$. Since $\lambda(x)-\bar\lambda=\frac{\pi}{4S}(E(x)-\bar E)$, to first order $w(x)=\bar w-\frac{\pi\sin\bar\lambda}{8S}(E(x)-\bar E)$. Then $W=\bar w$, because $\mathbb{E}_{p_\pi}[E-\bar E]=0$, and $\sum_{x:\,x_i=1}p_\pi(x)w(x)=\pi_i\bigl[\bar w-\frac{\pi\sin\bar\lambda}{8S}(\mathbb{E}[E\mid x_i=1]-\bar E)\bigr]$. As $\bar E$ is affine in each $\pi_i$, $\mathbb{E}[E\mid x_i=1]$ is $\bar E$ evaluated at $\pi_i=1$, so $\mathbb{E}[E\mid x_i=1]-\bar E=(1-\pi_i)\,g_i$. Dividing by $W$ gives Eq.~\eqref{eq:app-mf}.
\end{proof}

The neglected terms are smaller than the kept ones by a factor of order the spread of the phases, $\frac{\pi}{4S}\,\mathrm{sd}_{p_\pi}(E)$. On our instances this is about $0.6/n$ at the uniform start, and it falls as the marginals leave $1/2$. Iterated from the uniform start until the marginals settle, Eq.~\eqref{eq:app-mf} ends at the same string as the exact map in 59 of 60 instances at $n=8$ and $12$, after the same median number of rounds to within 4\%. Three consequences follow.
\begin{itemize}
\item \emph{A gradient method.} In the logits $\theta_i=\ln[\pi_i/(1-\pi_i)]$, Eq.~\eqref{eq:app-mf} reads $\theta_i'=\theta_i-\eta\,g_i+O(\eta^2g_i^2)$, the gradient method of Sec.~\ref{sec:gradient}. Proposition~\ref{prop:monotone} is its exact counterpart: $W\approx\cos^2(\bar\lambda/2)$ grows as $\bar E$ falls.
\item \emph{Step size.} At $\bar\lambda\approx\pi/4$, $\eta\approx0.33/S$, and $S\approx0.56\,n^2$ on our instances. From the uniform start, round 1 gives $D_i=2\pi_i'-1=-\eta g_i/2$, which is Eq.~\eqref{eq:app-round1}.
\item \emph{End points.} At a vertex $x$, $(1-2x_i)\,g_i(x)=E(x\oplus e_i)-E(x)$, so a small deviation of coordinate $i$ is multiplied by $1-\eta\,[E(x\oplus e_i)-E(x)]$ per round. This is the first-order form of $r_i$ in Proposition~\ref{prop:stability}, and the stable vertices are again the strict 1-flip local minima. The Hessian of $\bar E$ is $J$, which has zero trace, so every interior stationary point is a saddle and the iteration ends at a vertex. We call this vertex the end point $x_\infty$.
\end{itemize}

\paragraph{Near the centre: from $D_i$ to the couplings.} Write $\pi=\frac12+u$, $g^0=g(\frac12)$ and $\eta_0=\eta(\frac12)$. To first order in $u$, Eq.~\eqref{eq:app-mf} is linear, $u'=u-\frac{\eta_0}{4}(g^0+Ju)$, so from $u=0$
\begin{equation}
u(t)=-\bigl(I-B^t\bigr)J^{-1}g^0=-\sum_k\frac{1-a_k^t}{\mu_k}\,(v_k^Tg^0)\,v_k,\qquad B=I-\frac{\eta_0}{4}J,\quad a_k=1-\frac{\eta_0\mu_k}{4},
\label{eq:app-linear}
\end{equation}
where $\mu_k$ and $v_k$ are the eigenvalues and unit eigenvectors of $J$. With $\tau=4/(\eta_0|\mu_{\min}|)$, the marginals pass through two regimes.
\begin{itemize}
\item For $t\ll\tau$, $(1-a_k^t)/\mu_k\approx\eta_0t/4$ and $u(t)\approx-\frac{\eta_0t}{4}\,g^0=\frac t2\,D$. Every qubit keeps moving in the direction of its round-1 difference $D_i$, the sign of its gradient at the centre.
\item For $t\gtrsim\tau$, the components along eigenvectors with $\mu_k>0$ level off at $-(v_k^Tg^0)/\mu_k$, while those with $\mu_k<0$ grow like $(1+\eta_0|\mu_k|/4)^t$. The marginals turn toward the eigenvectors of $J$ with the most negative eigenvalues, with signs set by $g^0$.
\end{itemize}
This is the power iteration of the product state described in Sec.~\ref{sec:couplings}, with the linearized iteration matrix $B$. On our instances $J$ is a Wigner matrix with entry variance 2, so $\mu_{\min}\approx-2\sqrt{2n}$ and $\tau\approx2.5\,n^{3/2}$ for large $n$. The measured median of $\tau$ grows from 275 rounds at $n=20$ to 2\,610 at $n=100$, or $3.1$ to $2.6\,n^{3/2}$ (Table~\ref{tab:app-flow}). The linear regime ends when the $|u_i|$ approach $\frac12$; the factor $\pi_i(1-\pi_i)$ then slows the coordinates near 0 and 1, and the map settles at its end point.

\paragraph{Where the unlocked map goes.} Table~\ref{tab:app-flow} gives the end points of the unlocked runs of Sec.~\ref{sec:unlocked}, with the overlap $m(t)$ of Eq.~\eqref{eq:overlap}. Every run that settled ended at a strict 1-flip local minimum, and no run ended below $x_{\mathrm{ref}}$. The runs that miss level off at an overlap of about $0.7$ to $0.85$. The median run settles after 54 to 98 times $S$ rounds, and $x_{\mathrm{ref}}$ first reaches probability $1/2$ after $2.0\times10^3$ rounds at $n=20$ and $1.6\times10^5$ at $n=100$. Between 1 and 20 runs per $n$ were still moving after $3\times10^6$ rounds, close to a saddle; for them the end point is the nearest vertex.

\begin{table}[ht]
\centering
\small
\caption{The unlocked product-state map, Eq.~\eqref{eq:app-mf}, from the uniform start until every marginal is within $10^{-3}$ of a stable vertex (at most $3\times10^6$ rounds), 100 instances per $n$. $\tau=4/(\eta_0|\mu_{\min}(J)|)$ of Eq.~\eqref{eq:app-linear}, median. Misses: runs that end elsewhere than $x_{\mathrm{ref}}$, with their median distance in bits and median energy gap $(E-E_{\mathrm{ref}})/|E_{\mathrm{ref}}|$. Settle: median rounds until the run settles. $\Pr(x_{\mathrm{ref}})\ge\frac12$: median round in which $x_{\mathrm{ref}}$ first reaches probability $1/2$ after the filter, over the runs in which it does. Still moving: runs not settled after $3\times10^6$ rounds.}
\label{tab:app-flow}
\begin{tabular}{rrcccrrr}
\toprule
 & & & \multicolumn{2}{c}{misses} & & & \\
\cmidrule(lr){4-5}
$n$ & $\tau$ & ends at $x_{\mathrm{ref}}$ & bits & gap & settle & $\Pr(x_{\mathrm{ref}})\ge\frac12$ & still moving \\
\midrule
20 & 275 & 73\% & 3 & 1.3\% & 13\,714 & 1\,982 & 1 \\
30 & 475 & 58\% & 4 & 1.3\% & 44\,762 & 6\,572 & 3 \\
40 & 704 & 53\% & 4 & 0.9\% & 60\,400 & 11\,937 & 6 \\
50 & 966 & 48\% & 5 & 0.8\% & 106\,826 & 19\,790 & 5 \\
60 & 1\,250 & 41\% & 5 & 0.9\% & 165\,967 & 28\,992 & 6 \\
70 & 1\,557 & 34\% & 6 & 0.7\% & 182\,482 & 46\,472 & 8 \\
80 & 1\,894 & 38\% & 7 & 0.7\% & 275\,060 & 69\,658 & 15 \\
90 & 2\,256 & 19\% & 7 & 0.6\% & 359\,390 & 170\,479 & 20 \\
100 & 2\,610 & 24\% & 7.5 & 0.6\% & 563\,404 & 155\,226 & 20 \\
\bottomrule
\end{tabular}
\end{table}

\subsection{What locking adds}
\label{app:lockflow}
This section gives the details of the locked runs of Sec.~\ref{sec:locksigns}. Each run continues until every qubit is locked or the free ones have settled at a vertex that is stable for them.

\paragraph{The sign that a lock reads.} Table~\ref{tab:app-read} follows the unlocked map and records, for each threshold, the sign of every qubit in the round in which it first passes the threshold; Fig.~\ref{fig:app-flow}(c) plots the same shares. At $n=100$ the thresholds are passed in round 1 from $10^{-4}$ down, and after a median of 8 rounds at $10^{-3}$, 72 rounds at $10^{-2}$ and 1\,390 rounds at $0.2$, about $0.003\tau$, $0.03\tau$ and $0.5\tau$. Eq.~\eqref{eq:app-linear} predicts 97 to 100\% of all signs read and 94\% of those that differ from $D_i$, so the turn is the one toward the low eigenvectors of $J$. From $10^{-3}$ to $0.2$, the share of signs that agree with the unlocked end point rises from 81 to 94\%, and the share that agree with $x_{\mathrm{ref}}$ from 79 to 90\%: a larger threshold reads the signs later and closer to where the unlocked map ends.

\begin{table}[ht]
\centering
\small
\caption{Signs that locking reads, on the unlocked map at $n=100$ (100 instances). For each threshold: the sign of $2\pi_i-1$ in the round in which $|2\pi_i-1|$ first exceeds it, compared with the sign of the round-1 difference $D_i$, with the sign of the linear prediction $u_i(t)$ of Eq.~\eqref{eq:app-linear} in the same round, with the unlocked end point $x_\infty$ and with $x_{\mathrm{ref}}$, as shares of all qubits. Round: median over instances of the per-instance median round, in units of $\tau$ (median $\tau=2\,610$). At $n=20$ to $90$ the shares are within about 2 points of these.}
\label{tab:app-read}
\begin{tabular}{lccccc}
\toprule
 & & \multicolumn{4}{c}{sign read agrees with} \\
\cmidrule(lr){3-6}
$p_{\mathrm{diff}}$ & round$/\tau$ & $D_i$ & $u_i(t)$ & $x_\infty$ & $x_{\mathrm{ref}}$ \\
\midrule
$10^{-6}$ & round 1 & 100\% & 100\% & 80\% & 78\% \\
$10^{-4}$ & round 1 & 99.5\% & 100\% & 80\% & 78\% \\
$10^{-3}$ & 0.003 & 98\% & 100\% & 81\% & 79\% \\
$10^{-2}$ & 0.03 & 95\% & 99.8\% & 84\% & 82\% \\
$0.05$ & 0.14 & 90\% & 99\% & 89\% & 86\% \\
$0.1$ & 0.27 & 87\% & 98\% & 91\% & 88\% \\
$0.15$ & 0.41 & 85\% & 98\% & 93\% & 89\% \\
$0.2$ & 0.54 & 84\% & 97\% & 94\% & 90\% \\
\bottomrule
\end{tabular}
\end{table}

\paragraph{Success and rounds.} Table~\ref{tab:app-lockflow} and Fig.~\ref{fig:app-flow}(b, d) give the outcome of the locked runs. Below $10^{-2}$ the share of runs that reach $x_{\mathrm{ref}}$ drops quickly: at $n=20$ it is 15\% at $10^{-3}$ and 2\% at $10^{-4}$. The thresholds $0.05$ to $0.2$ gain at most 7 points over $10^{-2}$. At $p_{\mathrm{diff}}=0.2$, of the 1.1 wrong locks per run at $n=20$ and 7.8 at $n=100$, 0.7 and 3.4 also disagree with the unlocked end point $x_\infty$; of the instances in which the unlocked map ends at $x_{\mathrm{ref}}$, the run with $p_{\mathrm{diff}}=0.2$ still passes in 77\% at $n=20$ and 12\% at $n=100$. For $p_{\mathrm{diff}}\ge10^{-2}$ the number of rounds grows like $n^{1.9}$ to $n^{2.0}$, as $S$ does.

\paragraph{Check against sampled VQPM.} Table~\ref{tab:app-sweep} repeats the locked runs with sampled marginals: exact at $n=20$, and estimated with Eq.~\eqref{eq:app-weighted} for $n\ge30$, with $5\times10^4$ draws per round for $p_{\mathrm{diff}}\ge10^{-3}$ and $2\times10^5$ below. These runs use the stop rule of Appendix~\ref{app:large} and a budget of $\lceil1.5\,n^{3/2}\rceil$ rounds, which ran out in at most 1\% of them. On whether a run reaches $x_{\mathrm{ref}}$, sampled VQPM and Eq.~\eqref{eq:app-mf} agree on 99.0\% of the instances at $p_{\mathrm{diff}}=10^{-2}$, 98.3\% at $10^{-3}$, and all of them at $10^{-4}$ and $10^{-5}$. The table also lists the share of qubits that lock in round 1: 63 to 97\% at $10^{-4}$, and all of them from $10^{-6}$ down. For $n\ge30$ the sampled $\hat D_i$ has a standard deviation of about $2\times10^{-5}$ at $n=30$ and $6\times10^{-6}$ at $n=100$, so thresholds of $10^{-5}$ and below lie at or under the sampling noise there. This does not change the outcome, which is the same at $n=20$, where the marginals are exact, and in the first-order map, which has no noise.

\begin{table}[ht]
\centering
\small
\setlength{\tabcolsep}{4pt}
\caption{Sampled VQPM without rounding for a range of locking thresholds (exact marginals at $n=20$, Eq.~\eqref{eq:app-weighted} for $n\ge30$), 100 instances per $n$, at most $\lceil1.5\,n^{3/2}\rceil$ rounds, stop rule of Appendix~\ref{app:large} with $x_{\mathrm{ref}}$. It checks the first-order runs of Table~\ref{tab:app-lockflow}. Reached: runs in which $x_{\mathrm{ref}}$ reaches probability $1/2$; all other runs end in a wrong lock, except at most 1\% that use up the budget. Locked in round 1: share of the qubits locked in the first round.}
\label{tab:app-sweep}
\begin{tabular}{r cccc cccc}
\toprule
 & \multicolumn{4}{c}{reached $x_{\mathrm{ref}}$, $p_{\mathrm{diff}}=$} & \multicolumn{4}{c}{locked in round 1, $p_{\mathrm{diff}}=$} \\
\cmidrule(lr){2-5}\cmidrule(lr){6-9}
$n$ & $10^{-2}$ & $10^{-3}$ & $10^{-4}$ & $\le10^{-5}$ & $10^{-3}$ & $10^{-4}$ & $10^{-5}$ & $\le10^{-6}$ \\
\midrule
20 & 53\% & 15\% & 2\% & 0\% & 67\% & 97\% & 100\% & 100\% \\
30 & 29\% & 13\% & 0\% & 0\% & 43\% & 93\% & 99\% & 100\% \\
40 & 25\% & 13\% & 0\% & 0\% & 24\% & 91\% & 99\% & 100\% \\
50 & 19\% & 12\% & 0\% & 0\% & 10\% & 86\% & 99\% & 100\% \\
60 & 14\% & 11\% & 0\% & 0\% & 2\% & 83\% & 98\% & 100\% \\
70 & 5\% & 4\% & 0\% & 0\% & 0\% & 78\% & 97\% & 100\% \\
80 & 8\% & 6\% & 0\% & 0\% & 0\% & 73\% & 97\% & 100\% \\
90 & 3\% & 2\% & 0\% & 0\% & 0\% & 69\% & 97\% & 100\% \\
100 & 3\% & 3\% & 0\% & 0\% & 0\% & 63\% & 96\% & 100\% \\
\bottomrule
\end{tabular}
\end{table}

\end{document}